\documentclass[review,authoryear]{elsarticle} 

\usepackage{lineno} 
\modulolinenumbers[5]
\journal{Neural Networks}
\usepackage{fancyhdr}   

\usepackage{graphicx} 
\usepackage{booktabs} 
\usepackage{boldline}    
\usepackage{amsmath}  
\usepackage{amssymb}
\usepackage{amsthm}
\usepackage{bm}           

\usepackage{tikz}

\usetikzlibrary{backgrounds}
\usetikzlibrary{topaths}    
\usetikzlibrary{shapes}

\usetikzlibrary{calc}

\usetikzlibrary{arrows.meta, decorations.markings}
\usetikzlibrary{shapes,decorations}

\usepackage{setspace}
\usepackage[utf8]{inputenc}

\usepackage{tabularx}

\usepackage{threeparttable} 

\usepackage{xcolor,colortbl} 

\usepackage{layouts}     

\usepackage{subcaption} 

\usepackage{comment} 

\usepackage{float} 

\usepackage[font=normal]{caption}

\usepackage{adjustbox}

\usepackage{enumitem} 

\usepackage{pdfsync}

\newcommand{\beqn}{\begin{equation}}
\newcommand{\eeqn}{\end{equation}}

\newcommand{\xvec}{{\textbf x}}

\newcommand{\Xvec}{{\textbf X}}

\newcommand{\one}{{\mathbf {1}}}

\newcommand{\wvec}{{\textbf w}}

\newcommand{\ebar}{{\bar e}}

\newcommand{\ra}[1]{\renewcommand{\arraystretch}{#1}} 

\usepackage[short,12hr]{datetime}  
\mdyyyydate
\usdate

\usepackage[margin=1in]{geometry}

\newcommand{\noimage}{%
  \setlength{\fboxsep}{-\fboxrule}%
  \fbox{\phantom{\rule{10pt}{10pt}}File missing\phantom{\rule{10pt}{10pt}}}
}
\let\includegraphicsoriginal\includegraphics
\renewcommand{\includegraphics}[2][width=\textwidth]{\IfFileExists{#2}{\includegraphicsoriginal[#1]{#2}}{\noimage}}

\usepackage{array}
\usepackage{multirow}
\newcommand\MyBox[2]{ 
  \fbox{\lower0.75cm
    \vbox to 1.7cm{\vfil
      \hbox to 1.7cm{\hfil\parbox{1.4cm}{\centering #1\\#2}\hfil}
      \vfil}%
  }%
}

\begin{document}

\bibliographystyle{elsarticle-harv}

\pagestyle{fancy}
\fancyhf{}  
\lhead{\footnotesize Levy \& Baxter}  
\rhead{\scriptsize Submitted to Neural Networks  \hspace{0.01in} July 20, 2022 } 
\cfoot{\thepage}
\renewcommand{\headrulewidth}{0pt}  
\renewcommand{\footrulewidth}{0pt}    

\begin{frontmatter}

\title{
Growing dendrites enhance a neuron's computational power and memory capacity
      }

\author[uvaadr]{William B Levy\fnref{WLfn}\corref{ca}}
\cortext[ca]{Corresponding author. UVa Health Sciences Center, Charlottesville, VA, 22908. Email: wbl@virginia.edu}
\author[uvaadr]{Robert A. Baxter\fnref{RBfn}}

\address[uvaadr]{Department of Neurosurgery, University of Virginia School of Medicine, Charlottesville, VA 22908}
\address[RBfn]{Baxter Adaptive Systems, Bedford, MA 01730}
\address[WLfn]{Informed Simplifications, Earlysville, VA 22936}

\begin{abstract}
Neocortical pyramidal neurons have many dendrites, and such dendrites are capable of, in isolation of one-another, generating a neuronal spike. It is also now understood that there is a large amount of dendritic growth during the first years of a humans life, arguably a period of prodigious learning.  These observations inspire the construction of a local, stochastic algorithm based on an earlier stochastic, Hebbian developmental theory.  Here we investigate the neuro-computational advantages and limits on this novel algorithm that combines dendritogenesis  with supervised adaptive synaptogenesis. Neurons created with this algorithm have enhanced memory capacity, can avoid catastrophic interference (forgetting), and have the ability to unmix mixture distributions. In particular, individual dendrites develop within each class, in an unsupervised manner, to become feature-clusters that correspond to  the mixing elements of class-conditional  mixture distribution.  Although discriminative problems are used to understand the capabilities of the stochastic algorithm and the neuronal connectivity it produces, the algorithm is in the generative class, it thus seems ideal for decisions that require generalization, i.e., extrapolation beyond previous learning.

\end{abstract}


\end{frontmatter}

\begin{large}
\newenvironment{block}{\noindent}{\par\noindent}

\newpage

\section{Introduction}

The energy efficiency of brain function must heed both the costs of computation and long-distance communication. In fact, axon-based communication costs far outweigh computational costs \citep{Levy2021}. Then, when   increasing computational power by increasing the number of neurons in a network, the energy operating costs will increase largely due to the additional long distance communication accrued by axonal and presynaptic functions.  Indeed, there will not only be the additional costs for these new axons and synapses, but as neurons and their synapses, dendrites and axons proliferate, at a certain point  the average distance between communicating neurons will increase adding further to long-distance communication costs.  That is, the operating costs of enhancing computation via more neurons leads to  super-additive cost increases of communication (see PNAS supplement in Levy and Calvert, 2021). Perhaps Nature has found a more energy-efficient way to increase computational power.

Inspired by (i) recent evidence that perisomatic dendrites can produce distinct individual spikes on a single neuron 
\citep{Schiller2000,Jadi2012}, (ii) the long ago discovered \citep{White1990} and more generally argued idea that local dendritic regions
\citep{Levy1990,Colbert1993}, and as well individual dendrites \citep{Mel1994,Mel1999,Poirazi2003,Polsky2004,Behabadi2014} should be considered computational units, and (iii) the large amount of dendritogenesis and growth that occurs in early postnatal human development \citep{Petanjek2008,Petanjek2011}, the present paper develops a stochastic algorithm based on dendrite creation and growth (dendritogenesis). Compared to  the neurogenic approach to enhance computation, one can expect a much reduced increase of long-distance communications costs incurred by this alternative enhancement of  computational abilities.

Across the life-span, and especially at younger stages, there is a need, and in a certain sense opportunity  for greater computational power and memory. That is, as an organism develops, it experiences progressively more complexity. Co-occurring with this increasing complexity, the neocortex develops its  connectivity. Here our concern begins  after  pyramidal neuron  neurogenesis halts. From 7.5 months of gestation up to two years post-natal, there is tremendous dendritic growth accompanied by synapse formation \citep{Petanjek2011,Petanjek2019}. 

Further influencing the work here are issues arising in bio-, and not so bio-inspired machine learning. Grossberg \citep{Grossberg1976,Carpenter1987,Carpenter1991} and others \citep{McCloskey1989,French1991,French1999,McClelland1995} have long recognized the issue of catastrophic interference, and recently several groups have successfully addressed the catastrophic interference problem.
\citet{Delange2021} has categorized the existing approaches into three distinguishable methods: (i) replay methods
(i.e., replay previous tasks while learning a new task), (ii) regularization-based methods (i.e., modifying the loss function), and (iii) parameter isolation methods (e.g., using task-specific synaptic weights). 
\citet{Kirkpatrick2017} demonstrated an approach based on task-specific elastic weight consolidation (EWC) in which the plasticity of synaptic weights deemed to be important to previously learned tasks is reduced.
The synaptic intelligence approach of \citet{Zenke2017} estimates the importance of each synapse in solving previous tasks and penalizes changes to the important synapses when learning new tasks.
\citet{Masse2018} found that synaptic intelligence along with context-dependent inhibition provides the best performance when used together. Their context-dependent inhibition mechanism consisted of zeroing the activity of ca. 83\% of the hidden units selected randomly while leaving the remaining hidden units unchanged.

Our goal here is to describe, in detail, and to document the characteristics of a novel biologically plausible algorithm that uses dendritogenesis to enhance the computational power of single neurons and to enhance a simple feedforward network of such neurons. The basic idea is to produce dendrites, and  thus dendritic targets for synaptogenesis, as the number of novel experiences, and the underlying latent variables, increase.

Critical to a useful dendritogenesis algorithm is the supporting synaptogenesis algorithm, and critical to a synaptogenesis algorithm is its compatibility with a Hebbian/anti-Hebbian synaptic modification rule. The stochastic algorithm here uses the supervised extension \citep{BaxterLevy2019,BaxterLevy2020} of adaptive synaptogenesis \citep{LevyColbert1991,Adelsberger-Mangan1993B,Adelsberger-Mangan1994A,Adelsberger-Mangan1994B,Colbert1994,Thomas2015,Levy2016}, which is an algorithm based on its compatibility with   Hebbian/anti-Hebbian physiology \citep{Levy1985,Foldiak1990,Tazerart2020}. As in the case of all of the earlier work using adaptive synaptogenesis (AS) and supervised adaptive synaptogenesis (SAS), information for the algorithm is local, as is required for biological reality.

We are highly motivated by Mel's ideas (\citealt{Mel1991,Poirazi2001,Behabadi2014,Mel2017,Caze2013}), and there are two commonalities with the model proposed here: the neuron's output is determined by multiple dendrites each having a set of synaptic weights, and these dendritic weights can grow or shrink or shed. However, the adaptive flavor of Mel's model is different from the model proposed here primarily because (a) it uses gradient descent with annealing to determine the dendritic weights, (b) the dendrites do not compete to determine the neuron's output, (c) the synaptic modification rule is driven by derivatives and error signals, (d) synaptogenesis is not driven by missed detection errors, and (e) the number of dendritic branches is fixed; i.e., there is no explicit dendritogenesis rule driven by missed detection errors.

In what follows,  Methods presents (i)  the computationally local equations that define  the algorithm, (ii) the parameters that control the algorithm, and (iii)  three different environments used to challenge the algorithm. These three differ in how a network encounters novel inputs. 

The Results demonstrate that the algorithm resists catastrophic interference, has a significantly enhanced memory capacity for explicit as well as latent prototypes, and quantifies the distinctions between learnable and unlearnable situations. Also the Results reveal that the algorithm can unmix certain mixture distributions as it performs locally conditional but unsupervised clustering of more than 50\% overlapped feature vectors. Finally, the Results produce empirical asymptotic stability results for the algorithm while also performing  a partial dissection of the algorithm's two novel feature, dendritogenesis  \textit{per se} and cross dendritic suppression (CDS). The point of this dissection is to show the critical role played by CDS in creating stable  connectivity. 

The Discussion is rather interdisciplinary since it interprets the algorithm and the stable connectivity of the developed dendrites and synapses from perspectives that include  biology, behavioral psychology, statistical inference, signal processing and machine learning.

\section{Methods}

\subsection{The network and updating equations}

This section describes the dynamical equations for  a layer of neurons that  receive class information (a class present signal vs no signal) and some form of  a feature vector nominally arriving just a little before the class-signal. The   equations that modify the connections are (i) modification of an existing synaptic strength, (ii) discarding a synapse here called shedding,  (iii) synaptogenesis, and (iv) formation of  new perisomatic dendrites, i.e, dendritogenesis.

The network is  single layer, feedforward with one inhibitory feedback interneuron mediating a winner-take-all competition between primary neurons. There are as many neurons as classes. Each neuron receives a class-specifying, binary  $\{0,1\}$ signal. With training experience, these neurons acquire synaptic connections associated with the feature vector; these synapses are restricted to excitatory. SAS \citep{BaxterLevy2019,BaxterLevy2020} controls the acquisition of  feature-input lines. CDS is a weakened version of  spike-timing synaptic modification \citep{LevySteward1979,Levy1983}. The full algorithm consists of 
(i) synaptogenesis (ii) associative modification of existing synapses (iii) synaptic shedding (iv) dendritogenesis, and (iv) cross-dendritic  suppression (CDS) of synaptic modification  In developing the full algorithm, the minimum error-rate constraint is supplemented by adaptively implemented, stabilization requirements.

In the spirit of making the catastrophic interference as hard to overcome as possible, the number of decision-making neurons is minimized. That  is, the number of supervised, decision-making neurons  exactly equals the number of classes that  must be discriminated in the problem set. In fact arising from the data sets used here, this number is either two or three neurons (although the algorithm can always accept more than the minimum number). The feature space input dimensions are either four dimensional (4-D), eight dimensional (8-D), or 256 dimensional (256-D). The geometric structure of the feature vectors  and  the structure of the learning problems that are themselves constructed from the prototypes of the feature vectors are described as they are introduced in Results. 

In what follows the phrase 'the algorithm', and as well CD\&SAS, refers to the full set of dynamical equations and  functions just outlined. This fundamental set  is dendrite localized modification of existing synapses with CDS (Eq  3),  synaptogenesis (Eq 4 and 5), synaptic shedding (Eq 6), and dendritogenesis (Eq 7 or its variant Eq 8).

The accompanying table summarizes the parameter and variable notations.

To simplify notation, time is denominated in  training trials but several computational and communication events occur within one such timestep. Specifically, the input feature vector $\xvec(t)$ generates a non-negative value within every dendrite, generically $y_{dj}(t)$. For each neuron $j$, the largest one of these dendritic excitations  is transmitted  to a competition-judging neuron. This interneuron determines  which of the primary neurons $j$ has the  most excited excited dendrite. The  neuron with this largest $y_{dj}$ wins the competition and nominally fires. That is, the interneuron sends an inhibitory signal tailored to inhibit all but the winning primary neurons, and this winning neuron $j$ produces a second positive output, $z_j(t)=1$ while for all the losers $z_j(t)=0$ (see Discussion that presents a mechanism making this dendritic competition local and thus neurally plausible). About the same time that the feature-based competition ends, the class input signal, $z^*_k(t)=1$, is received by one neuron; i.e, when $k=j,\; z^*_j(t)=1$ while for the other neuron(s) $k\neq j$, i.e.,  no class-signal is received, i.e.,$\; z^*_j(t)=0$. At this point, the various neural modification equations are executed where permitted. The order of execution is (i) modification of existing synaptic weights, (ii) shedding where dictated, (iii) dendritogenesis, and (iv) synaptogenesis. The details are found in the dynamical equations that follow.   

\begin{singlespace}
	
	\begin{table}[H]   
		\label{vardef}
		\ra{1.0}
		\begin{center}      
			\scalebox{1.0}{
					\begin{tabular}{ l l }  
						
						\midrule
						\textbf{Indices} \\
						$i$ & index for input lines, $i = 1, 2, \ldots, N$\\
						$N$ & number of feature vector input lines \\
						$j$ & index for supervised neurons\\
						$k$ & index for input pattern classes\\
						$D_j(t)$ & the set of dendrites on neuron $j$ at time $t$ \\
						$d = d_j$ & index for dendrites on neuron $j$, $d \in \{1, 2, \ldots, |D_j(t)|\}$ \\
						
						\midrule
						\textbf{Inputs} \\
						$\xvec(t)$ & input (exemplar) vector at time t\\
						$\Xvec$ & random input vector, $\Xvec = [ X_1, X_2, \ldots,X_i \ldots X_N ]^T$  \\
						$E[X_i]$ & expectation value of $X_i$, it theoretical average firing rate \\
						$z_j^*(t)$ & binary supervision signal for neuron $j$, $z_j^*(t) = 1$ if $k=j$ and 0 otherwise \\
						\midrule
						\textbf{Dendrite} &\\
						\textbf{Excitation} \\
						$y_{dj}(t)$ & when positive, excitation value of dendrite $d$ on neuron $j$ at $t$\\
						$y_j(t)$ & excitation value of $j$ equals its most excited dendrite $y_{dj}(t)\one_{d_{\max}}\cdot(d_j(t))$\\
						$\one_{d_{\max}}(d_j(t))$ & indicator function valuing to one when $d_j$ is the most excited dendrite of $j$  \\
						$\one(x)$ &  generic indicator (no subscript), $\one(x) = 1$ only if $x>0$ otherwise  $\one(x)=0$ \\
						\midrule
						\textbf{Neuron}&\\
						\textbf{Outputs}  \\
						$z_j(t)$ & $\in\{0,1\}$ output of neuron $j$ after the competition or via simple thresholding \\
						\midrule
						\textbf{Connectivity}  \\
						$c_{idj}(t)$ & binary connectivity indicator between input line $i$ and dendrite $d$ on neuron $j$ \\
						$w_{idj}(t)$ & synaptic strength between input line $i$ and dendrite $d$ on neuron $j$ \\
						$p_{idj}(t)$  & probability of forming a synapse between input line $i$ and dendrite $d$ on neuron $j$  \\
						$\gamma_{dj}(t)$ & $p_{idj}(t)$, but only if other conditions are true; see Eq (4) \\
						\midrule 
						\textbf{Parameters}&\\	
						$W_0$        & initial (\textit{de novo}) synaptic weight (0.1)  \\                                  
						$\theta_w$   & synaptic shedding threshold (0.005)   \\
						$\epsilon_w$ & synaptic modification scaling factor (0.025 or 0.002) \\
						$\theta_\gamma$  & synaptogenesis and dendritogenesis threshold (0.05) \\ 
						$\epsilon_\gamma$ & synaptogenesis decrementing factor (0.05 or less; see Methods) \\   
						$\alpha$  & missed detection rate constant (1.0 with one exception) \\
					\end{tabular} 
			}			
		\end{center}
		\caption{Notation for variables and parameters}\label{parameter_table}
	\end{table}
	
\end{singlespace}

\subsection{Dendritic  excitation and neuron firing}

To keep close to neocortical  neurons,   only excitatory input connections are permitted for the inputs to the class supervised neurons. The excitation of dendrite $d$ on neuron $j$ at time $t$, $y_{dj}(t)$, is the inner product of the input pattern (the exemplar $\xvec(t)$) and the dendrite's synaptic weights divided by the sum of the dendrite's weights, i.e.
\beqn
y_{dj}(t) = \left\{ \begin{array}{cl}
	\xvec^T(t) \wvec_{dj}(t) / \sum_i w_{idj}(t)  & \mbox{if } \sum_i w_{idj}(t) > 0 \\
	0     & \mbox{otherwise} 
\end{array} \right.
\eeqn
\subsection{Neuron and dendrite spike generation}
In all cases, a feature based class prediction takes the form $z_j(t)=1$ when neuron $j$ at time $t$ predicts the generator of its input is in class $j$. When $z_j(t)=0$, this is an implicit prediction by $j$ at $t$ that the feature vector generator is out of class.

For the four dimensional XOR, deterministic problem sets, labeled 4-D, there is a shared spike-threshold applicable to all dendrites. The threshold is selected so that only one dendrite on each neuron can produce an output pulse (spike). As is true for the more general problem sets, a dendritic spike in neuron $j$ at time $t$ implies neuron $j$ produces an output spike, $z_j(t)=1$.

More generally in what follows, a competitive mechanism is used to determine the most excited dendrite. This neuron-specific dendritic competition precedes the neuron vs neuron competition mediated by the interneuron. A local and biological interpretation of both  competitions is found in the Discussion. Here just the appropriate notation is introduced.

The most  excited dendrite on neuron $j$ is identified via $\one_{d_{\max}}(d_j(t))$ valued one for that  dendrite and zero for the other dendrites of $j$. The excitation of this dendrite becomes $j$'s excitation, $y_j(t)$. Of all the neurons, the neuron $j$ with the largest $y_{j}$ produces an output pulse, $z_j(t) =1$ while the other neurons have their output valued zero. 

\subsubsection{A comment on the notation used below.} When the indicator (characteristic) is used without a subscript, it means that any positive value it operates on is valued one and any other value, zero or negative, is valued zero; in notation, 
$\one(u)=1$  only if $u>0$, otherwise $\one(u)=0$.

\subsection{A missed detection is local information}
All the problems here include a supervising class signal. Thus each neuron receives the local information $z^*_j(t)=1$ when its class is present at time $t$ and no signal, i.e., $z^*_j(t)=0$,  when its class is not present at time $t$. As a result this local supervising signal, there are two local calculations that it is useful for each neuron to make: (i) $\one (z^*_j(t)-z_j(t))  $, which is a missed detection (MD) when evaluates as one and (ii) $z^*_j(t)\cdot z_j(t)  $, which is a successful prediction when evaluates to one.

\subsection{Cross dendritic suppression (CDS) alters the synaptic modification rules}   
The next few subsections describe the synaptic modification rules. Because the Results examines the algorithm with and without CDS, some equations have two forms, one being implemented when CDS is present and the other holding when CDS is absent.

\subsection{Updating synaptic weights of existing connections}

When there is no CDS, the class supervised synaptic weight modification rule for existing connections is the  previously implemented one from \citet{BaxterLevy2020},
\beqn
\wvec_{dj}(t+1)  = \wvec_{dj}(t) + \epsilon_w ( \xvec(t) - E[ \Xvec ] - \wvec_{dj}(t) ) y_{dj}(t) z^*_j(t)
\eeqn

and when CDS is operating, there is an additional multiplier,
\beqn
\wvec_{dj}(t+1)  = \wvec_{dj}(t) + \epsilon_w ( \xvec(t) - E[ \Xvec ] - \wvec_{dj}(t) ) y_{dj}(t) z^*_j(t) \one_{d_{\max}}(d_j)
\eeqn

where $z^*_j(t)$ and $\one_{d_{\max}}(d_j)$ are defined above. The addition of the $\one_{d_{\max}}(d_j)$ multiplier limits weight modification to a single dendrite. Thus, this multiplier implements  dendrite selective modification at neuron $j$. 

Additionally, note that this $\Delta w$ modification rule implies each $w_{idj}(t)$ is bounded from above since when $X_i=1$, its largest value,  $\Delta w_{idj}(t)<1-w_{idj}(t)$. The shedding rule bounds each synaptic weight from below.

\subsection{Creation of a new synapse,  synaptogenesis}

With or without CDS, the probability that input line $i$ makes a connection  onto dendrite $d$ of neuron $j$ is
\beqn
p_{idj}(t)  
=  \gamma_{dj}(t) x_i \one(z^*_j(t)-z_j(t)) (1 - c_{idj}(t))
\eeqn
In words: At time $t$, given that there is no connection from input line $i$ to dendrite $d$ of neuron $j$, $c(i,dj)=0$, then for there to be a positive probability, $0<\gamma_{dj}\leq 1$, for the formation of a new synapse from $i$ to $j$, requires that the  a presynaptic input, $i$, must be co-active, with a missed detection by $j$; that is, there is the requirement for the joint event  $x_i(t)=1$, $z_j^*(t)=1$, and $z_j(t)=0$. Note that all dendrites of $j$ are eligible for synaptogenesis.

The factor $\gamma_{dj}(t)$, for the particular dendrite $d$ on neuron $j$, adapts as a function of a dendrite's successful detections. With each such success, it is decremented as
\beqn
\gamma_{dj}(t+1) = \gamma_{dj}(t)(1 - \epsilon_\gamma \cdot \one_{d_{\max}}(d_j) \cdot z^*_j(t) z_j(t))
\eeqn
Thus, whenever $\one_{d_{\max}}(d_j) \cdot z^*_j(t) z_j(t) = 0$, there is no change, i.e., $\gamma_{dj}(t+1) = \gamma_{dj}(t)$. 

In the event that  $\one_{d_{\max}}(d_j)\cdot z^*_j(t) z_j(t)  = 1$, i.e., a correct in-class prediction by $d$ on $j$, $\gamma_{dj}(t)$ decreases,

$ \gamma_{dj}(t+1) = \gamma_{dj}(t)  \cdot (1-\epsilon_\gamma)$. 

We have investigated the effect of the rate of decrement, $1 - \epsilon_\gamma$, on performance with values of $1 - \epsilon_\gamma$ ranging from 0.001 to 0.999; unless otherwise stated, $\epsilon_\gamma = 0.05$. 

The value of a \textit{de novo} synapse is the same for all simulations presented here, $W_0=0.1$

\subsection{Shedding synaptic weights}

Because of historical usage, arising from the decision to remove a synapse originates postsynaptically, the term shedding --- rather than pruning --- is used \citep{Levy1985}; \textit{viz}, a tree \textit{sheds} its leaves in the fall while the gardener, a force external to the tree, \textit{prunes} a tree's limbs).
After the synaptic weights are modified, the synaptic shedding rule ensures that there are only positive weights since neocortical  computation is dominated by positive synapses that never convert to inhibitory synapses. The shedding threshold  is $\theta_w >0$. In notation and given that $c_{idj}(t-1)=1$, the dynamic expression is
\beqn
\{c_{idj}(t), w_{idj}(t)\} = \left\{ \begin{array}{cl} 
	\{0,0\}  & \mbox{if } w_{idj}(t) < \theta_w \\
	\{1, w_{idj}(t)\} & \mbox{otherwise}
\end{array} \right.
\eeqn
where $c_{idj}$ indicates whether or not there is a connection from input $i$ to the $d^{\textrm{th}}$ dendrite of neuron $j$. Note that, when present, CDS limits shedding to the maximally excited dendrite.

For all the simulations presented here $\theta_w=0.005 $ is a constant.

\subsection{Creation of a new dendrite,  dendritogenesis}

Dendritogenesis is the process of creating a new dendrite. The process is local to each neuron and requires two conditions for its occurrence. Formally, define the dendritogenesis event $\{0,1\}$ at time $t$ as either

\beqn
\xi_j(t) = \one\left( \theta_\gamma - \gamma_{\delta j}(t) \right) \one\left(z^*_j(t)-z_j(t)) \right))
\eeqn

which is predominant, but occasionally the variant

\beqn
\xi_j(t) = \one\left( \theta_\gamma - \gamma_{\delta j}(t) \right)  \one\left( \ebar^{MD}_j(t) - \theta_{MD} \right)
\eeqn
with

\beqn
\ebar^{MD}_j(t+1) = \ebar^{MD}_j(t)+ \alpha( \one(z_j^*(t)-z_j(t))- \ebar^{MD}_j(t) )
\eeqn
Then dendritogenesis occurs when $\ebar^{MD}_j(t) > \theta_{MD}$ where $\theta_{MD}$  is the preset error-threshold. 

Recall from the  above the $\gamma_{dj}$ is a monotonically decreasing function so that when its value drops below $\theta_\gamma$, the first indicator function shifts from zero to one. Also, the term $ \one\left( \theta_\gamma - \gamma_{\delta j}(t) \right)$ refers to the most recently formed dendrite of $j$. Indeed, this dendritogenesis equation guarantees that dendritic development is sequential; only one dendrite is added at a time and the previously added dendrite is reasonably mature. The second term of the equation tracks $j$'s missed detections. In it simplest form, a single missed detection at $t$ is enough for the indicator function to value one for just this one trial. The alternate form investigated uses a moving average of errors.

Note when $\alpha = 1$ the second version of synaptogenesis reverts to the first version. Most simulations use $\alpha=1$ with the exception of 

In two places $\alpha = 0.01$, specifically sections 3.3.3 and 3.5.2.

A variety of values of $\theta_\gamma$ were investigated. In all the simulations here, the value is the same $\theta_\gamma=0.05$.

\subsubsection{Miscellaneous}
In general empirical convergence is defined as 500 epochs with no change in the number of connections. Under conditions that appear to prevent convergence, simulations are continued for 3000 or sometimes 5000 epochs and then ended.

\subsection{Inputs and Training methods}

In addition to the class supervising signal, the inputs to the network are feature vectors, each one defined by a class associated prototype. 

The are a variety of learning problems. They are referred to as problem sets. Different problem sets consist of different groupings of prototypes and classes. These different problem sets are explained in the Results as they are introduced.
To point out the significant contribution  this algorithm  adds to a single neuron's computational prowess, the learning problems emphasize exclusive-or (XOR) input worlds as well as more complicated problems. Except for the initial portion of Results, where the feature vectors are pure prototypes, the prototypes are never seen by the network and are thus latent variables. The actual feature vectors used as inputs, both for training and testing, are called  exemplars. When prototypes are latent, the exemplars are randomly perturbed versions of these  prototypes. All prototypes and exemplars are binary, $\{0,1\}$ vectors. Perturbations are of two types: (i) occlusion, in which a specified fraction of a prototype's positively valued dimensions are complemented to  zeros, and (ii) on-noise, in which a  specified fraction of a prototype's zeros are complemented to ones.

There are three different  training paradigms: concurrent, progressive, and segregated.

The number of exemplars presented during training is expressed as blocks of exemplars called 'epochs'. An epoch consists of a pseudo-randomly mixture of  exemplars generated from a prescribed set of prototypes. 

For concurrent training, all prototypes are present in each successive epoch of training.

Segregated training has multiple phases. In each phase, exemplars are based on a single,  novel prototype.

\begin{table}[h]
	\begin{minipage}{0.33\linewidth}
		\centering
		(a) 2$|$3$|$3 \\
		\vspace{0.1in}
		\begin{tabular}{| c | l | l | l |} 
			\hline
			2$|$3$|$3 & \multicolumn{3}{c|}{Prototypes Present} \\ \cline{2-4} 
			Phase &  Class 1 & Class 2 & Class 3 \\ \hline
			1      &    1 & 3 & 6 \\
			2      &    1,2 & 3,4 & 6,7 \\
			3      &    1,2 & 3,4,5 & 6,7,8 \\ 
			& & & \\
			& & & \\
			& & & \\ \hline
		\end{tabular}
	\end{minipage}
	\hfill
	\begin{minipage}{0.26\linewidth}
		\centering
		(b) 4$|$4 \\
		\vspace{0.1in}
		\begin{tabular}{| c | l | l |} 
			\hline
			4$|$4 & \multicolumn{2}{c|}{Prototypes Present} \\ \cline{2-3} 
			Phase &  Class 1 & Class 2  \\ \hline
			  1      &    1 & 5 \\
			2      &    1,2 & 5,6 \\
			3      &    1,2,3 & 5,6,7 \\
			4      &    1,2,3,4 & 5,6,7,8 \\ 
			& & \\
			& & \\ \hline
		\end{tabular}
	\end{minipage}
\hfill
\begin{minipage}{0.3\linewidth}
\centering
(c) 2$|$6 \\
\vspace{0.1in}
\begin{tabular}{| c | l | l |} 
\hline
2$|$6 & \multicolumn{2}{c|}{Prototypes Present} \\ \cline{2-3} 
Phase &  Class 1 & Class 2  \\ \hline
1      &    1 & 3 \\
2      &    1,2 & 3,4 \\
3      &    1,2 & 3,4,5 \\
4      &    1,2 & 3,4,5,6 \\
5      &    1,2 & 3,4,5,6,7 \\
6      &    1,2 & 3,4,5,6,7,8 \\ \hline
\end{tabular}
\end{minipage}
\caption{Prototypes present in each phase for progressive training.}
\label{progressive-proto-vs-phase-tbl}
\end{table}

Progressive training is a little more complicated and is described here for the 8-D and 265-D inputs, each with their own eight prototypes. At each successive training phase,  one new prototype from each class is combined with the  prototypes introduced previously. If a class is exhausted of its novel prototypes its prototypes continue into the next phase until all prototypes of all classes are part of the, necessarily, final training phase. Table \ref{progressive-proto-vs-phase-tbl} visualizes the progressive training scheme for three different problem sets: the three class 2$|$3$|$3 problem set has three training phases; the two class 4$|$4 problem set has four phases; and the two class 2$|$6 problem set has six training phases. In the table the successive training phases begin at the top, with phase 1, and end at the bottom with the last training phase, which is different for each problem set listed but is always characterized by introduction of prototype eight.  Note that each prototype is associated with one and only one class as indicated by the visualization; nevertheless, as will be quantified in Results, the eight prototype severely overlap.

The total number of training epochs is treated as a research variable in the convergence section of Results, but minimally training continues until there are at least 500 or more epochs with no connectivity changes.

In all cases, testing is performed with all modification equations inactivated.

\begin{table}[H]
	\centering
	\resizebox{1.02\textwidth}{!}{%
		\begin{tabular}{ll}
			{\textbf {Term}}  & {\textbf {Description}}                 \\
			
			model         & The collection of the all the equations governing the activities \\
			& of the dendrites, neurons, and their connections \\
			network      & Input lines, dendrites, neurons, and their connections \\
			
			SAS            &  Supervised Adaptive Synaptogenesis   including  $\Delta w$ equation, synaptogenesis, and shedding \\
			
			dendritogenesis & the creation of a new dendrite  \\
			
			CDS      &Cross Dendritic Suppression of plasticity\\
			
			DC\&SAS &  Dendritogenesis, CDS and SAS; the full synaptic modification algorithm  \\
			
			non-functional dendrite&  a dendrite that never wins a max competition during testing\\
			
			redundant dendrites  &  two functional dendrites strongly excited by exemplars from one prototype  \\
			\midrule
			Training and Testing\\
			\quad Concurrent  &   training procedure\\
			\quad Progressive &   training procedure \\
			\quad Segregated  &   training procedure \\
			
			\quad trial      &  a single input vector and ensuing update of all equations and network states \\
			\quad epoch         &  one trial for each available prototype (varies by training procedure) \\
			\quad phase          & distinctive training epochs that define progressive and segregated training \\
			
			\quad XOR &  Exclusive OR problem \\
			\quad  $\cdot|\cdot$ or $\cdot|\cdot|\cdot$    & prototypes per class, 4$|$4, 2$|$6, 2$|$2$|$4 or 2$|$3$|$3 \\ 
			\quad  problem set  &  XOR or, for 256-D, either  4$|$4, 2$|$6, 2$|$2$|$4 or 2$|$3$|$3\\
			\quad convergence criterion &  trials with no connection changes \\
			
			\midrule
			Input Vectors \\
			
			\quad 4-D, 8-D, 256-D &  feature set dimensions \\

			\quad prototype     &  a $\{0,1\}$ feature vector that is associated with a single class, often a latent variable\\
			
			\quad exemplar &  a single input vector presented to the network during  training or testing \\
			
			\quad prototype perturbation &   exemplar generation by randomly complementing a prototype \\
			
			\quad occlusion & randomly complements a fraction of a prototype's  1's  to 0's \\
			
			\quad on-noise  & randomly complements a fraction of a prototype's  0's  to a 1's \\

			\quad 10/20, 50/30, etc. &  occlusion fraction/on-noise fraction, e.g.,   20/30 means\\
			&    complementing 20\% of 1's and 30\% of a prototype's 0's\\

		\end{tabular}
	}
	\caption{Glossary}
	\label{glossary-tbl}
\end{table}

The accompanying Table \ref{glossary-tbl} presents  idiosyncratic jargon and abbreviations used in describing the methods 
A
\section{Results}

\subsection{Fixed feature vectors: The prototypes are the exemplars}
\subsubsection{The network requires dendritogenesis to learn the 4-D XOR problem }

The  Results begins with our simplest XOR learning problem. The problem set to be learned is four-dimensional instead of the classic two-dimensional (see Table \ref{xor_table1}), and of necessity, these are class supervised neurons that receive the appropriate class information, $z^*_j \in \{0,1\}$, whenever an exemplar is selected during training as input to this two neuron feedforward network. Although simple because of the required decision-boundaries, it still challenging enough to  prevent learning when there are a limited number of single dendrite supervised neurons. Given these highly overlapped feature vectors (the prototypes of table \ref{xor_table1} ), overwriting, i.e., catastrophic interference is inevitable under the current circumstances, the current circumstances being no dendritogenesis and no extra reserve neurons. Thus we start two neurons that suffer catastrophic interference.  Thus, adaptive synaptogenesis alone cannot solve the XOR problem under these circumstances

  \begin{table}[H]
  	\centering
\begin{tabular}{|c|c|c|c|}
\hline
 Pattern  &   4-D     & 2D     & Class \\
 Number &    Vector  &    \;\quad Equivalent       & Label  \\ \hline
      1       &    0011    & 00 &    1  \\
      2        &   1100    & 11 &  1  \\
      \hline
      3        &   0110    & 01 &  2  \\
      4        &  1001     & 10 &  2  \\
      \hline
\end{tabular}
\caption{4-D XOR codes and corresponding class}\label{xor_table1}
\end{table}  

The different training methods succumb to the XOR problem in slightly different ways. But in any event, the error-rates are about 50\%; i.e., just the guessing rate. Moreover, with concurrent training, synaptogenesis never stabilizes. With segregated training, the synapses (connectivity and weights) stabilize but can only get the right answer for the last prototype presented, the one prototype (per neuron) experienced at the end of training. 

On the other hand instead of just supervised AS, enhancing the same two supervised neurons with dendritogenesis and CDS,  is enough to banish catastrophic interference. With the full DC\&SAS algorithm, each neuron develops a second dendrite, and because an appropriate connectivity also develops, one neuron for each class is sufficient to solve the learning problem. In particular each dendrite of each neuron's pair develops the connectivity and weights that exactly match just one of the four input vectors. Notably, such a within-class dendritic pair  is an  orthogonal pairing in connectivity space. The functional weights  (i.e., normalized in the calculation of y) go to 0.5. As a result, any threshold greater than 0.5 and less than one allows the appropriate  selective responding after training: Neuron one fires to its two orthogonal inputs and does not fire to the other two  inputs (feature vectors). Exactly the same result is true for the other neuron and its class. 

This  learned, perfect  performance applies to all three training paradigms, even segregated learning, which, \textit{a priori}, should be the most predisposed to catastrophic interference.

The immediate results to follow further illustrate this resistance to catastrophic interference but with  more complex problem sets. Moreover, as a function of an exemplar's randomizations away from its parent prototype, the simulations (i) demonstrate the degree to which these neurons can generalize while (ii) bounding  the limits of learnability  when using  the DCf\&SAS supervised neuron. Following these results is a dissection of the algorithm yielding a more detailed explication of why and how DC\&SAS works. Finally, various dynamical aspects of development are illustrated, including empirical convergences.

\subsubsection{Harder problems: 8-D multiple XORS with two or three classes and with  equal or unequal class probabilities}
A more interesting version of the XOR problem begins with a set of eight prototypes in an eight dimensional binary space. The first column of Table \ref{8D_prototypes_problems}  displays the eight $\{0,1\}^8$ prototypes, labeled p1 through p8. Here each prototype overlaps 50\% with six of the seven other prototypes.

\begin{table}[h]
\adjustbox{trim={.0\width} {.5\height} {0.0\width} {.0\height},clip}
{\includegraphics[width=\textwidth]{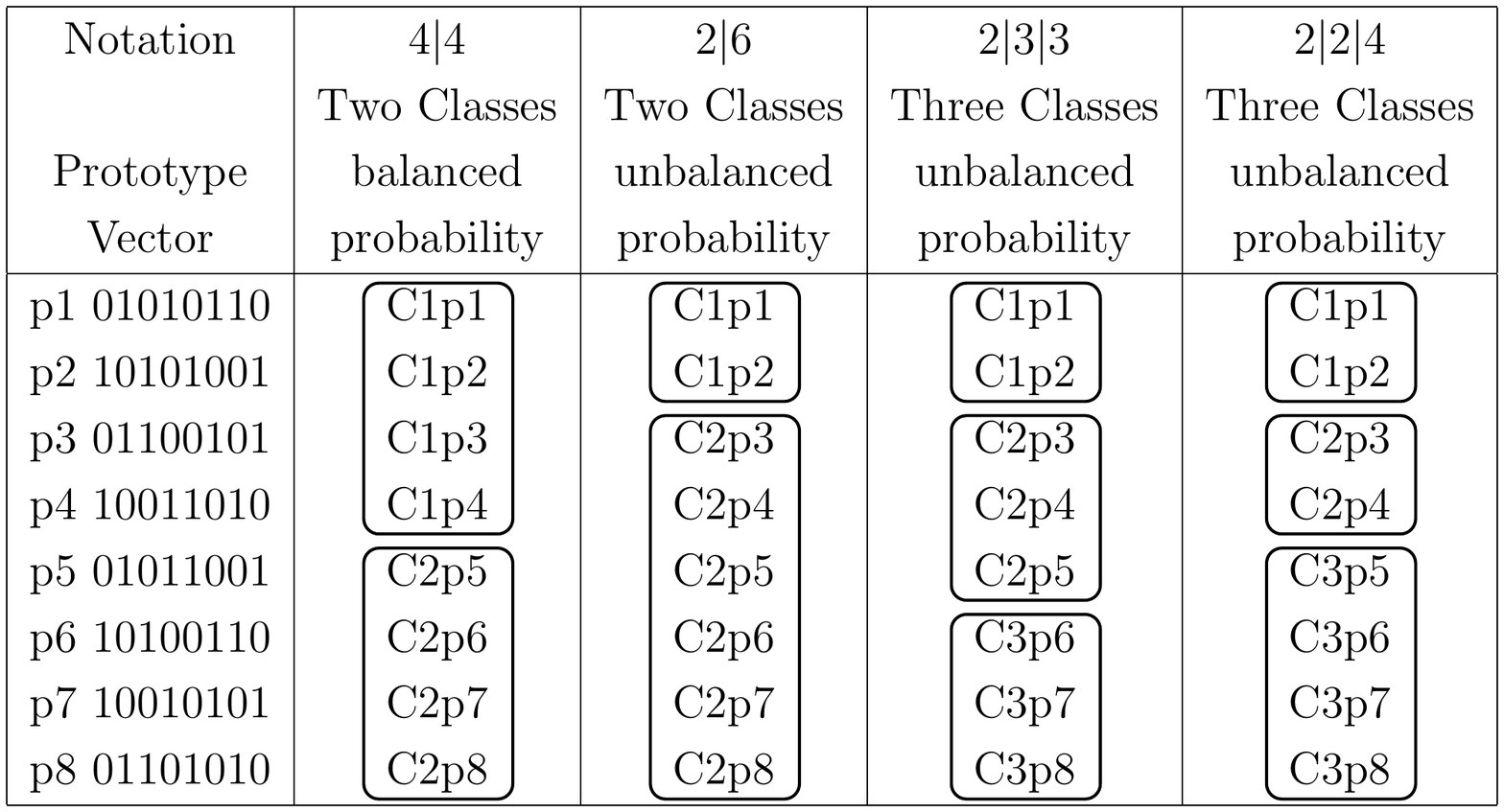}}
	
\caption{Prototypes and problem sets}
\label{8D_prototypes_problems}
\end{table}	

 For comparison purposes, Fig \ref{angle-matrix-4d-vs-8d-fig} illustrates the angles between every prototype pair. Each 60 degree angle is a fifty-percent overlap and the possibility of contradiction between such a prototype pair. Note the much larger fraction of overlapping  8-D prototypes there are compared to the previous 4-D XOR world. Of course, the closer the prototypes to each other, the harder it is to separate them to different dendrites and still harder to  separate their perturbed exemplars.  
 
  Building  problem-sets from  eight prototypes  allows  greater complexity of the classes compared to the previous four-prototype XOR version. These eight prototypes are configured into four learning problems via the number of prototypes per class. As Table \ref{8D_prototypes_problems}  delineates, the problem sets are (i) the well-balanced, two class, 4$|$4 problem; (ii) the rather imbalanced, two class, 2$|$6 problem (one neuron has to learn six different prototypes); (iii)  the three class 2$|$3$|$3 problem; and  (iv) the three class 2$|$2$|$4 problem. Note how these problem sets extend the XOR problem. For example, the 4$|$4 problem requires each neuron to learn not two but four distinct XOR vectors. In the 2$|$6 problem the class-1 supervised neuron only needs to learn a single XOR (2 prototypes); however, the class-2 supervised neuron must learn six prototypes and do so in spite of the strong overlap across these six. which is in addition to the overlap with the prototypes in the other class. The  three class 2$|$3$|$3 and the 2$|$2$|$4 problem sets illustrate how one must, minimally, expand the network  as the number of classes increase. However, here to increase the chance opportunities for  catastrophic interference,  we always require just the minimal number of neurons per class, i.e., one. Naturally, there can be no less than one postsynaptic supervised neuron per class, so these three class problems require three such neurons. 
 
 \begin{figure}[H]
 	\scalebox{1.0}{    
 		\begin{subfigure}{0.4\textwidth}
 			\includegraphics[width=\textwidth]{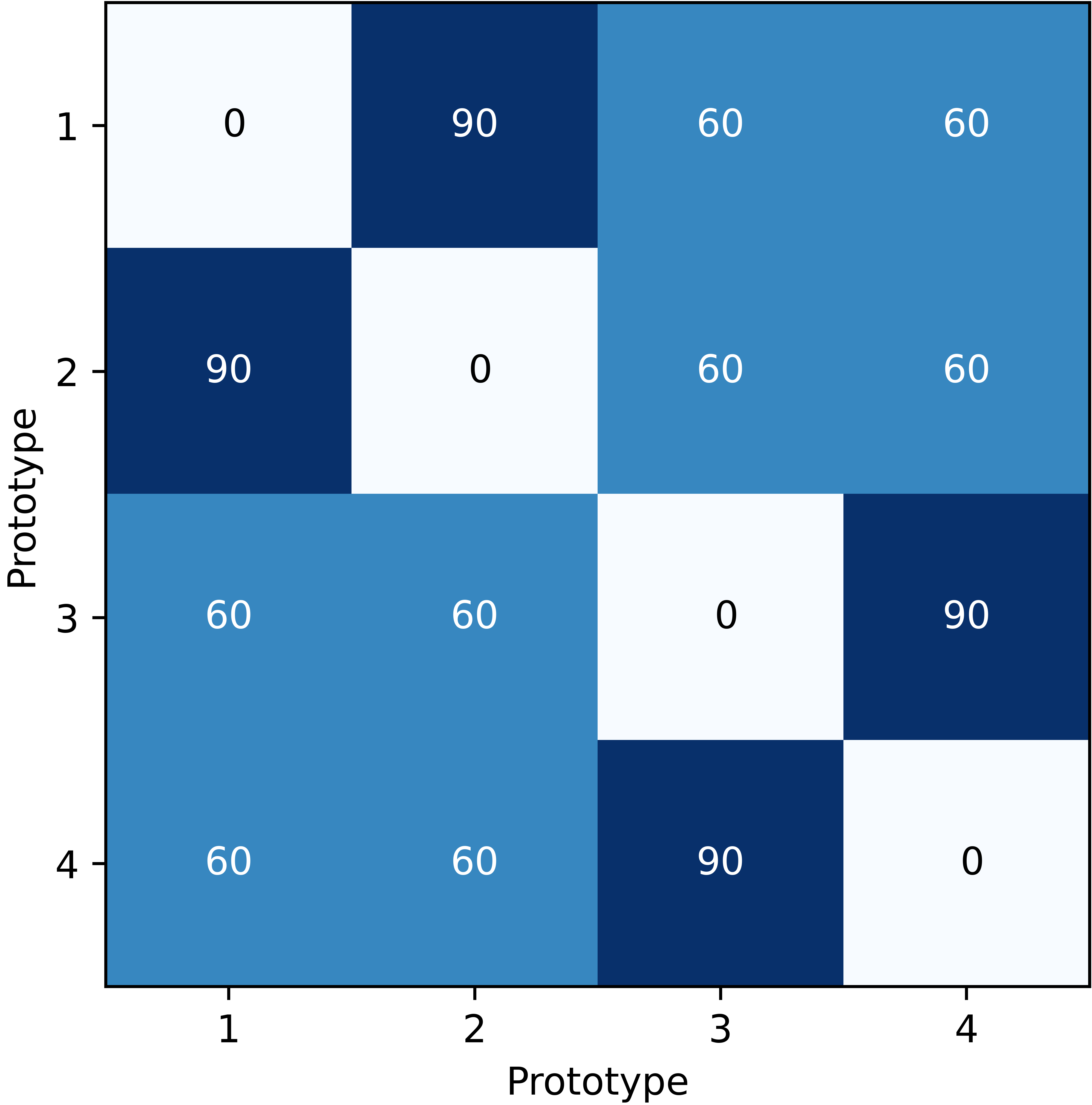}
 			\caption{Angles between the 4-D Prototypes.}
 			\label{angle-matrix-4d}
 		\end{subfigure}
 		\hfill
 		\begin{subfigure}{0.5\textwidth}
 			\includegraphics[width=\textwidth]{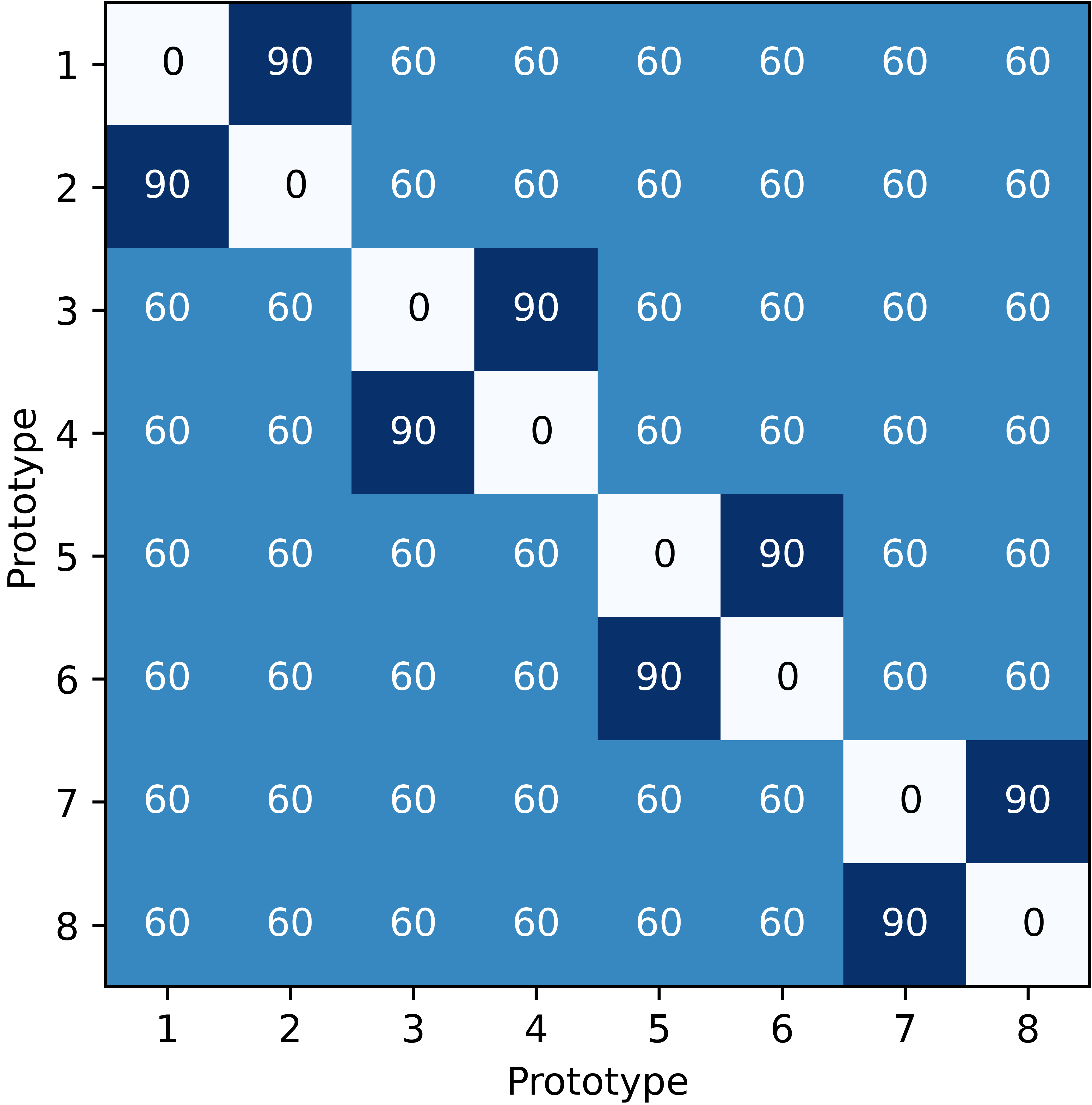}
 			\caption{Angles between 8-D Prototypes.}
 			\label{angle-matrix-8d}
 		\end{subfigure}
 	} 
 	\centering
 	\caption{
 		The 4-XOR problem vs the 8-D extended XOR has many more overlapping prototypes surrounding each prototype. Thus the 8-D problem can fail in many more ways. In either problem set
 		prototypes are either 60 or 90 degrees apart, but in the 8-D case
 		86\% of the prototypes are 60 degrees apart and only 14\% are 90 degrees apart.
 		Each square indicates the angle between the two prototypes (in degrees).
 	}
 	\label{angle-matrix-4d-vs-8d-fig}
 \end{figure}
 
All of these problem sets are unlearnable by single dendrite neurons operating under supervised SAS alone. 

On the other hand, endowing these class supervised neurons with the full algorithm dendritogenesis, CDS and SAS, i.e., DC\&SAS, overcomes catastrophic interference. For all 
problem sets and all training methods, neurons and networks train to error-free performance. The data in figure \ref{8D_bar_graph}  relies on  10 simulations of each problem set, and these 10 simulations are repeated for each  of the three training paradigms. In every simulation, the single dendrite neurons suffer  catastrophic interference. On the other hand for every simulation over every problem set, neurons equipped with the full algorithm,  dendritogenesis combined with CDS and  the supervised adaptive synaptogenesis, are one hundred percent successful in obtaining error-free performance to every prototype. The only differences over training type is that the phased training methods create neurons which usually match the number of prototypes per class while concurrent training tends to create neurons with just two dendrites. Regardless of the developed neurons, with unperturbed prototype training and testing, the DC\&SAS algorithm uniformly obviates  catastrophic interference.
  
\begin{figure}[H]
\includegraphics[scale=1]{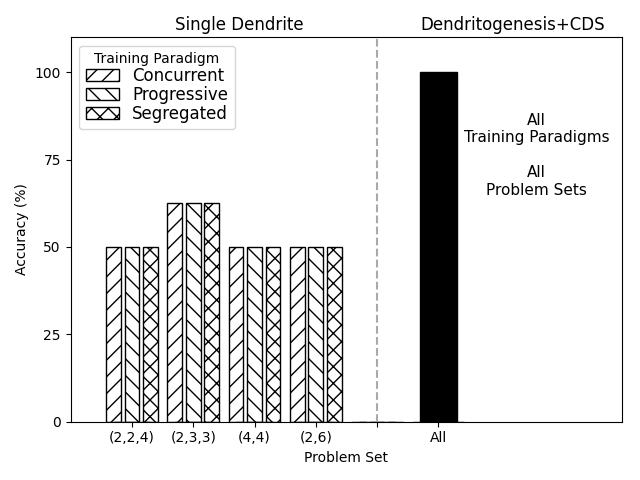}
\caption{Denritogenesis avoids catastrophic interference for all four problem sets. Training uses the noise-free eight dimensional prototype set associated as either: the 2$|$2$|$4 three-class problem; the 2,3,3 three-class problem;  the 4$|$4 two-class problem; or the 2$|$6 two-class problem.   Regardless of the training method, single dendrite neurons cannot solve this problem due to  anti-Hebbian synaptic competition on this one dendrite. With dendritogenesis and CDS, all four problems are solved perfectly no matter which of the three training paradigms used, concurrent, progressive, or segregated. The number of supervised neurons corresponds to the number of classes in any problem set.  Each simulation is distinguished by a random number seed which affects the random adaptive synaptogenesis and affects the within epoch orders of exemplar presentations.}
\label{8D_bar_graph}
\end{figure}

\subsection{Learning the prototypes as latent variables: generalization, abstraction, and the limits on robustness}

Although the above low dimensional problems incontrovertibly demonstrate the utility of DC\&SAS, these noise-free problem sets do not require two of the most prized properties of the neural paradigm, the abilities to conditionally cluster and to generalize. To demonstrate these properties, the eight 8-D   prototypes are extended to eight, 256 dimension (256-D) prototypes that become latent variables via randomly perturbing them to create each training or testing exemplar. Thus each single dimension of the eight dimensional space maps, without overlap, its value to 32 dimensions of the 256 dimension space, but there still remains just eight prototypes. Now however, via the exemplar generation method of randomized prototype perturbation, the prototypes are latent. That is, prototypes are never seen during training or during testing when there is prototype perturbation.  

The problem sets remain the same (4$|$4; 2$|$6; 2$|$3$|$3; 2$|$2$|$4) but now 128 dimensions of each prototype are valued one and 128 dimensions  are valued zero, nevertheless, the angles between prototype pairs, but not exemplar pairs, remain the same. That is, the angles between prototypes are just as in Fig \ref{angle-matrix-4d-vs-8d-fig}b.

\begin{figure}[H]
	\includegraphics[width=.98\textwidth,trim= 1.1cm 0cm 0cm 0cm]{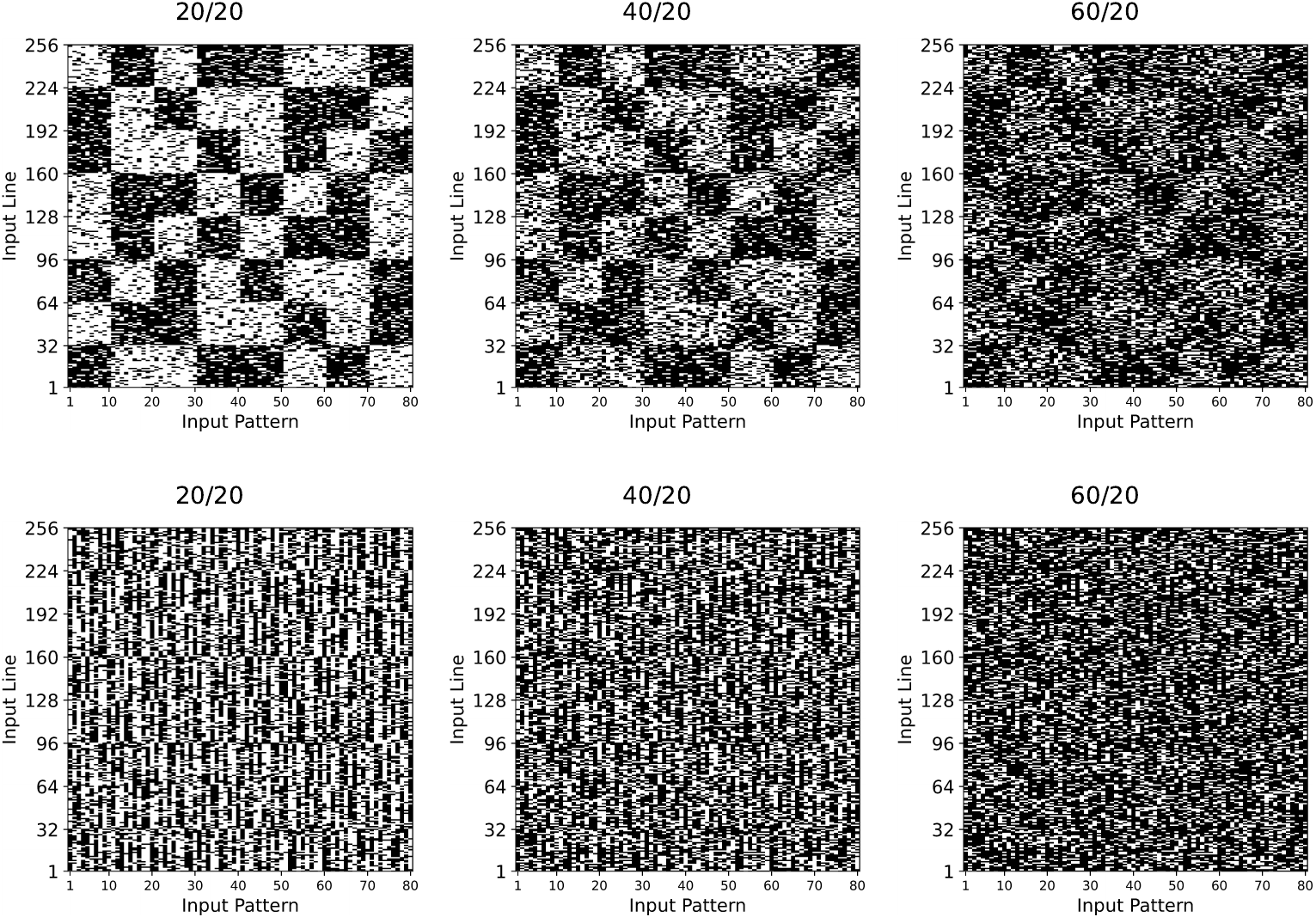}
	\caption{
		Visualizing representative exemplars as a function of fixed on-noise and varying the fractional occlusion perturbation of prototypes. With fixed 20.3\% on-noise, the three diagrams illustrate exemplars constructed from successively greater occlusions:  20.3\%, (20/20), 39.8\%, (40/20), and 60.2\%, (60/20). 
		In the top row, each checkerboard-like diagram contains 10, 256-D exemplars as column vectors for each of the eight prototypes, thus 80 column vectors. Note how much more difficult it is to guess the underlying prototypes as the fractional occlusion increases. For these $\{0,1\}^{256}$ vectors, white pixels correspond to a one and a black pixel corresponds to zero. Randomizing the positions of the column vectors make it impossible to identify a prototype at 60/40. Random shuffling of the top-row destroys perception of the different latent prototypes. Each figure in the bottom row is created by randomizing the position of the column of the corresponding figure just above.}
	\label{exemplars-with-diff-occ-fig}
\end{figure}

Associated with each prototype are the exemplars it generates. Such exemplars are generated, for training and testing, via two types of randomization: (i) random occlusion of a prototype (a fixed fraction of the 128 one's is randomly complemented to  a zero) and (ii) random on-noise (a fixed fraction of a prototype's zero-valued dimensions is randomly complemented to a one). For  three such randomizations, the Figure \ref{exemplars-with-diff-occ-fig} visualizes 80 column vector exemplars, ten for each prototype. The top row places the sibling exemplars next to each other while the bottom row randomizes exemplar positions, pointing out the difficulty one has in perceiving the different sets of exemplars as randomizations increase.  Three fractional randomizations are shown. For simplicity, the occlusion perturbations are rounded. For example  20.3\% is denoted  20\%. A white pixel corresponds to a one and a black pixel is a zero. Note that with 60.3\% occlusion and 20.3 \% on-noise, designated 60/20 contains only 51 firing neurons  from the prototype's 128, while there are 27 non-prototype neurons being turned on.   In this labeling scheme occlusion fraction always precedes on-noise fraction.

For another visualization of perturbation effects on exemplars see Fig \ref{exemplars-prototypes-fig}.

As Fig \ref{error-vs-occ-vs-problem-set-fig} shows for progressive training, each of the four problem sets is  learned with zero errors for perturbations up to 30/20. As fractional occlusion increases beyond 30/20, error-rates rise very slowly, until something greater than  the 50/20 perturbation.

\begin{figure}[H]
	\centering
	\includegraphics[width=.45\textwidth]{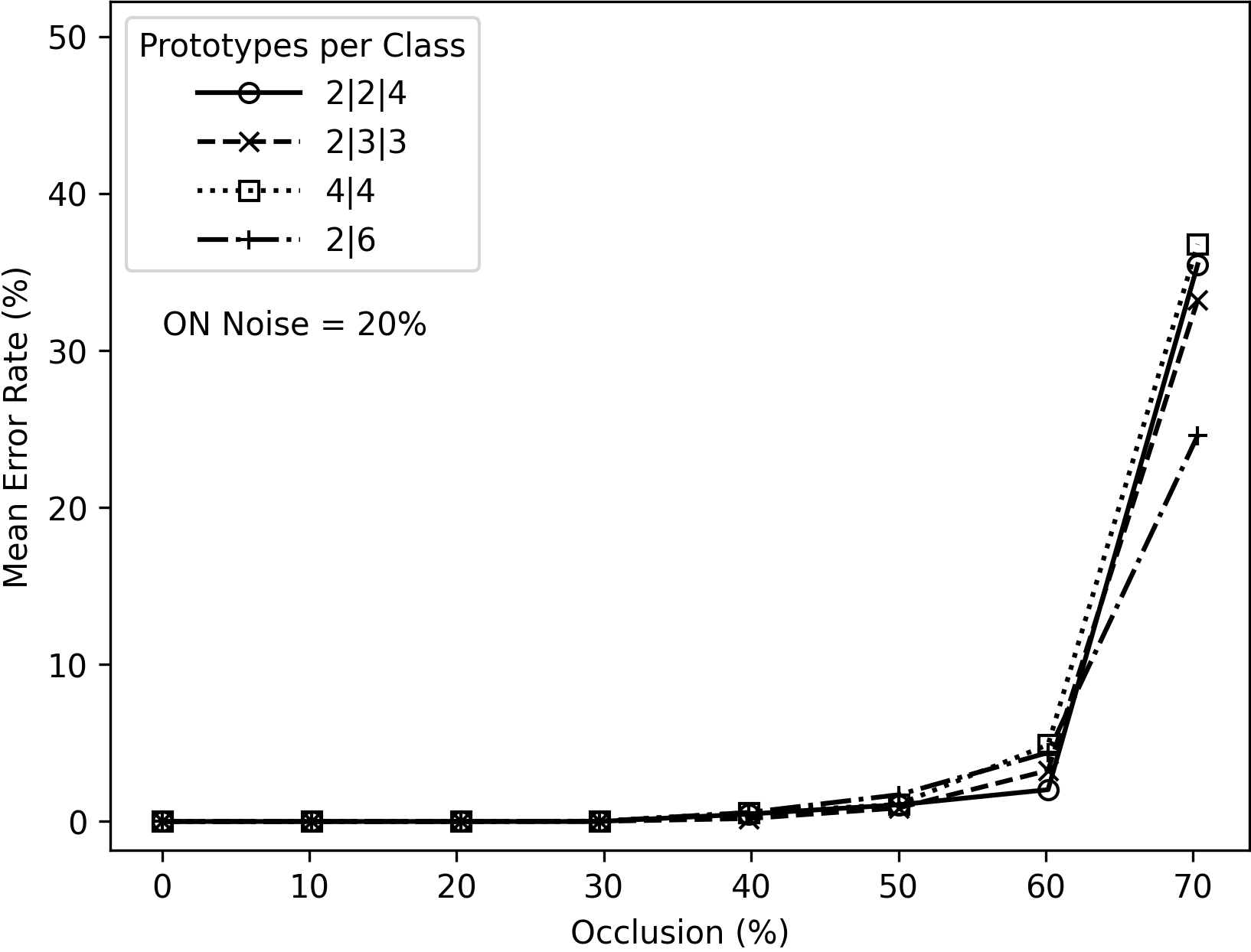}
	\caption{For progressive training, each of the four problem sets is learned with zero errors when the prototype perturbations are 30/20.
		As occlusion increases beyond 30\%, the error-rates rise slowly up to 50\% and then accelerate. 
		Nevertheless, even with 60\% occlusion the patterns are learnable to a significant extent with an error-rate of about 10\%. See Table \ref{8D_prototypes_problems} for the defined problem sets indicated by the inset legend. On noise is 20\% throughout.$\alpha=1; \; \epsilon_w=0.002;\;\epsilon_\gamma=0.05$. Empirical stability criterion halts training at 500 consecutive  epochs with no connection changes, or if this does not occur (which is  the case for larger perturbations), the maximum training epochs is 3000.}
	\label{error-vs-occ-vs-problem-set-fig}
\end{figure}

 Somewhere above this perturbation level, there is an acceleration of  error-rate. Nevertheless, even the 60/20 perturbations are learnable to a significant extent with an error-rate of just less than 5\%.
 
Fig \ref{error-vs-occ-and-on-fig}, with its pair of figures, gives a fuller picture of the interaction between occlusion and on-noise while limiting our attention to the 4$|$4 problem and progressive training.  Quantitatively, fractional on-noise seems slightly less harmful to performance than random occlusion. Comparing between types of randomization, both the 30/20 and 20/30 randomizations perform with no errors in ten out of ten simulations. However, performance deteriorates a little faster for occlusion.
Specifically, the 50/20 error-rate of 1.1\% is more than double the 0.42\% error-rate when 20/50 prototype perturbations are used. Beyond these perturbation values, performance deteriorates rapidly, although still below chances levels of 50\% even for the 70/20 and 20/70 prototype perturbations.

\begin{figure}[H]  
	\scalebox{1.0}{    
		\begin{subfigure}{0.49\textwidth}
			\includegraphics[width=\textwidth]{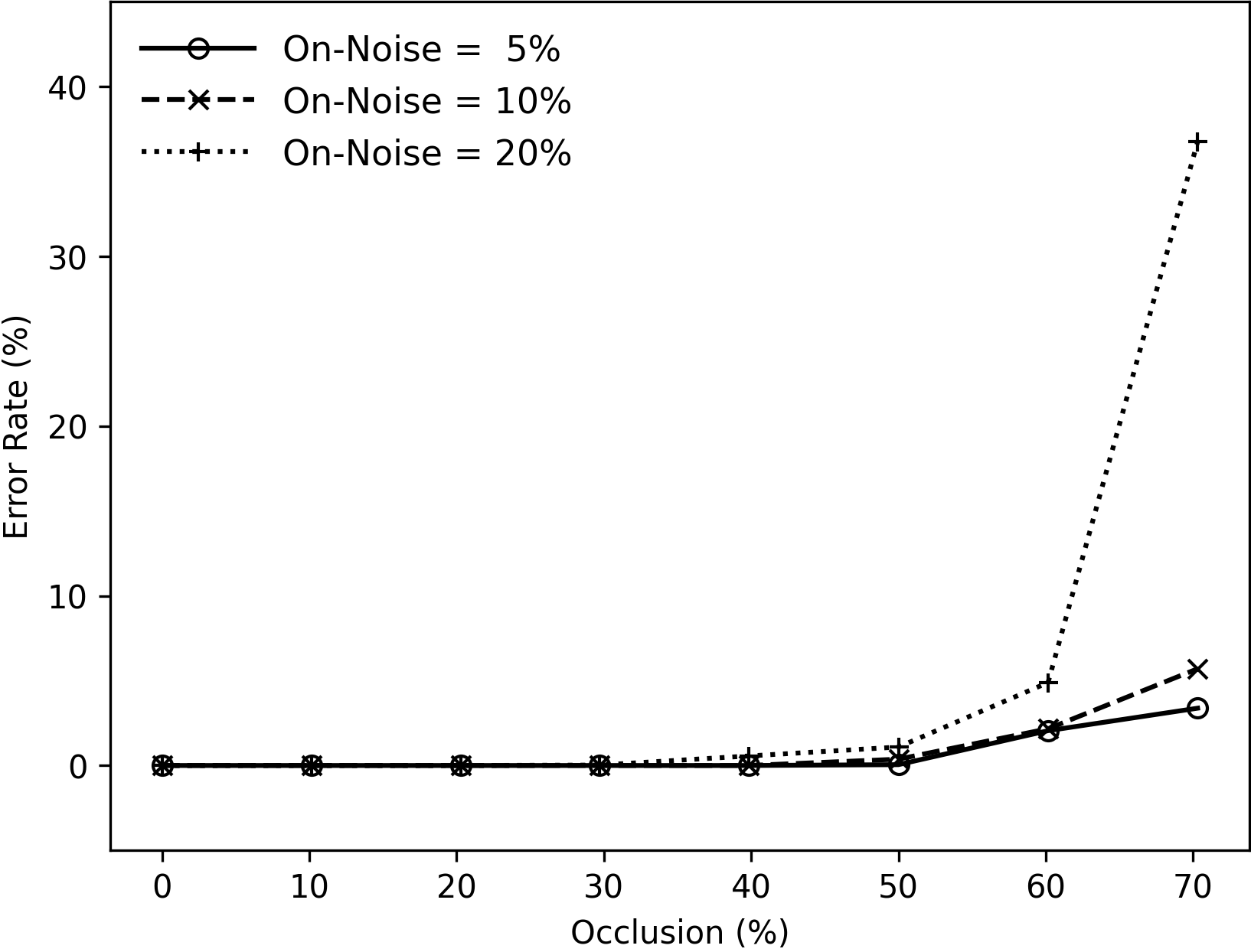}
			\caption{}
		\end{subfigure}
		\begin{subfigure}{0.49\textwidth}
			\includegraphics[width=\textwidth]{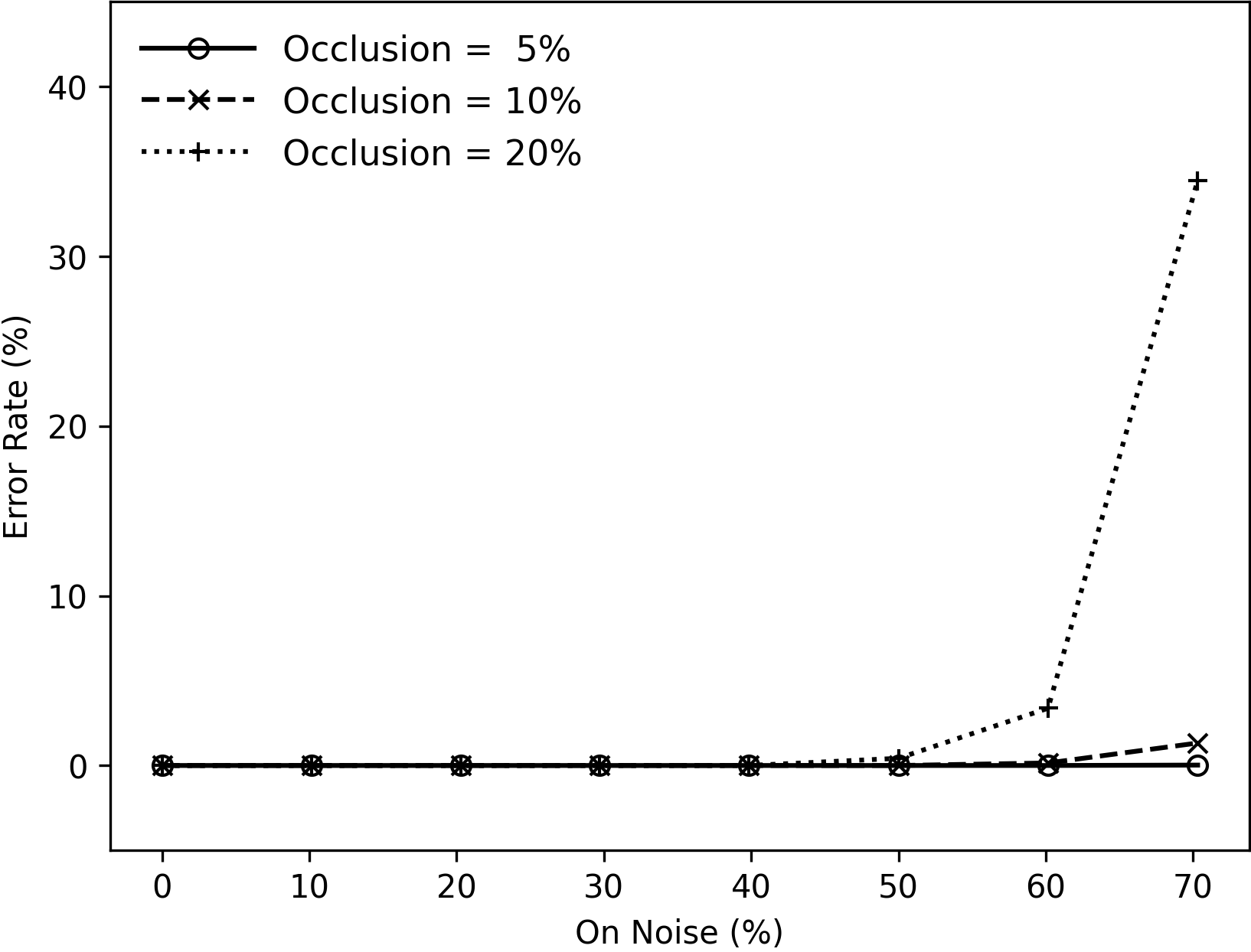}
			\caption{}
		\end{subfigure}
	} 
	\centering
	\caption{Perfect performance, i.e., error-rate zero, is sustained for up to 30\%  occlusion and 20\% on-noise.  On-noise appears to be slightly less harmful
		to classification performance than occlusion. In any event, occlusion and on-noise combine to make the task more difficult. The inset legend of the left graph indicates the three fixed levels of on-noise and similarly for fractional occlusion for the graph on the left. The number of complemented input lines for each fractional randomization of
		10\%, 20\%, 30\%, 40\%, 50\%, 60\%, and 70\% corresponds to  13, 26, 38, 51, 64, 77, and 90, respectively. The progressive paradigm is used for all training. Same parameter values as Fig \ref{error-vs-occ-vs-problem-set-fig}
	}
	\label{error-vs-occ-and-on-fig}.
\end{figure}

Two small calculations point out the satisfactory nature of this performance under the perturbation procedure.
Recall Fig 1b: Six out of seven prototypes share half their active input-lines, i.e., they share 64 out of the 128  neurons valued one for each such pair. As in the XOR problem, this fact alone creates a competitive situation between  prototype pairs or even among multiple prototypes. Further recall, a correct discrimination requires the appropriate neuron to have a dendrite that  produces  the largest, $y_{dj}$ value. This desired maximum excitation must occur in spite of  the  inherent overlap of exemplars, and with random perturbations this overlap can be extensive. Consider the extent to which the 50/20 and 20/60 perturbations can produce exemplars that are far from their parent-prototype. It is even  possible for such an exemplar to be  even closer to a non-parent prototype than its own parent. For example, the 50/20 complements 64 ones to zeros for an exemplar so this portion of the perturbation by itself might generate an exemplar equally close to the parent and non-parent, but then 20 percent of the parent's zeros are complemented to ones which implies the generated exemplar can end up much closer to a non-parent than a parent prototype. A similar possibility exists for the 20/60 perturbation. That is, the on-noise complements 77 zeros of the parent latent variable to ones which again, this perturbation alone can move an exemplar closer to a non-parent latent variable. Moreover, this movement toward a non-parent is only exacerbated by the 26  of the parent's active input-lines being complemented to off. Recall that for the 50/20  perturbations the error-rate is only about 1.1\%.

Not surprisingly, error-rates are parameter sensitive. If one speeds up convergence by raising $\epsilon_w$ by 12.5-fold, from 0.002 to 0.025, then the 50/20 error-rates increase about four-fold. Nevertheless, with this larger $\epsilon_w$ and larger error-rates the qualitative shapes of the curves and general conclusions of this section remain the same.

In sum, if zero error is required, then up to  30/20 or 20/30 randomizations can be tolerated (even 30/30 perturbations are learnable to error-free performance; see below). If a maximum of 5\% error-rates are acceptable, then the acceptable learning occurs even when prototype perturbations are as large as   60/20  and 20/60. Considering how far exemplars can range away from parent prototypes in these later two cases, such a five percent error-rate is not at all disappointing.

\subsubsection{Progressive training is robust against unequal prototype probabilities}

The results in previous subsections are based on equal prototype probabilities within a class while the class probabilities are often unequal. Thus one way of varying prototype probability has been examined. However there is another way to examine prototypes that are not equally probable, i.e., within a class. Even in this case, the model can still learn and produce  error-free performance .

The data in Table~\ref{unequal-prob-tbl} document the appropriate comparisons when, for each class, one of the four  prototypes  appears at one-tenth the frequency of the other prototypes in that class (progressive training with 20/30 perturbations). There are two control training-conditions. First, there is an equal frequency set of prototypes, designated 1,1,1,1. This is the standard progressive paradigm with all prototypes equally probable although the prototypes are introduced in the usual progressive manner. The second control, designated 1,10,10,10 introduces the low probability prototypes (one for each class) in the first of the four phases of training. Obviously, there is no within class competition going on during this first phase. However, on every later phase, each of the pairs (one for each class) of low probability prototypes is trained one-tenth as much as any other prototype in its class. Thus, in later phases, a low probability prototype is associated with  a smaller and smaller fraction of exemplars per epoch; in the final phase of progressive training, such an exemplar occurs once out of the 31 exemplars (of its class) per epoch. The important training-challenge  is the third set, designated 10,10,10,1. In this case the low probability prototypes do not get trained until the final phase of progressive training. Note that in each epoch of this last phase, there are 30 occurrences total for the three other  prototypes of the same class for every single training trial devoted to the low probability prototype.  
 
 \begin{table}[H]
 	\centering
 	\resizebox{1.0\columnwidth}{!}{%
 		\begin{tabular}{c|c|c|c|c|c}
 			
 			Prototype              &  Class     &  Dendrites &   Connections    &   D vs P$^{\star}$   &  D vs D$^{\dagger}$         \\ 
 			Occurrences           &  Error      & Count/Nrn  &  Count/Dendrite  &   Max   &  Min \\ 
 			per epoch per class &   Mean (SEM) & Median    &  Median           &  Angle (deg)    &   Angle (deg)  \\ \hline
 			
 			1,   1,   1,   1            & 0.00 (0.00)         & 4              &  102            &   28.8            &   62.7   \\
 			1, 10, 10, 10           & 0.00 (0.00)          & 4              &  102            &   29.0            &   62.8   \\
 			10, 10, 10,   1           & 0.00 (0.00)          & 4              &  102            &   28.7            &   62.9   \\ \hline
 			
 		\end{tabular}
 		
 	}
 \footnotesize{$^{\star}$ every dendrite against every prototype $\quad  ^{\dagger} $ all six dendritic pairs}
 	\captionsetup{width=0.98\textwidth}
 	\caption{Classification error and connectivity are unaffected by unequal prototype probabilities with progressive training.
 		The mean and standard error of the mean are provided for the class error.
 		The dendrite and connection counts are median values rounded up to the nearest integer and were the
 		same across all dendrites and neurons.
 		``Max Angle D-P'' is the maximum angle between dendrites and their preferred prototype across all dendrites,
 		both neurons, and 10 simulations;
 		``Min Angle D-D'' is the  minimum angle between dendrites across all dendrite pairs (six pairs per neuron), 
 		both neurons, and 10 simulations. 
 		Dataset: 8, 256D, nonorthogonal patterns, 4 prototypes per class, 30\% ON noise and 20\% Occlusion. 
 		( $\alpha = 1,\;\epsilon_w = 0.025$,\; $\epsilon_\gamma = 0.05$, and convergence criterion 500 trials.)
 	}
 	\label{unequal-prob-tbl}
 \end{table}
 
Notably all three variants of  the progressive training paradigm result in 10 out of 10 simulations with perfect, error-free performance. Moreover, as indicated by the numbers in the table, there are no notable differences between the neurons that develop in each of the paradigms. These comparisons include number of dendrites per neuron, number of synapse per dendrite, angle between the low probability prototype and the dendrite that recognizes its this prototype's exemplars (`Max Angle D-P''), and finally, there is no meaningful difference across angles generated by dendritic pairs (``Min Angle D-D''). The results can be compared to other results presented previously and to be presented further below. In sum, the imbalanced relative frequency of 10:1 is well tolerated.

It is worth pointing out that the best settings used here for concurrent training are different than the settings for the progressive training paradigms. Also, as will be described later, the nature of concurrent developed neurons can be altered with changes of certain parameters. For concurrent training, the value of $\epsilon_\gamma$ is decreased from 0.05 of phased training to 0.03. Then concurrent training yields  zero error performance even when there are the two low probability prototypes.  Notably, with this parameterization,  neurons are producing zero errors with only  two dendrites per neuron. 

There is no need to test segregated training as it never simultaneously trains with exemplars from different prototypes.

\subsection{Resistance to catastrophic interference across two tasks}\label{two-tasks}
There are many recent algorithms that resist catastrophic interference \citep{Grossberg1976,Carpenter1987,Carpenter1991,McCloskey1989,French1991,French1999,McClelland1995,
	Kirkpatrick2017,Zenke2017,Masse2018,Limbacher2020,Delange2021}.
Some of these papers go beyond taxing the memory limits by explicitly creating inputs that will overwrite, and here such overwriting of previously stored memories is conceivable given the possibility of  shedding and severe weight modifications. 
One particular method that encourages overwriting first trains on one set of prototypes (task 1) and then permutes each prototype vector and then trains again (task 2). That is, each class specific neuron must learn a new but related prototype. For example, in the case of  images, the ordering of pixels are permuted. Thus the memory attack requires remembering  two types of inputs, the original unpermuted input vectors and the permuted input vectors.

Using $\epsilon_w = 0.025$, $\epsilon_g = 0.05$, and $\alpha = 1$, this method was applied to the eight, 256-D prototypes for the 4$|$4 problem without random perturbation (task 1) and  trained with  20/30 exemplars (task 2). The network is  progressively trained on task 1 for 1000 epochs and then the network was concurrently trained on task 2 for an additional 1000 epochs. Testing was performed on a different set of 800 unpermuted test patterns, and then on another set of 800 permuted test patterns. Without randomization, there were no errors on both the unpermuted and permuted test sets; with 20/30 randomization, the mean error on the unpermuted test set was 0.34\% (with a standard error from the mean of 0.072) and the mean error on the permuted test set was 0.30\% (0.092). Thus, the total mean test error across the two tasks was below one percent, i.e., 0.64\% implying that the two-task problem is successfully encoded and overwriting is avoided.

\subsubsection{ Further observations on the limits of generalization and robustness}
As already noted, the ability to abstract and generalize typifies the  neural paradigm, and as earlier prototype perturbation results indicate, the full algorithm, DC\&SAS, leaves this desirable property intact. This subsection explores perturbation effects in an arguably  more systematic way. That is, we quantify  the ability to generalize by training with a fixed 20/20 perturbation and then  testing under different amounts of prototype perturbations. (Recall earlier, that training and testing perturbations where altered in parallel).
\begin{figure}[H]
	\centering
	\includegraphics[width=0.55\textwidth]{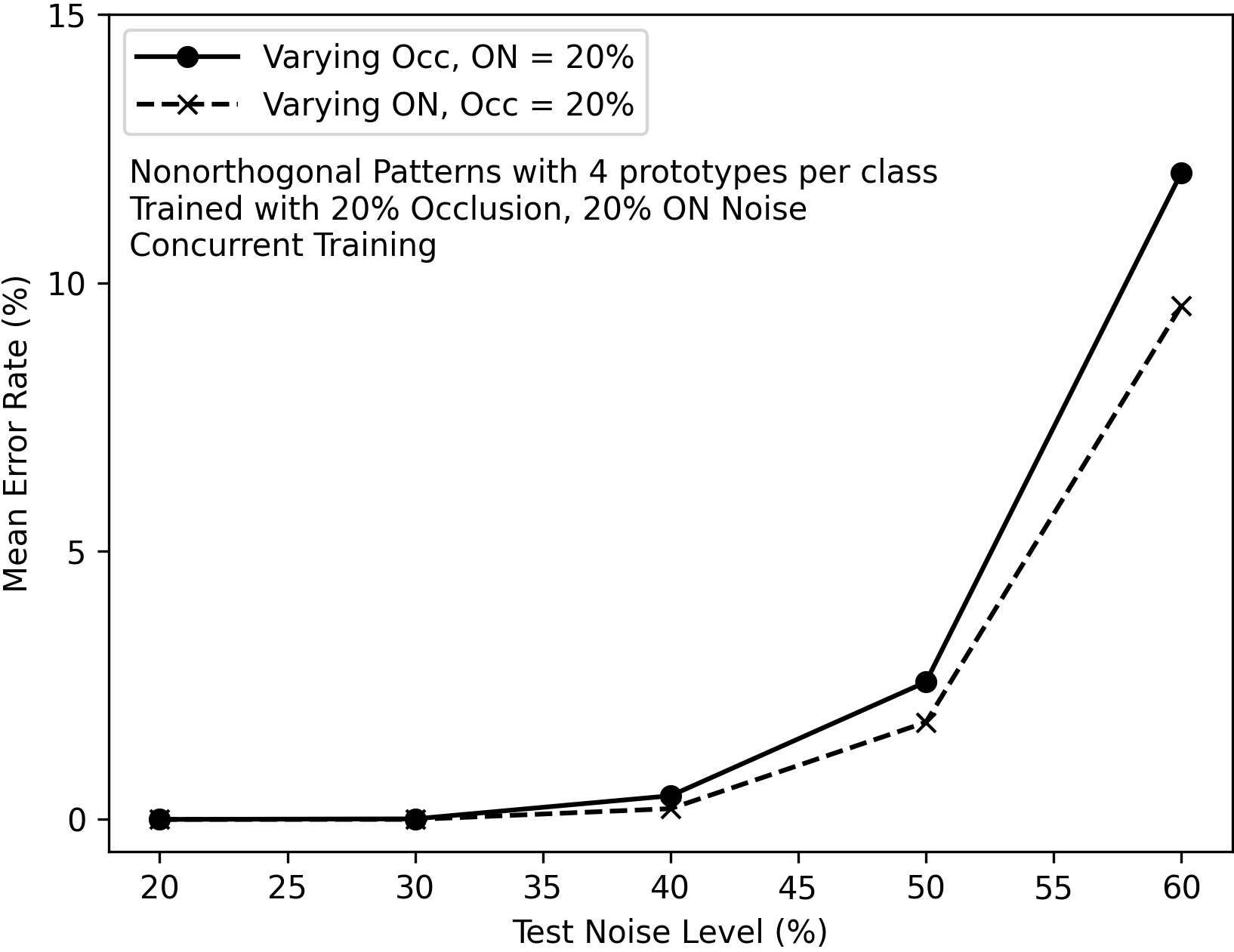}
	\caption{
		One measure of generalization is how much classification performance resists degradation with increased occlusion
		or increased on-noise during testing. This graph shows that
		classification performance is resistant to approximately twice the amount of occlusion or twice the amount of  on-noise during testing. 
		All training is  concurrent with 20\% occlusion and 20\% on-noise.  During testing one of these randomizations (occlusion or on-noise) remains
		fixed at 20\% while the other randomization is systematically varied. $\alpha=1;\; \epsilon_w=0.025;\; \epsilon_\gamma=0.05$; stability criterion is 800 epochs without a connection change.
	}
	\label{diff-test-noise-concurrent-fig}
\end{figure}

Again we  study the 4$|$4 problem with equiprobable prototypes  All training exemplars are generated by 20/20 perturbations. During testing either one of these randomizations (either occlusion or on-noise) stays  fixed at 20\% while the other randomization is systematically varied. Figure \ref{diff-test-noise-concurrent-fig} shows the resulting learned performance when concurrent training is used. 
The two different randomizations varied during testing produce error curves that are qualitatively similar. Quantitatively, fractional  occlusion during testing  is slightly more deleterious to performance than the same value of  fractional on-noise during testing. Importantly for either type of randomization, 40/20 and 20/40, error-rates are well under one percent. More pointedly, with the differing perturbations during  training vs at test, there is an absolute guarantee that the exemplars with the higher on-noise during testing are well beyond the  cluster of exemplars experience during training. Similarly for  high occlusion levels, there is a guarantee that testing exemplars, whose signals are remote from the training exemplars, can fall into a subspace of active input-lines equally shared by exemplars generated from two different prototypes. 

\begin{figure}[H]
	\centering
	\includegraphics[width=0.55\textwidth]{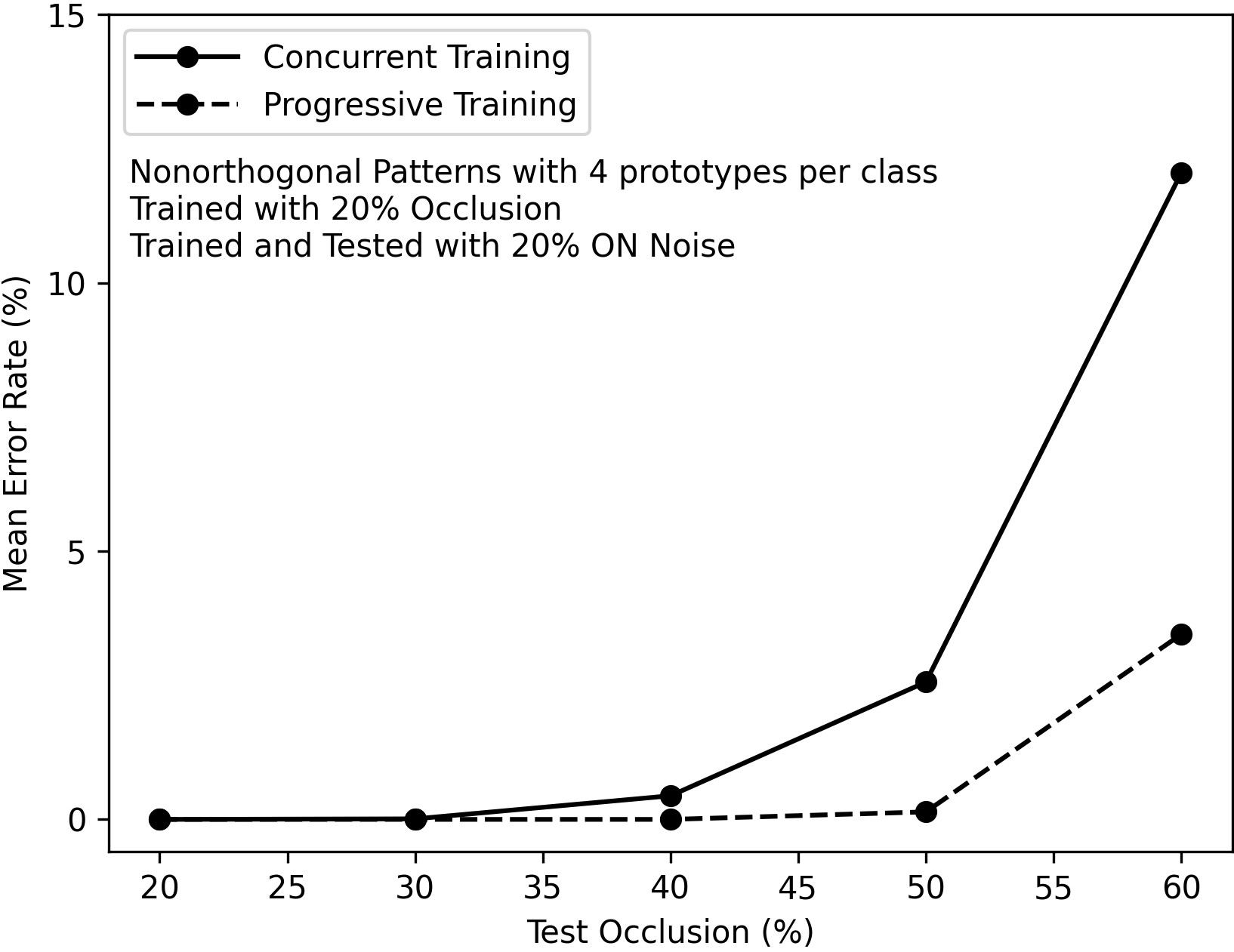}
	\caption{Progressive training at 20/20 produces better generalization than  concurrent training at 20/20. Generalization is judged by testing at occlusions greater than 20\%.
		The two-dendrites per neuron of concurrent training do not generalize as well as the four-dendrites per neuron the develop  with progressive training. The superior ability to generalize when using progressive training  is barely noticeable at prototype perturbation of 40/20 but is easily seen at higher levels of occlusion. The concurrent data are a replotting of data from the previous figure. Same parameterizations as this previous figure, Fig \ref{diff-test-noise-concurrent-fig}.
		.
	}
	\label{con-vs-pro-test-occ-fig}
\end{figure}

While on the topic of generalization, one might ask if the four dendrite solution produced by progressive training to the 4$|$4 problem is superior to     a two dendrite per neuron solution produced by concurrent training as occur using our standard parameterizations. To help answer this question, one compares  concurrently trained networks with their potentially zero-error, two functional dendrites per neuron solution to  progressively trained networks with their potentially zero-error, four dendrite per neuron solution. In fact when faced with higher perturbations during testing, progressively trained                                                                                                                                                                                                                                          simulations, which produce the four dendrite per neuron solution, yield better performance compared to concurrent training, with its  two functional dendrites per neurons. Fig  \ref{con-vs-pro-test-occ-fig} compares two networks, one progressively trained vs another concurrently trained; both training styles used 20/20 prototype perturbations. The testing challenged each of the two developed networks at larger occlusion values than used during training (replicated in 10 simulations for each training paradigm). Progressive training is superior with nearly error-free at 50/20 testing, 0.14\%,  while with concurrent training testing at 50/20 produces a larger,  2.56\% error-rate. Moreover, this  difference of error-rates grows as fractional  occlusion during testing increases.  

The next result continues  with the question of performance generalization, in particular, when testing perturbations differ from training perturbation. For the first such comparison, Fig \ref{diff-occ-pro-fig} replots  the  test-data of the progressively trained networks of Fig  \ref{con-vs-pro-test-occ-fig}. These replotted data (the dashed line) serve as a baseline for comparing the effect of training and testing with the same perturbations at both stages. (Note differences from the previous figure, the  y-axis scaling and the extension to 70\% occlusion, which is not plotted in the earlier figure.) Beginning at the 40/20, the deleterious effect of larger training perturbations appears, and this deterioration of performance accelerates with greater prototype perturbations during training. The specific percent error-rates for sequential occlusion values of $\{40, 50, 60, 70\}$ percent are $\{0 \,\textrm{vs}\,2.5,\;0.1 \,\textrm{vs}\,4.8,\;3.5 \,\textrm{vs}\,14.1,\;30.1 \,\textrm{vs}\,45.7,\;\}$, fixed 20/20 training vs perturbation-matched training and testing, respectively..

\begin{figure}[H]
	\centering
	\includegraphics[width=0.55\textwidth]{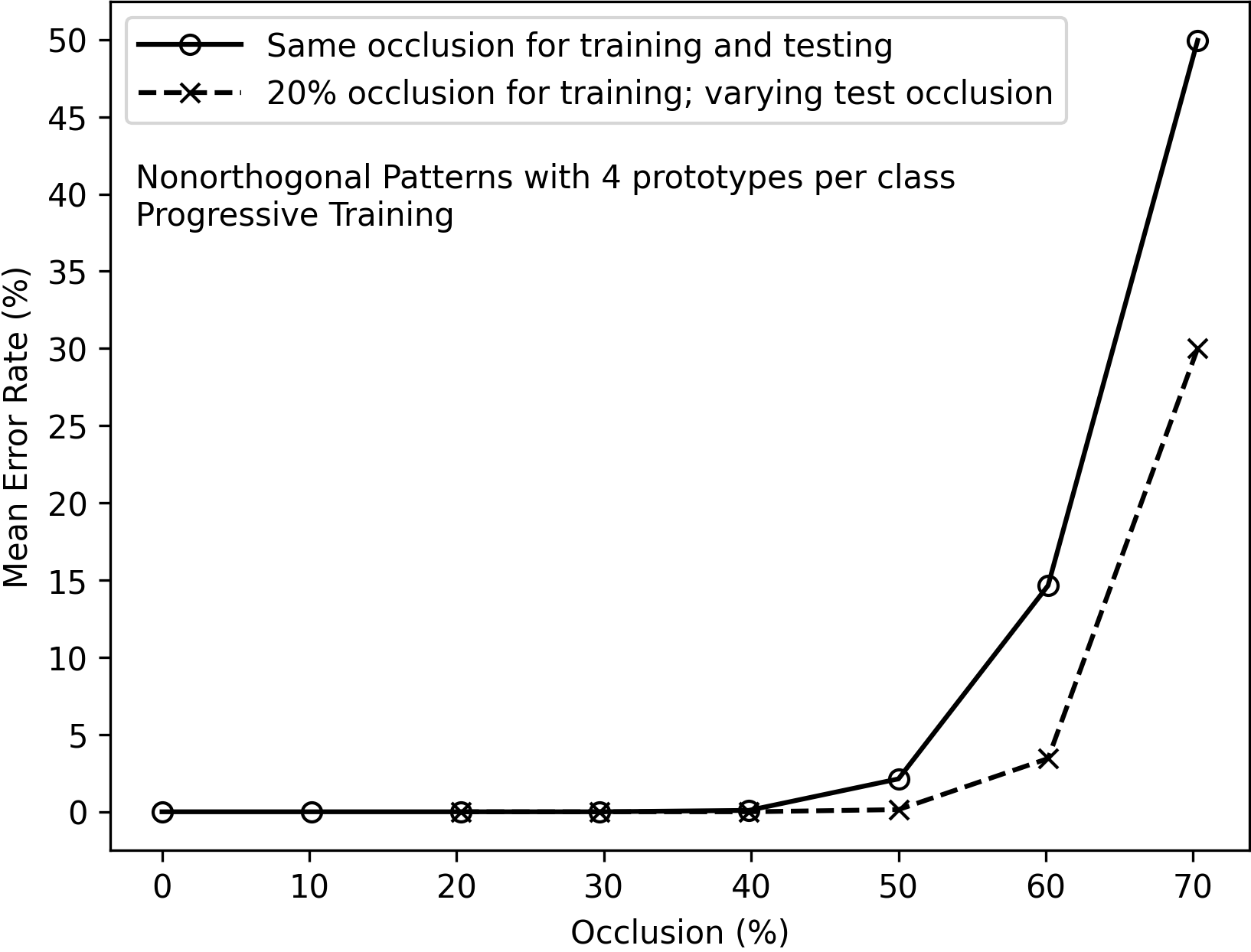}
	\caption{
		Training with greater prototype perturbations does not help generalization.
		Occlusion during training, not during testing, tends to impose a stronger limitation on the classification accuracy.
		This graph compares using the same occlusion for training and testing against using a fixed 20\% occlusion for training
		and varying the test occlusion. Beyond 30\% occlusion, using the same occlusion for training and testing results
		in higher error-rates and thus poorer generalization. The patterns were based on the nonorthogonal, 256-D prototypes with four prototypes per
		class (4,4) and progressive training. Same parameterizations as Fig \ref{diff-test-noise-concurrent-fig}.
	}
	\label{diff-occ-pro-fig}
\end{figure}

These comparisons call for a full comparison of training perturbation vs testing perturbation effects, especially a comparison that addresses the interaction effect directly. To examine the interaction effects the simulations are performed near the boundary of zero vs non-zero error-rates.

Table \ref{contingency_table}'s contingency table confirms the super-additive effect of training and testing at a prototype perturbation larger than 30/20. Here 50/20 perturbations are used. The predicted error-rate from an additive effect is  $0.08 + 0.19 = 0.27\%            $ while the actual error rate is 1.05\%. Thus the actual error-rate of 1.05 is about four times the error-rate of a purely additive effect.

These four different types of simulations with altered parameter settings of $\epsilon_w-0.025$ and 800 training epochs yields the super-additive effect except with higher error rates including 50/20 train 30/20 test having a higher error rate than 30/20 train and test.

\begin{table}
	\centering
	\begin{tabular}{c >{\bfseries}r @{\hspace{0.7em}}c @{\hspace{0.4em}}c @{\hspace{0.7em}}l}
		\multirow{10}{*}{\parbox[c]{1.1cm}{\bfseries\raggedleft \vspace{0.5in} \LARGE{Test}}} & 
		& \multicolumn{2}{c}{\bfseries \LARGE{Train}} & \\
		& & \bfseries 30/20 & \bfseries 50/20 &  \\
		& 30/20 & \MyBox{ \LARGE{0.06\% }}{\small{(0.06\%)}} & \MyBox{ \LARGE{0.08\%} }{\small{(0.05\%)}} &   \\ [2.4em]
		& 50/20 & \MyBox{ \LARGE{0.19\% }}{\small{(0.13\%)}} & \MyBox{ \LARGE{1.05\% }}{\small{(0.14\%)}} &  \\
	\end{tabular}
	\caption{Super-additive effect on error-rate when both training and testing exemplars are 50/20  prototype perturbations. Progressive training for 3000 epochs. $\epsilon_w=0.002;\;\epsilon_\gamma=0.05;\;\alpha=1$.
	}
	\label{contingency_table}
\end{table}
To summarize this section, we hypothesize that  (i)  DC\&SAS is as capable of generalization as any other neural paradigm that encodes pairwise statistics in its synaptic weights, that (ii) progressive training with its four-dendrite solution produces a more  robust result compared  to concurrent training with its two functional dendrite solutions; and finally  in the case of occlusive randomization (iii) training at higher perturbations does not prepare the developed network for testing at these same higher perturbations.

\subsection{Why does DC\&SAS work?}
Due to the highly overlapped nature of the feature vectors associated with different classes, the discrimination problems here require non-linear separations for error-free performance. A multiple dendrite allowance per neuron provides at least the possibility of such error-free performance. However, multiple dendrites per neuron is often not enough to guarantee such performance. In many situations, there needs to be a developmental mechanism that encourages a distinctively different connectivity for the dendrites of a single neurons. In this regard, CDS is a critical part of the developmental algorithm.
  
Before literally dissecting the DC\&SAS algorithm, this section first quantifies the distinctiveness of the different dendritic connectivities. In the cases when there is a dendrite for every prototype, this quantification suggests conceptualizing  each dendrite's converged synaptic weight vector as its estimate of one particular prototype. In the cases where training uses a moderate amount of random prototype perturbations, each dendrite is learning a conditional approximation of a cluster center, and this dendritic weight vector  is most strongly excited by the exemplars of one particular, latent prototype. That is, each such cluster center reflects the one prototype that a particular dendrite recognizes. Moreover, as the number of prototypes per class increases, the number of dendrites per relevant neuron tends to increase.  

\subsubsection{What are the dendrites learning when  the prototypes are not latent?}
The answer to this question can be simple, but only for the simplest problem set.
In the case of the 4-D XOR problem set, the two class-supervised neurons each end up with two dendrites. For each neuron, the dendritic pair is an  orthogonal set. That is, each dendrite of one neuron acquires two of the four possible synaptic connections, and these two connections are totally different for each dendrite of one neuron. Thus the angle between dendrites of one neuron is  ninety degrees.

In the case of progressive and segregated training, the 8-D problem sets are also easily interpreted since dendrites match prototypes. However in terms of dendrites vs prototypes, with concurrent training there is a wider variety of zero-error solutions with many dendrites recognizing two prototypes. 

In contrast to this simplest example, there is more complexity to the learned solutions for the 8-D problem sets. Challenging the algorithm with the noise-free 8-D problem sets can lead to a variety of solutions. The solutions vary with the problem set and the training method. In general, for progressive and for segregated training, there is a consistent tendency to develop more dendrites on a neuron when there are more prototypes to be learned by that neuron. 

\subsubsection{What are the dendrites learning when  the prototypes are  latent?}
Depending on the training style (concurrent vs either of the two phased methods) and depending on parameters setting ($\epsilon_w$ and $\epsilon_\gamma$), there are some quantitative differences in the dendrites that form when exemplars are randomly perturbed versions of the 256-D prototypes. However,  there is, arguably, one particular type of result that is of greatest interest, and this is where we start. Later parts of this subsection describe results that differ quantitatively from the result we find most engaging.

\subsubsection{DC\&SAS can selectively encode each latent variable with one and only one dendrite}

When using progressive or segregated training, there are a wide range of parameter settings  that create one and only one dendrite matched to every prototype for 30/20 and 20/30 perturbations. For example in the case of progressive training, $0.002\leq \epsilon_w \leq 0.05$ and $0.001\leq \epsilon_\gamma  \leq 0.05$. 
 
\begin{table}[h]
\centering
\begin{threeparttable}
\renewcommand{\arraystretch}{1.1}
\begin{tabular}{|c|c|c|}
	\hline
 Prototypes      &  Median Dendrites     &   Median Connections    \\
 per Class        & per Neuron           &   per Dendrite  per Neuron  \\
  \hline
    2$|$2$|$4       &     2; 2; 4                       &   90     \\ 
    2$|$3$|$3       &     2; 3; 3                       &   90     \\
    4$|$4           &     4; 4                        &    90    \\ 
    2$|$6           &     2; 6                        &    90    \\
    \hline

 \end{tabular}

\begin{tablenotes} \footnotesize
\item[]progressive training; 30/20; $\epsilon_w=0.002$; $\epsilon_\gamma=0.05$
\end{tablenotes}
	
\end{threeparttable}
\captionsetup{width=0.95\textwidth}
\caption{Dendrites match prototypes for each class, 30/20 prototype perturbations}
\label{nonortho8-256d-varying-camode-on20-off30-tbl}

\end{table}
Table \ref{nonortho8-256d-varying-camode-on20-off30-tbl} summarizes a result for progressive training on all four problem sets with exemplars formed through 30/20 perturbations with $\epsilon_w=0.002$ and $\epsilon_\gamma=0.05$. For each of the four problem sets, performance is perfect, zero-errors for ten out of ten simulations. 

\begin{table}[h]
	\centering
	\resizebox{0.7\columnwidth}{!}{%
		\renewcommand{\arraystretch}{1.1}
		\begin{tabular}{|c|c|c|}
			\hline 
			Prototypes      &  Median Dendrites     &   Median  Connections  \\
			per Class        &  per Neuron &   per Dendrite  per Neuron   \\ \hline
			\multicolumn{3}{|c|}{Progressive Training 20/30} \\ \hline
			$2|2|4$       & 2; 2; 4                       &   All 102   \\ 
			$4|4$       & 4; 4                           &  All 102 \\ 
			$2|6$       & 2; 6                           &  All 102   \\ 
			\hline
			\multicolumn{3}{|c|}{Segregated Training  20/30} \\ \hline
			$2|2|4$       &    2; 2; 4                        &   All 102   \\
			$4|4$     & 4; 4                            &   All 102  \\
			$2|6$           &    2; 6       &   \{103, 102\};\{102, 102; 102, 102, 102, 102\}\\
			\hline 
		\end{tabular}
	}
\caption{Dendrites match prototypes for each class, 20/30 prototype perturbations}
	\label{nonortho8-256d-tbl}
	
\end{table}

For a little variety, the perturbations are changed to 20/30 and similar results are obtained, Table~\ref{nonortho8-256d-tbl} using  $\epsilon_w=0.002$ and $\epsilon_\gamma=0.05$. Again performance is perfect for all simulations (10 out 10 for each problem set and each type of training), and again, each dendrite matches one of the latent prototypes.

Regarding the noted training dependence of the dendrite matching prototype development, the two-dendrite solution that the standard settings produce when using concurrent training  can be altered to a four-dendrite per neuron solution. In particular, such a result is accomplished by substantially speeding up  synaptogenesis via  $\epsilon_\gamma=0.999$ instead of $\epsilon_\gamma=0.05$ backed up by a reduced $\gamma_{dj}(t)$ decrement factor, 0.001 instead of 0.95, plus using an average  missed-detection signal via $\alpha=0.01$ instead of 1.0.  With these new parameters concurrent training finds an error-free, typically  four-dendrite per neuron solutions instead of a two-dendrite per neuron solution  to the $4|4$ 20/30 and 30/20 problems. Quantitatively the new parameter settings yield the four-dendrite solution 80\% of the time, and with such four-dendrite per neuron developments, each dendrite is recognizing one and only one prototype.

These results justify the idea that dendrites are learning conditional cluster centers. In the case of progressive and segregated training with moderate prototype perturbations, each cluster center is a mildly perturbed version of a single prototype. In the case of concurrent training with the typical parameter settings, there is a strong tendency to merge a pair of underlying cluster centers. As part of this merging, a smaller fraction, compared to phased training, of each prototype is combined to create such a merger of cluster centers.

\subsubsection{Quantifying dendrites as cluster centers}

\begin{figure}[H]
	\centering
	\begin{subfigure}{0.48\textwidth}
		\includegraphics[width=\textwidth]{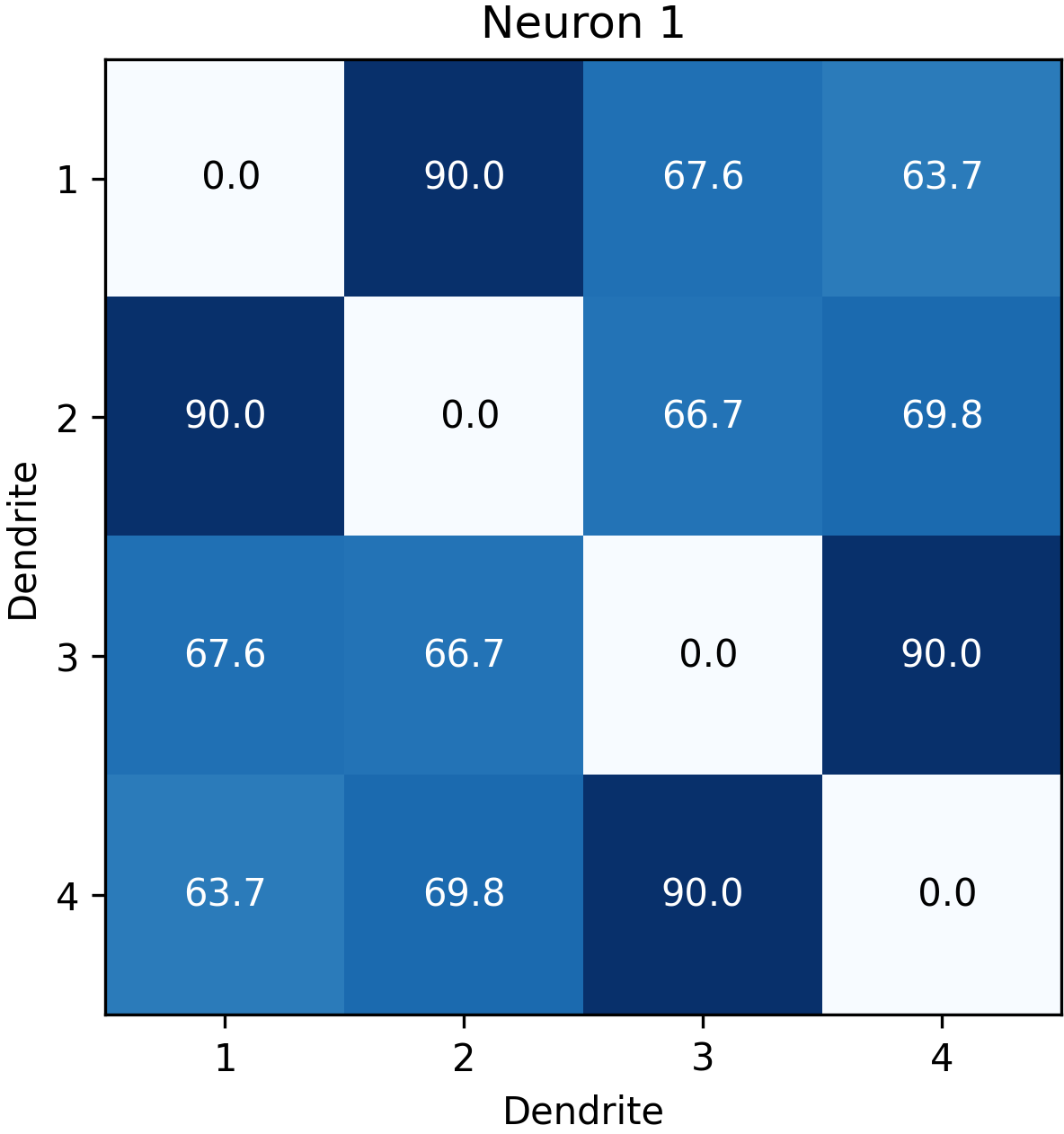}
		\caption{}
	\end{subfigure}
	\hfill
	\begin{subfigure}{0.48\textwidth}
		\includegraphics[width=\textwidth]{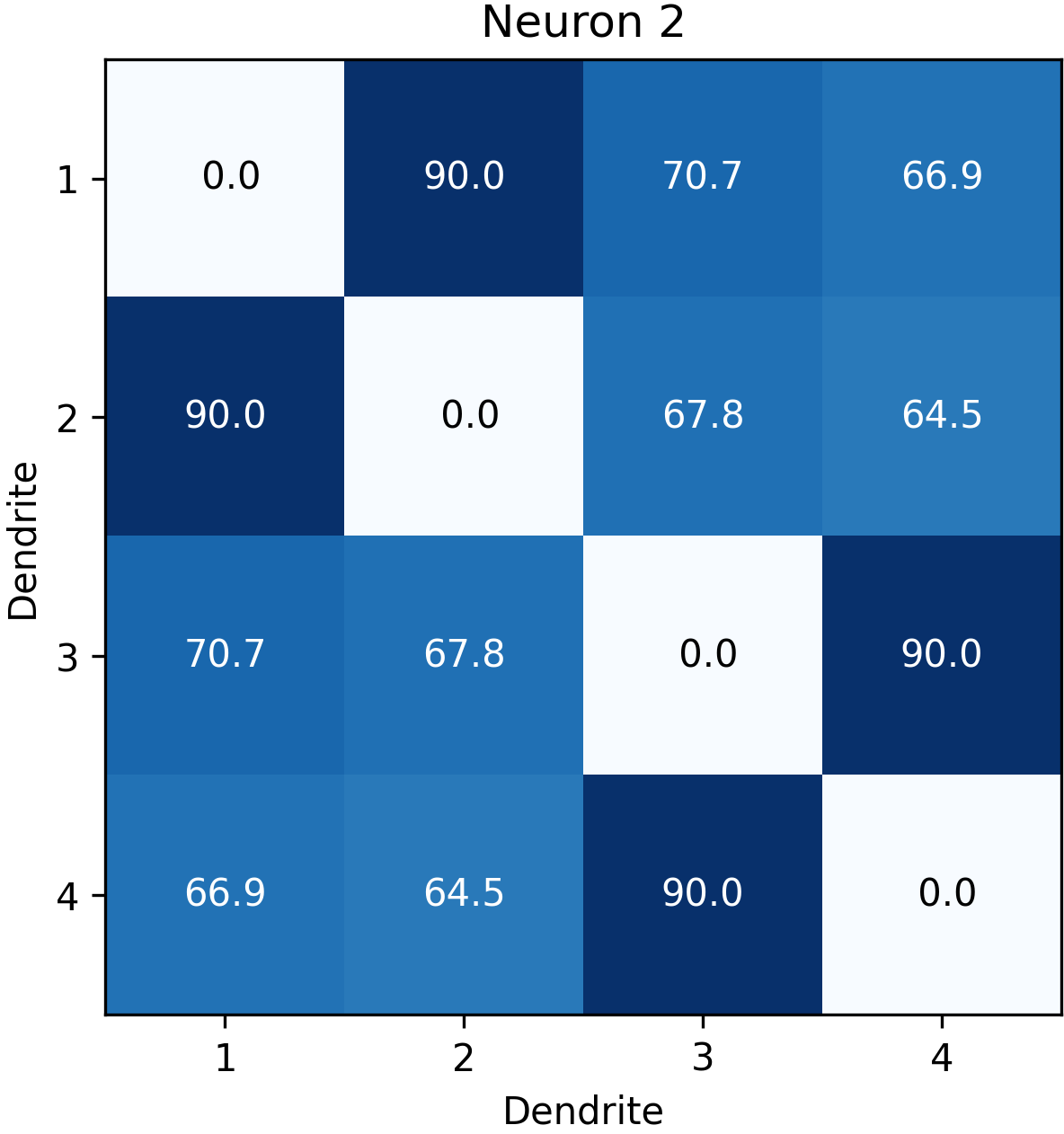}
		\caption{}
	\end{subfigure}
	\caption{
		The angles between dendrites nearly match the angles between prototypes (compare to Figure~\ref{angle-matrix-8d}). 
		The 90 $\deg$ angles are reproduced while  the dendritric pairs are  at a slightly greater angle to each other than the prototype pairs of 60 $\deg$.
		Each square indicates the angle between two dendrites on the same neuron.
		Dataset: nonorthogonal, 256-D patterns based on 4 prototypes in each of two classes (4,4) with 20/30 prototype perturbations.
		Progressive training.
		($\epsilon_w = 0.025$; $\epsilon_\gamma = 0.05;\;\alpha=1$.)
	}
	\label{dendrite-angles-44-pro-fig}
\end{figure}
This remarkable result, unmixing of all the conditional mixture distributions, deserves further quantification. This section quantifies all the dendritic pairs and dendrite-prototypes pairs for the $4|4$ and the $2|6$ problems using 20/30 perturbed prototypes.

First the dendrite-to-dendrite comparisons are made for these two problems. These figures, Fig \ref{dendrite-angles-44-pro-fig} and Fig \ref{dendrite-angles-26-pro-fig}b,  resemble the earlier Fig \ref{angle-matrix-4d-vs-8d-fig}b  that compares all pairs of  prototype angles for the 256-D generating set. Through this comparison, we see that the dendrites are packed about as tightly as the original prototypes. Of course the diagonals match but so do the orthogonal comparisons. The only quantitative difference  is that the dendrites are slightly farther apart, for the $4|4$ problem ca. 67 degrees for the dendrites and for the $2|6$ problem ca. 68 degrees, compared to the 60 degrees of the prototype pairings. The reason for this difference is that the dendrites  have only 102 inputs each, a proper subspace of the 128 positively valued dimensions of the prototypes.

\begin{figure}[H]
	\centering
	\begin{subfigure}{0.48\textwidth}
		\includegraphics[width=\textwidth]{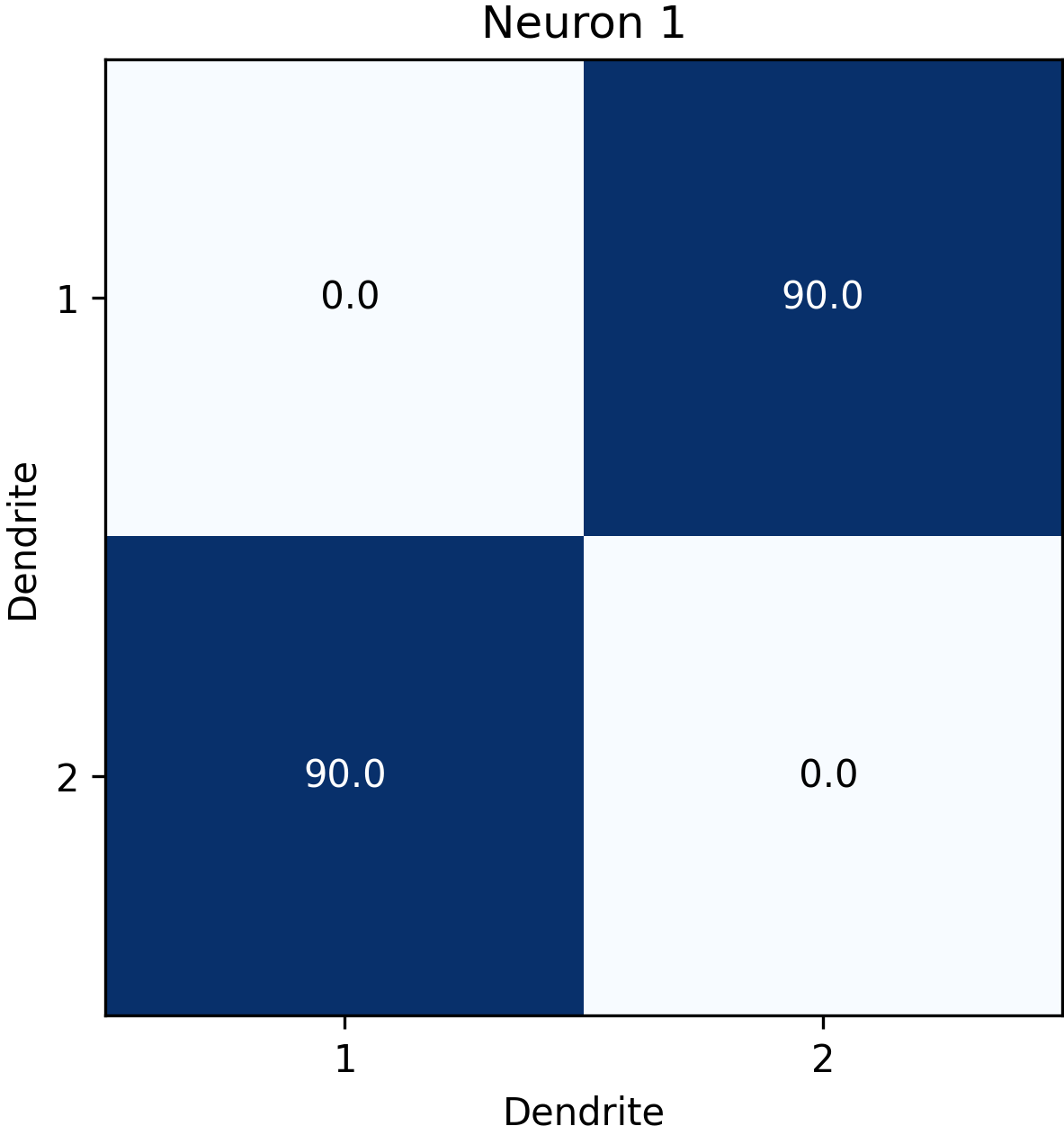}
		\caption{}
	\end{subfigure}
	\hfill
	\begin{subfigure}{0.48\textwidth}
		\includegraphics[width=\textwidth]{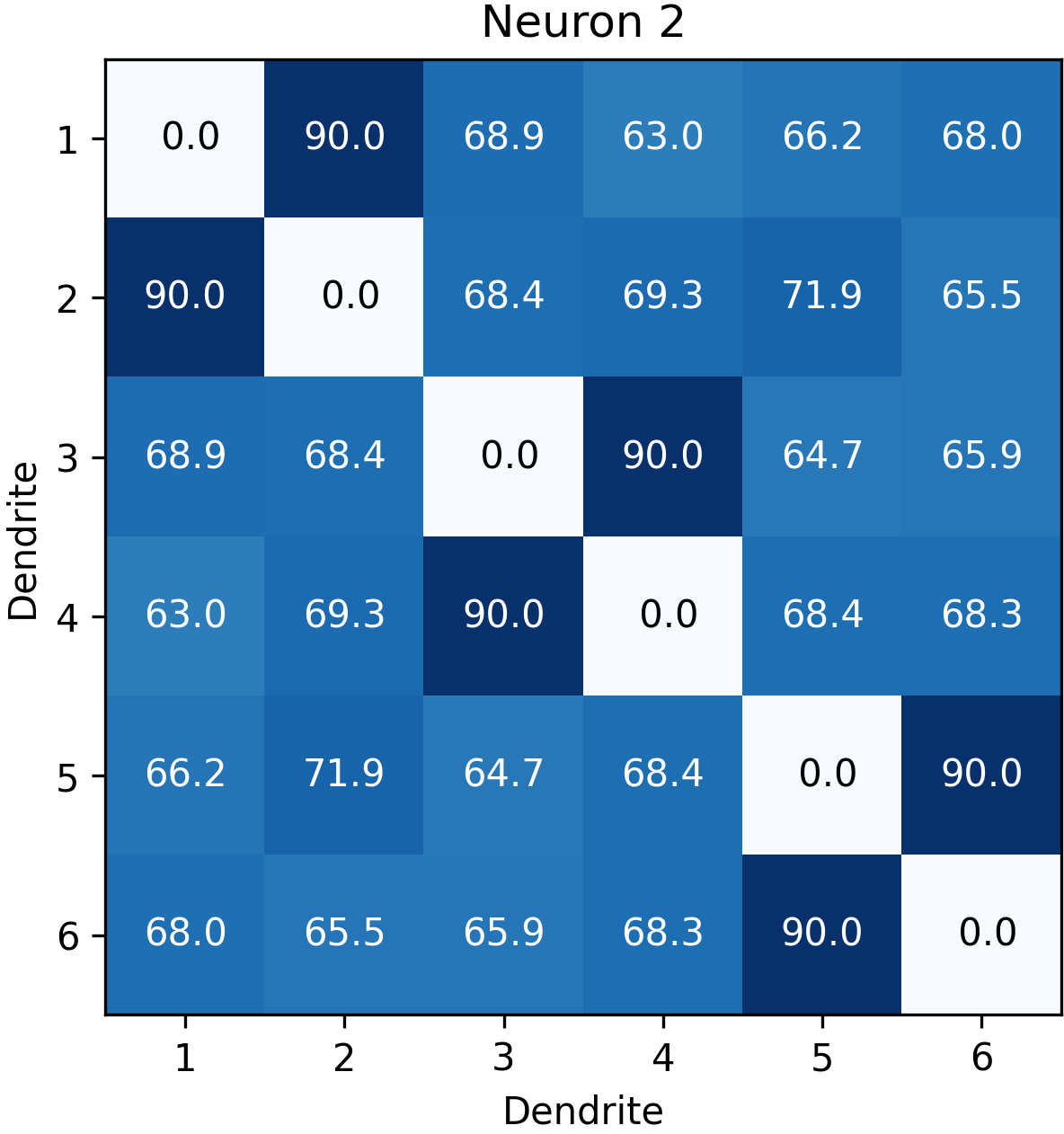}
		\caption{}
	\end{subfigure}
	\caption{
		Just as in the previous 4$|$4 problem, the 2$|$6 problem produces dendritic pairs with slightly greater angles (where possible) than prototype pairs. 
		Once again these angles should be  compared to Figure~\ref{angle-matrix-8d}.
		Each square indicates the angle between two dendrites on the same neuron.
		Dataset: nonorthogonal, 256-D prototypes, (2,6) with 20/30.
		Progressive training.
		($\epsilon_w = 0.025$;\; $\epsilon_\gamma = 0.05;\;\alpha=1$.)
	}
	\label{dendrite-angles-26-pro-fig}
\end{figure}

Having seen these dendritic angles, the dendrite-to-prototype angles of Figs \ref{dp-angles-44-pro-fig} and \ref{dendrite-angles-26-pro-fig} will not be surprising. The 90 degree angles match up to the prototype-against-prototype angles. The angles on the diagonals of Figs \ref{dendrite-angles-44-pro-fig} and \ref{dendrite-angles-26-pro-fig} are positive rather than zero because there are 102 connections per dendrite while each prototype has 128 dimensions valued one. For the same reason, the angles that are sixty degrees for the prototype-to-prototype comparisons are again slightly larger for the corresponding dendrite-to-prototype comparisons, about 64 degrees.

Thus the simulations show that each dendrite points to a subspace belonging to just one of the latent prototypes when trained with exemplars of moderately perturbed prototypes. To say it another way, each dendrite's input set is a proper subset of the $\{x_i=1\}$'s defining  one and only one prototype.

\begin{figure}[H]
	\centering
	\begin{subfigure}{0.70\textwidth}
		\includegraphics[width=\textwidth]{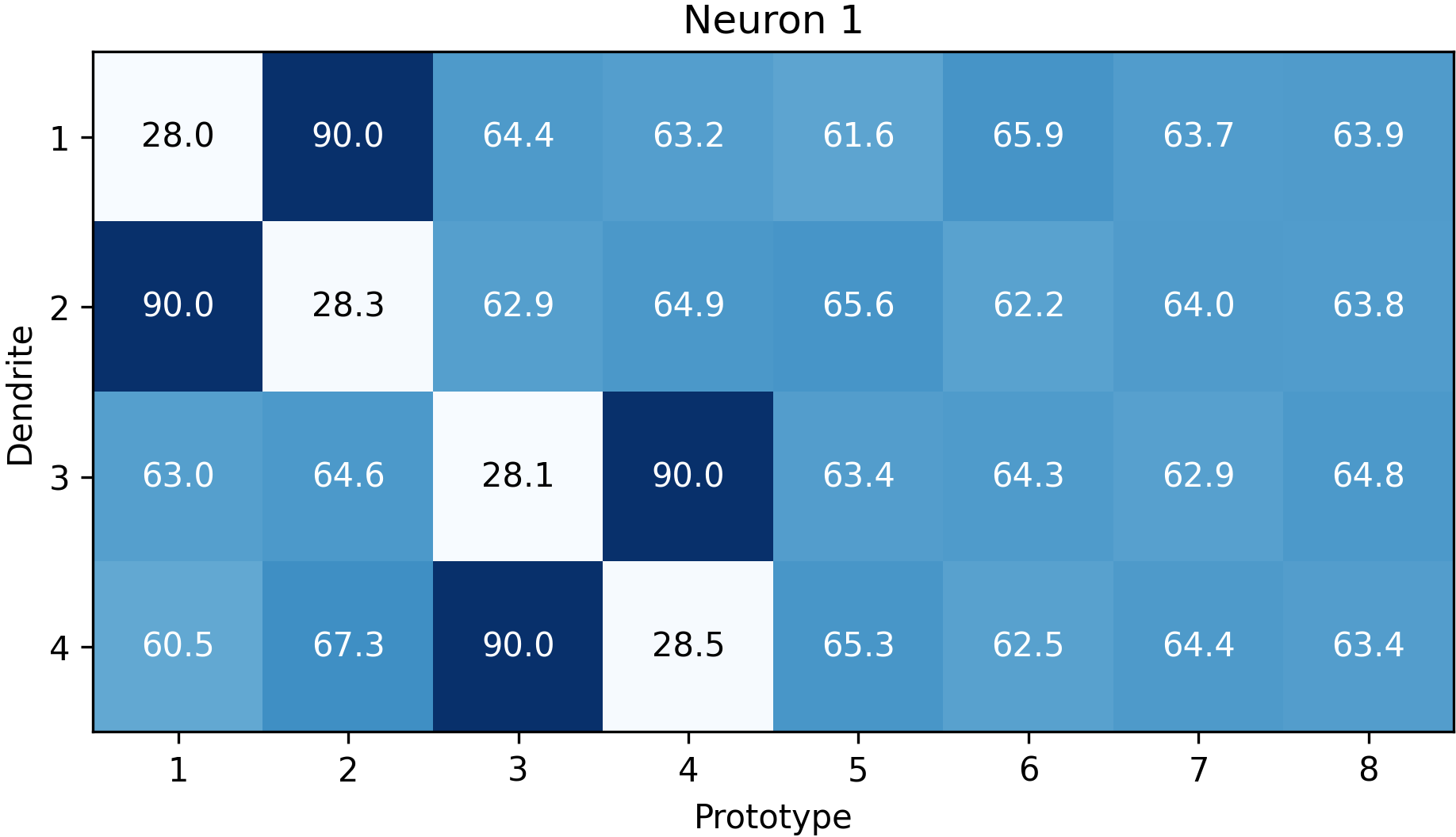}
		\caption{} \vspace{0.1in}
	\end{subfigure}
	\begin{subfigure}{0.70\textwidth}
		\includegraphics[width=\textwidth]{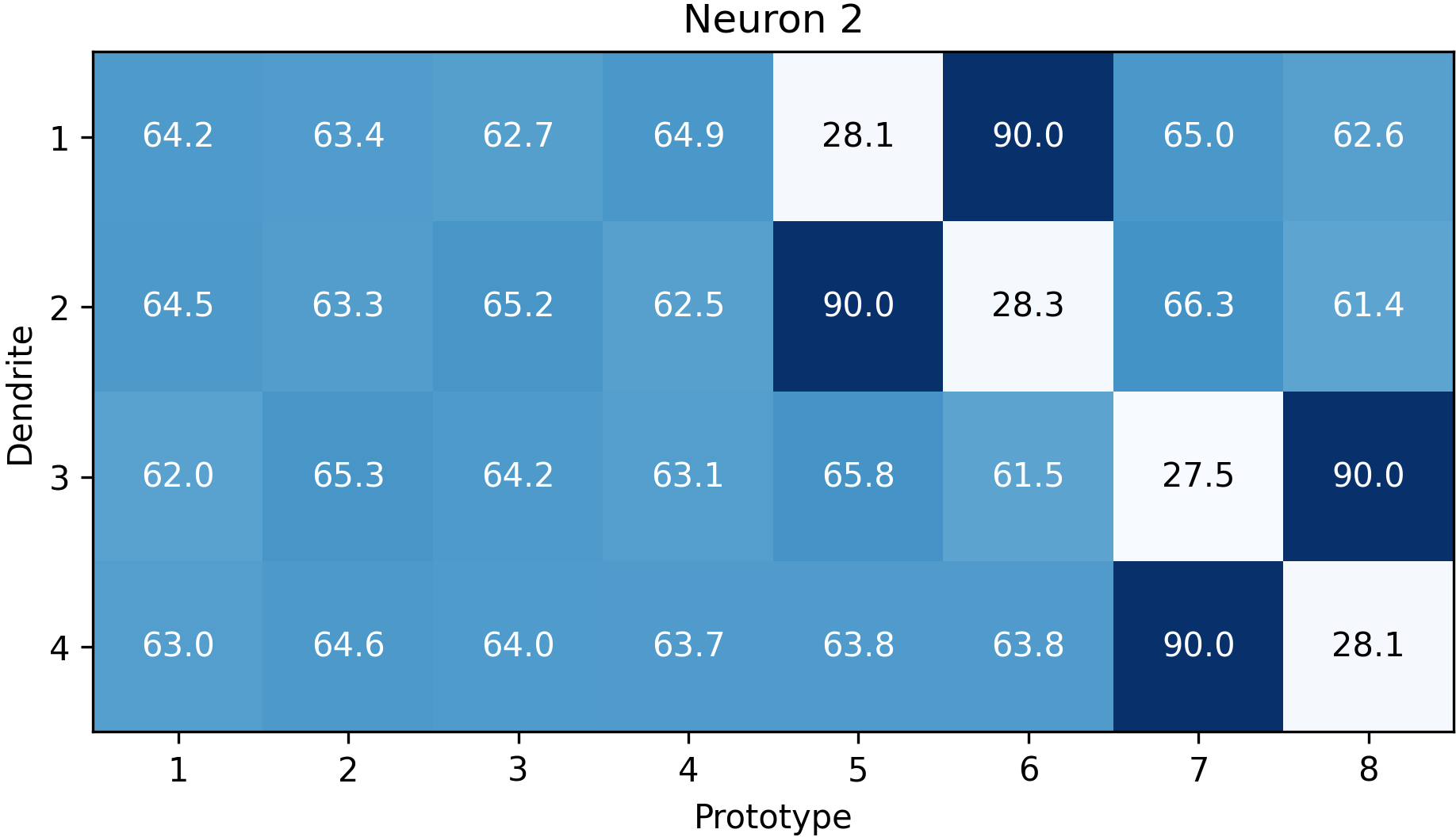}
		\caption{}
	\end{subfigure}
	\caption{
	Each dendrite points towards just one prototype, $~28.1\deg$, and away from the others, $~63\deg$. A dendrite fails to point exactly in the direction of its prototype because each dendrite has, on average, 102 of the prototype's 128 one-valued dimensions. Up to some small amount of randomness, for the same reason the angle between one dendrite and its non-recognized prototypes is slightly greater than the $60\deg$ between most prototypes. Progressive training on the $4|4$ problem with 20/30 prototype perturbation; $\epsilon_w = 0.025$;\; $\epsilon_\gamma = 0.05;\;\alpha=1$.
	}
	\label{dp-angles-44-pro-fig}
\end{figure}

\begin{figure}[H]
	\centering
	\begin{subfigure}{0.70\textwidth}
		\includegraphics[width=\textwidth]{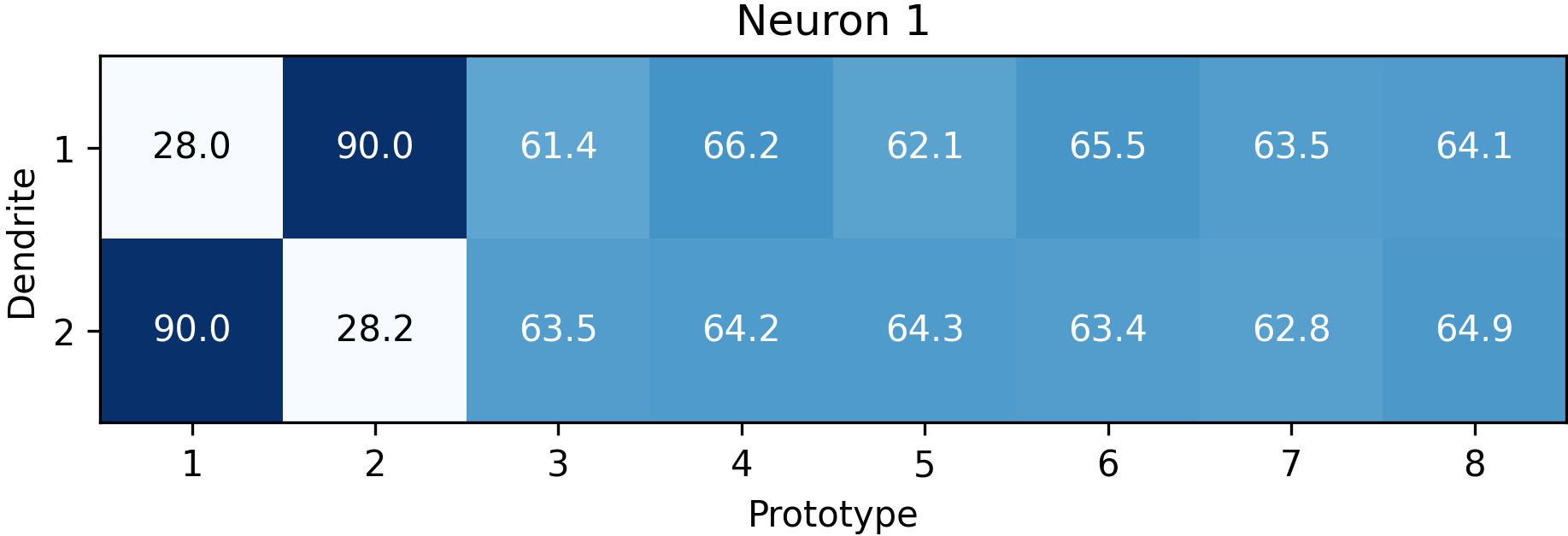}
		\caption{} \vspace{0.1in}
	\end{subfigure}
	\begin{subfigure}{0.70\textwidth}
		\includegraphics[width=\textwidth]{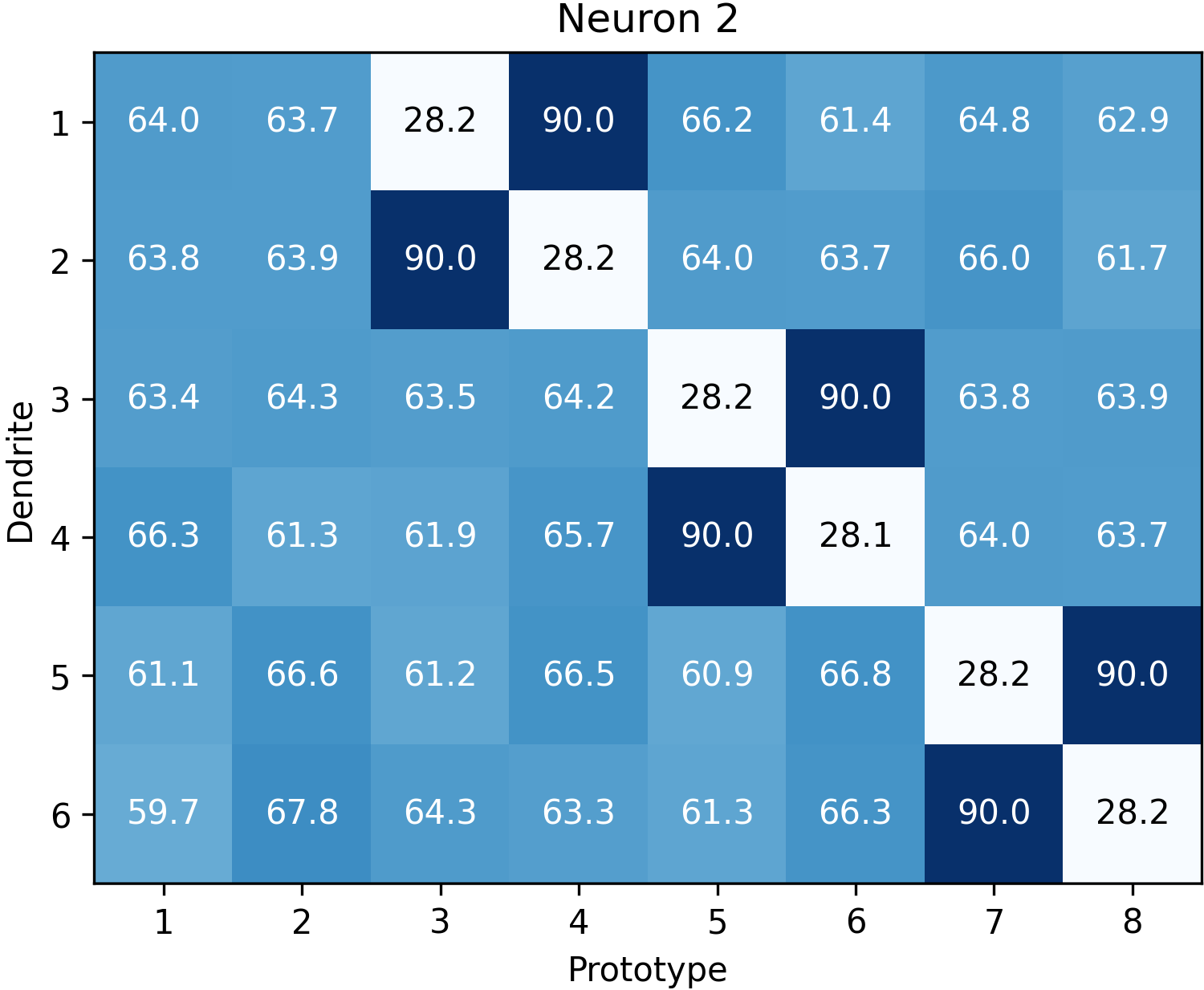}
		\caption{}
	\end{subfigure}
	\caption{ As in the previous figure each dendrite points more towards one prototype than any other, this time for the 2$|$6 problem. Progressive training, prototype perturbations
		($\epsilon_w = 0.025$;\; $\epsilon_\gamma = 0.05;\;\alpha=1$.)
	}
	\label{c1}
\end{figure}

The size of this proper subspace is under parametric control of  $\gamma_{dj}(0)$, the initial rate of synaptogenesis where permitted. Table \ref{gam0-tbl} quantifies this, nearly linear, control. Decreasing  $\gamma_{dj}(0)$, decreases connections per dendrite up to the point that error-free performance can barely be sustained. For example, lowering the initial rate of synaptogenesis by 40\%, $\gamma_{dj}(0)=0.6$, produces a 38\% reduction in connections, and the angle between a dendrite and its favored prototype increases to ca. 45 degrees. Given that most prototypes themselves overlap fifty percent, it is not at all surprising that error-free performance is lost when $\gamma_{dj}(0) =0.5$. At this parameter value the developed connectivity becomes even sparser than  64 connections out of the 128 possible when single prototype exclusivity is maintained by a dendrite.

\begin{table}[H]
	\centering
	\begin{tabular}{c|c|c|c|c|c|c}
		&  Class  Error  &  \multicolumn{2}{c|}{Dendrite Counts}    & \multicolumn{2}{c|}{Median Connection Counts} & Reduction \\
		&&\multicolumn{2}{c|}{}&\multicolumn{2}{c|}{per dendrite} & from 102 \\  
		$\gamma_{dj}(0)$  &   \%           &  Neuron 1 & Neuron 2 &   Neuron 1           & Neuron 2           & \%                         \\ 
		\hline
		
		1.0           & 0.00                    & 4 & 4               &  102 102 102 102   & 102 102 102 102      &    \\
		0.9           & 0.00                    & 4 & 4               &  94 93 91 92  & 89 92 93 90   & 10     \\
		0.8           & 0.00                    & 4 & 4               &  83 85 82 83  & 79 83 82 81   & 19     \\
		0.7           & 0.00                    & 4 & 4               &  74 76 72 70  & 69 71 71 72   & 30   \\
		0.6           & 0.01                    & 4 & 4               &  62 68 62 67  & 58 62 64 64   & 38   
		
	\end{tabular}
	\caption{Reducing the initial synaptogenesis rate parameter, $\gamma_{dj}(0)$, reduces connectivity. Progressive training; $4|4$ problem set; 20/30 prototype perturbations. At $\gamma_{dj}(0) \geq 0.7$ all simulations produce exactly four dendrites per neuron. At $\gamma_{dj}(0)=0.6$ eight simulations produced four dendrites per neuron but two out of ten simulations developed redundant dendrites. $\epsilon_w=0.025;\; \epsilon_\gamma=0.05;\; \alpha=1$. Stability criterion 500 continuous epochs of fixed connectivity; median total epochs of training ca. 1300.}
	\label{gam0-tbl}
 	
\end{table}

The desirable amount of connectivity depends on ones preferences.  More connections, up to the 128 defining features of one prototype, will provide greater resistance to prototype perturbations during testing. On the other hand, more connections imply greater energy use, and thus a lower energy efficiency under many modest prototype perturbations   .

\subsubsection{With large  perturbations not all dendrites  identify  just one prototype}

As seen above (Table \ref{contingency_table}), the effect of prototype perturbations during training alone can contribute more to errors than the same perturbations selective to just testing. To understand the structural effects of larger prototype perturbations during training, this section examines  dendrites and number of connections per dendrite at rates of perturbation that include but go  beyond 20/30 and 30/20. Above these moderate rates of perturbations, there are structural correlates associated with the onset of a positive error-rate. The most consistent result is the tendency to produce fewer functional dendrites per neuron and more non-functional dendrites.

Fig\ref{con_counts_dendrites} displays the results across perturbation levels during training. At 30/20 or 20/30 and below, neurons reliably  appear with four dendrites, one for each prototype. Beginning at 40/20 and 20/40 prototype perturbations, this number of functional dendrites-per-neuron drops. Slightly more than half the neurons have fewer than four dendrites, and by 50/20 and 20/50, many  more than half the neurons have less than four functional dendrites. Nevertheless error-rates under such perturbations are near or  below 1\% (Fig \ref{error-vs-occ-and-on-fig}). Such low error-rates imply that at least one of the three functional dendrites on each neuron is recognizing exemplars from two different prototypes. 

\begin{figure}[H]
	\centering
	\begin{subfigure}{0.48\textwidth}
		\includegraphics[width=\textwidth]{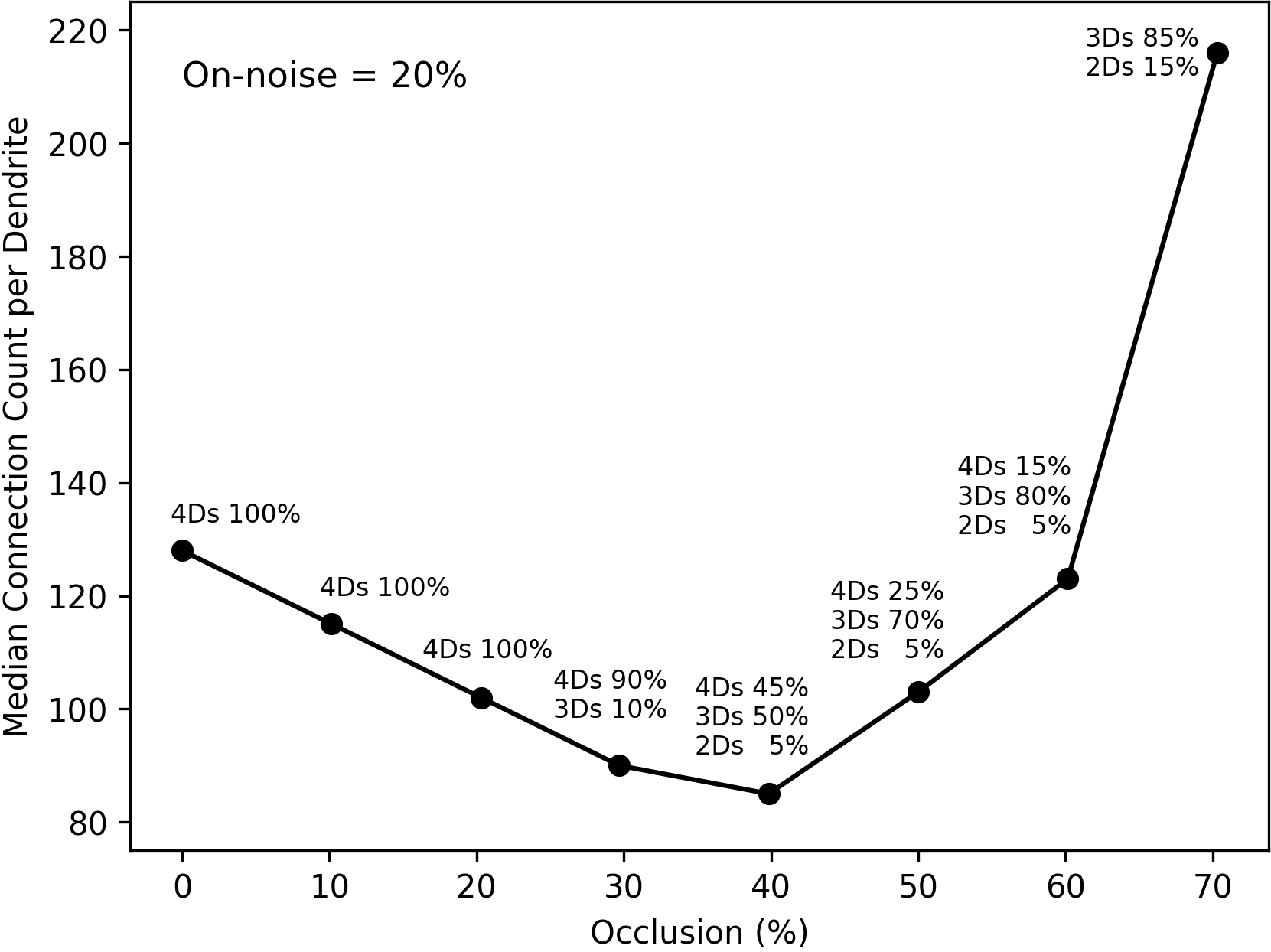}
	\end{subfigure}
	\hfill
	\begin{subfigure}{0.48\textwidth}
		\includegraphics[width=\textwidth]{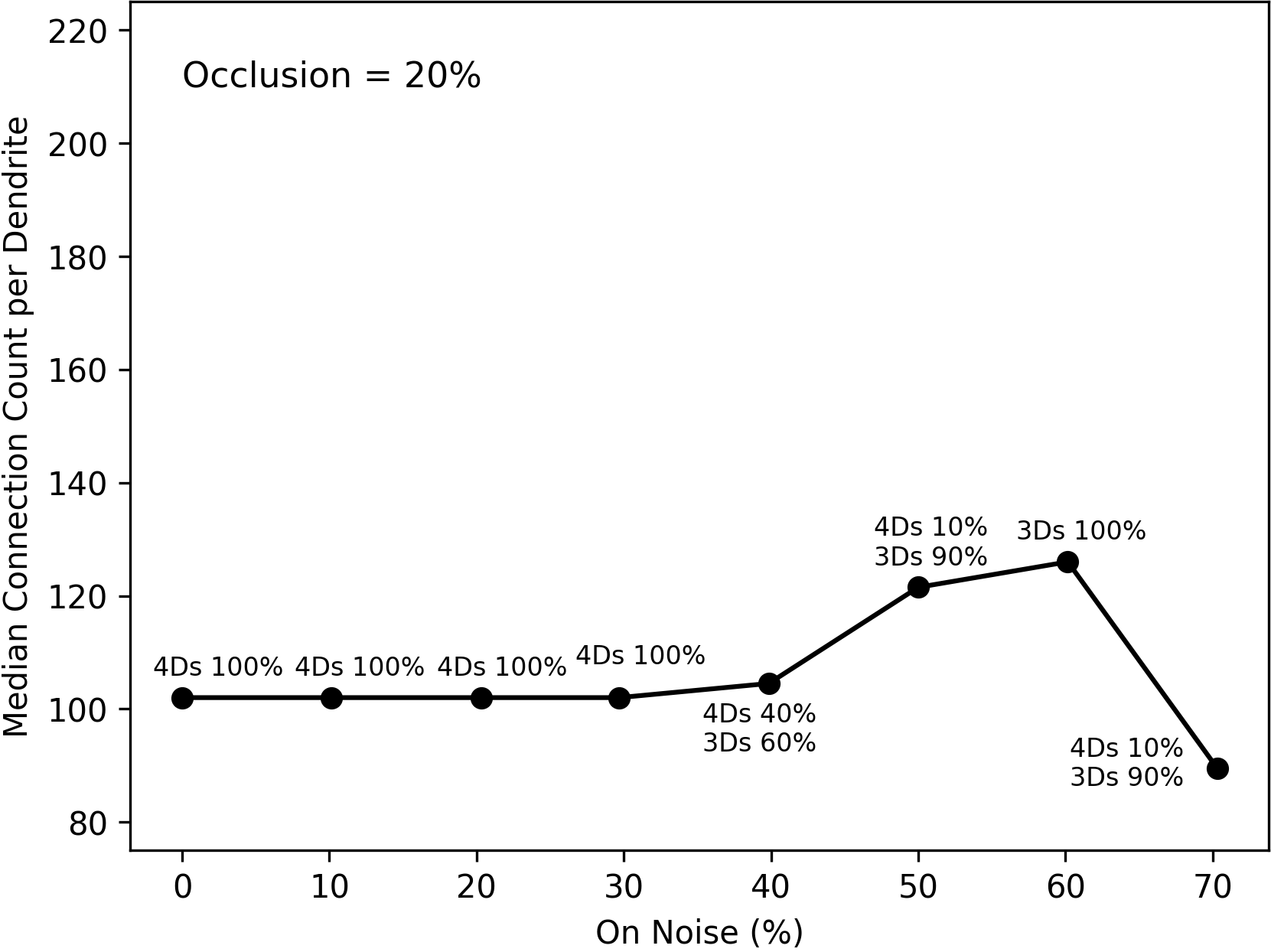}\\
	\end{subfigure}
	\caption{Larger prototype perturbations, associated with non-zero error-rate, lead to fewer dendrites per neuron but more synapses per dendrite. Four dendrites per neuron, one per prototype, dominate from 0\% through 30\% prototype perturbations. At 40\% or greater perturbations, the three dendrite per neuron dominates. At occlusions greater than 40\%, there is a strong trend for increased connections per dendrite. With 40/60 or 60/40 perturbations, the median synapse count per dendrite is about the same. The median are over 10 simulations, that is, 20 neurons. Progressive training on the 4$|$4 problem set. ($\alpha=1; \epsilon_w = 0.002$; $\epsilon_\gamma = 0.05$. Empirical stability criterion halts training at 500 epochs with no connection changes, or if this does not occur (which is  the case for larger perturbations), the maximum training epochs is 3000.)
	}
	\label{con_counts_dendrites}
\end{figure}

To confirm and better define the boundary of reliable four dendrite-per-neuron solutions vs three or less, simulations for the $4|4$ problem with progressive training are examined at 30/30 perturbations ($ \epsilon_w = 0.002$;\; $\epsilon_\gamma = 0.05;\;\alpha=1$). Typically, training for 800 epochs is required to achieve the 500 epoch stability criterion. Error-free performance is nearly always obtained under 30/30 perturbations (train and test), with an the average error-rate of 0.29\% over twenty simulations (40 neurons). However, the number of functional dendrites per neuron is sometimes four but most of the time it is three, with one or more non-functional dendrites per neuron. Once again, the very small error rate along with fewer than four functional dendrites per neuron implies  that some single dendrites are reliably recognizing exemplars from two different prototypes.

Regarding connection counts per dendrite at 30/20 and smaller occlusions, there is the systematic relationship connection counts equals 128 times the quantity one minus occlusion fraction. In the case of varying on-noise at 20/30 or below, connections counts are essentially constant at $128\times (1- 0.2)\approx 102$.  For the larger perturbations (but below 70/20 and 20/70) with the small but positive possibility of an error, the median number of connections per functional dendrite approaches or reaches the full prototype matching value of 128.

\begin{figure}[H]
	\centering
	\begin{subfigure}{0.70\textwidth}
		\includegraphics[width=\textwidth]{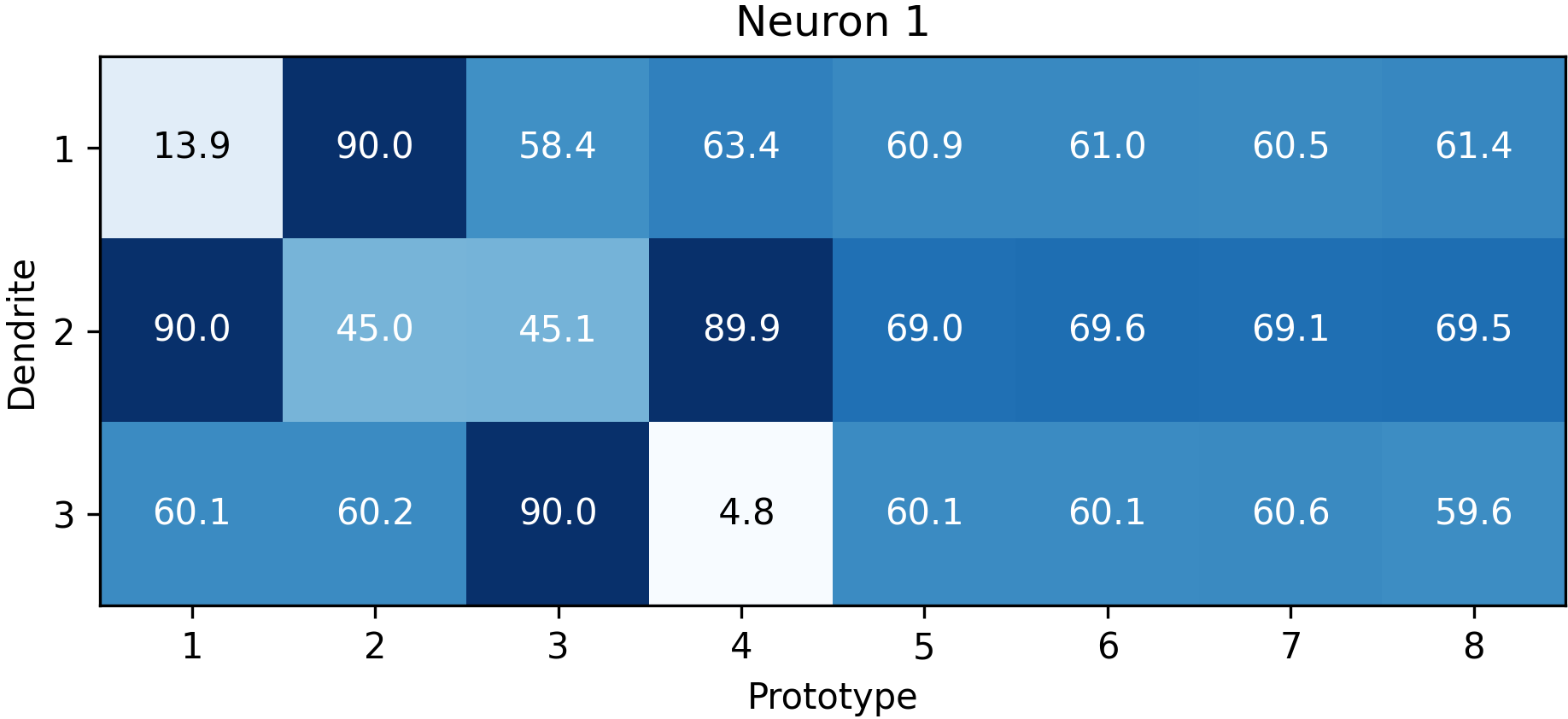}
		\caption{} \vspace{0.1in}
	\end{subfigure}
	\begin{subfigure}{0.70\textwidth}
		\includegraphics[width=\textwidth]{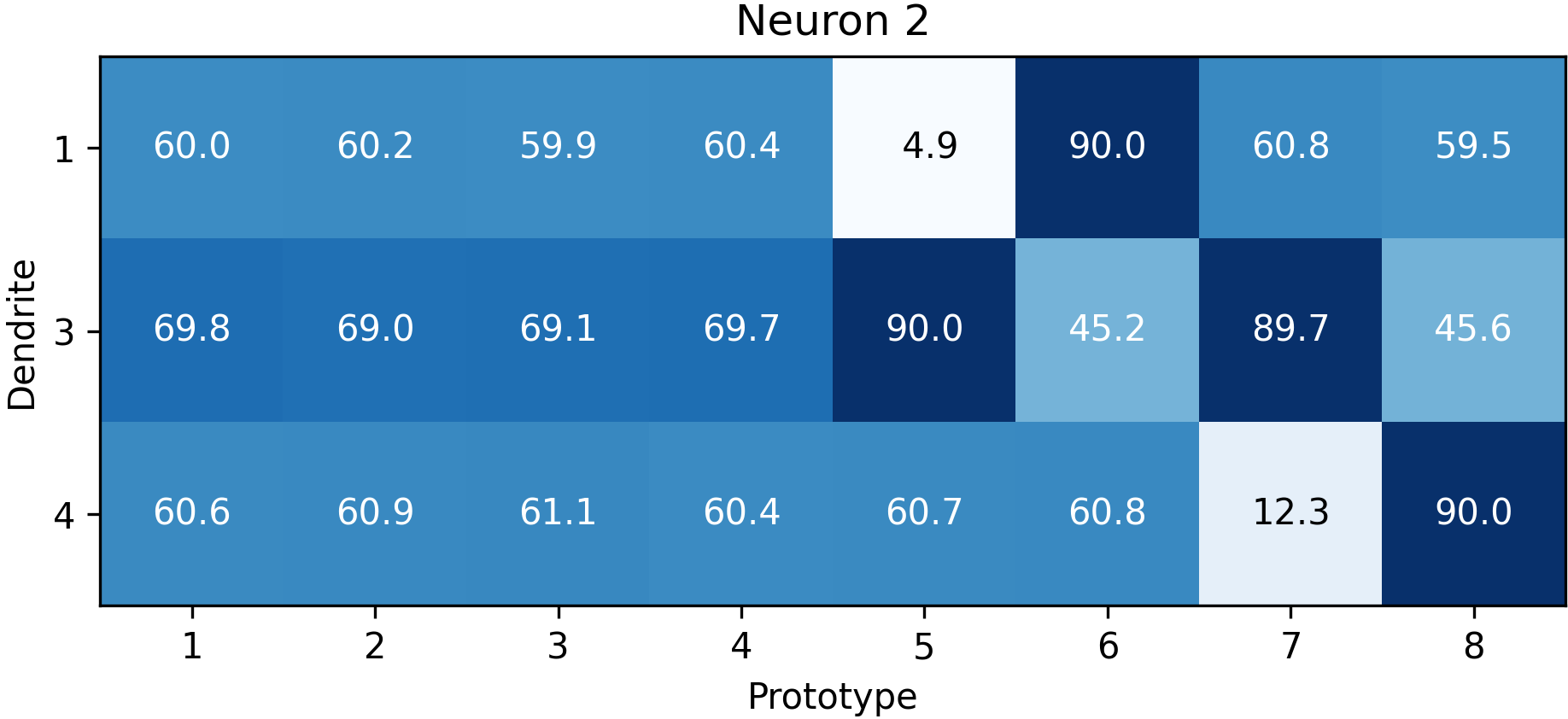}
		\caption{} \vspace{0.1in}
	\end{subfigure}
	\caption{The  20/50 prototype perturbations during progressive training produces neurons with three functional dendrites. This representative simulation has an error-rate of 1.96\% nearly an exact match  of  the 2\%  rate averaged over all ten simulation. Note the dendrites doing double-duty for discrimination, dendrite three of neuron one, which recognizes prototypes two and three, and dendrite two of neuron two, which recognizes prototypes six and seven. The numbering of dendrites takes into account non-functional dendrites. $\epsilon_w = 0.002$; $\epsilon_\gamma = 0.05$; $\alpha=1$, Training for  3000 epochs.  
	}
	\label{dp-angles-44-pro-50on-fig}
\end{figure}

To get a better idea about those dendrites doing double-duty relative to prototypes being discriminated, let us look  at the angles between dendrites and prototypes for a representative simulation.  The angles of Fig \ref{dp-angles-44-pro-50on-fig}, which is the outcome of 20/50 training, should be compared to those of  Fig \ref{dp-angles-44-pro-fig}, which is the outcome of 20/30 training on the same $4|4$ problem. The first thing one notices is that the two neurons of Fig \ref{dp-angles-44-pro-fig} have four dendrites each while each of the two neurons of Fig \ref{dp-angles-44-pro-50on-fig} have just three dendrites. 

In Fig \ref{dp-angles-44-pro-50on-fig}, it is easy to find the dendrites doing double-duty. On neuron one,   note that functional dendrite three has an angle of 45.0 degrees relative to prototype two and an angle of  45.1 degrees relative to prototype three and no smaller angle. Similarly on neuron two,  functional   dendrite three is doing double duty. This time the two poorly but  still, nearly always, successfully discriminated prototypes are  six and eight again with ca. 45 degree angles  relative to each of the two  prototypes.

Another way to view the effects of prototype perturbations on development is to compare minimum angles between dendrites, particularly between dendrites of different neurons, which comparisons predict discrimination errors. With increasing perturbation, these angles decrease. The effect is particularly dramatic for occlusive perturbations; see Fig \ref{min-angle-pro}. 

In sum, the larger  prototype perturbations do not just alter dendritic angles but lead to the development of fewer functional dendrites. With fewer dendrites than prototypes, the excitations provided by dendrites doing double-duty will be weaker than when a dendrite is just pointing at a large fraction of the subspace of just one prototype.

Having quantified the kinds of neurons that the algorithm builds in response to its training experience, it is time to get a better understanding just how the algorithm builds these dendrites, in particular, the role played by CDS.

\subsection{Dissecting the algorithm  shows that Cross Dendritic Suppression is critical for robust performance}

As the initial data of Results point out, dendritogenesis can protect against catastrophic interference. However, as the data just below demonstrate, dendritogenesis without CDS is often useless or nearly so. In particular when training with perturbed prototypes, (i) CDS is a necessary partner, working along with dendritogenesis, to obtain the simulations with perfect, error-free performance; and   (ii)  CDS is necessary for fully stabilized connectivity.

Table \ref{error-stability-onset-cds-tbl} presents a quick characterization of  error-rates and time to stabilize with and without CDS for both concurrent and progressive training on the 4$|$4 problem with 20/30 perturbations.  CDS improves both error-rates and appears necessary for empirical convergence. 

Concurrent training without CDS never attains a low error-rate and never stabilizes.  Progressive training error-rates without CDS are imperfect but attain a possibly acceptable of 5.6\% error-rate after 800 epochs. However, this error-rate is a function of  training duration. As examined in one simulation, stopping just around epoch 400 yields  a 1.0\% error-rate but this error-rate quickly increases with more training, $\{5.6,\;6.6,\;18.6\}$ percent errors at the end of training epochs 800, 1600, and 3600, respectively.  Halting the simulation at epoch 3600, produces an error-rate of 18.6\%. Correlated with this increasing error-rates is a decline in connections for the functional dendrites. Also on-going during this prolonged training is the formation of many non-functional dendrites.

\begin{table}[h]
	\centering
	\begin{tabular}{c | c c c | c c c |}
		\cline{2-7}
		& \multicolumn{3}{c}{CONCURRENT} & \multicolumn{3}{|c|}{PROGRESSIVE} \\ \cline{2-7}
		&  ERROR & STABILITY  & EXTRA           &  ERROR  & STABILITY  & EXTRA \\
		&   RATE    & ONSET       & DENDRITES   &    RATE    &   ONSET    & DENDRITES \\
		&     (\%)     &  (epochs)    &  per neuron     &   (\%)        &  (epochs)   &  per neuron  \\ \hline
		\multicolumn{1}{|c|}{Without CDS}  &    13.1    &  NEVER       &       2.5            &    10.6      &   NEVER     &      2.75              \\
		\multicolumn{1}{|c|}{With CDS}        &    0.0      &       261         &      0.0            &   0.0         &    337          &      0.0              \\
		\hline
	\end{tabular}
	\\ \hspace{0.85in} {\footnotesize {median values across 10 simulations}} 
	\caption{Error, stability onset, and extra dendrite counts with and without CDS. Without CDS, error rates increase with additional training. $4|4$ problem set. 20/30 prototype perturbations. $\alpha=1$.}
	\label{error-stability-onset-cds-tbl}
\end{table}

When CDS is included in the simulations,  the error-rate  improves all the way  to perfect performance for both training paradigms. Similarly with CDS present, the training time to best  error-rate shortens, and there is empirical stabilization of errors and stabilization of connectivity.

Using the $4|4$ problem with 20/30 prototype perturbations, Table \ref{cds-tbl} presents a much more detailed comparison  that substantiates the importance of CDS. Not only does this table present simulations  without vs with CDS, but using the retrospective trick of endowing neurons with the exact number of dendrites usually found under the full algorithm, one can actually compare without vs with dendritogenesis. (In this table's left-hand column the letter D=dendritogenesis and the presence of CDS is indicated by the letter C; DC\&SAS = Dendritogenesis (without dendritogenesis, each neuron had four dendrites); \newline Conc = Concurrent training; Prog = Progressive training.) 

\begin{table}[H]
	\newcommand{\unstable}{\cellcolor{lightgray}}
	\centering
	\resizebox{0.8\columnwidth}{!}{%
		\begin{tabular}{|c|c|c|c|} 
			\hline
			Paradigm  & Training  &  Error (\%)   &  Comments \\ 
			\hline 
			\multicolumn{4}{c}{ Without CDS} \\
			\multicolumn{2}{l} { 8D exemplar=ptototype} \\ \hline
			SAS-only  &  Conc &   50.0           &      redundant dendrites \\
			SAS-only  &  Prog  &   50.0           &     redundant dendrites \\
			D\&SAS &  Conc &     0.0           &     2 \& 3-dendrite solutions  \\
			D\&SAS &  Prog  &     0.0           &    3 \& 4-dendrite solutions \\ \hline
			\multicolumn{2}{l}{  256D 20/30 prototype perturbations} \\ \hline
			SAS-only  & Conc &    50.4          &      redundant dendrites \\
			SAS-only   & Prog  &    50.6          &     redundant dendrites \\
			D\&SAS  & Conc &    13.6          &       unused \& redundant dendrites \\
			D\&SAS  & Prog  &     10.7          &    unused \& redundant dendrites \\ 
			\hline 
			\multicolumn{4}{c}{ With CDS}\\
			\multicolumn{2}{l}{ 8D exemplar=ptototype} \\ \hline
			C\&SAS  & Conc  &   0.0             &    2, 3 \& 4-dendrite solutions, unused dendrites \\
			C\&SAS  & Prog   &   0.0             &    2, 3 \& 4-dendrite solutions, unused dendrites \\
			DC\&SAS & Conc  &   0.0            &    2-dendrite solution \\
			DC\&SAS & Prog   &   0.0            &    2, 3 \& 4-dendrite solutions \\ \hline
			\multicolumn{2}{l}{ 256D  20/30 prototype perturbations} \\ \hline
			C\&SAS  & Conc  &   0.0             &    2 \& 3-dendrite solutions, unused dendrites \\
			C\&SAS  & Prog   &   0.0             &    2 \& 3-dendrite solutions, unused dendrites \\
			DC\&SAS & Conc  &   0.0            &    2 \& 3-dendrite solutions, unused dendrites \\
			DC\&SAS & Prog   &   0.0            &    4-dendrite solution \\
			\hline
		\end{tabular}
	}
\vspace{0.2 cm}
	\\ {\footnotesize {unused dendrites $:=$ did not win the competition during testing}}
	\\ \vspace{-0.13in} {\footnotesize {redundant dendrites $:=$ represents the same prototype as another dendrite}}
	\captionsetup{width=0.95\textwidth}
	\caption{$4|4$ problem set. C=CDS. D=dendritogenesis. When no D, then four dendrites per neuron \textit{ab initio}. $\alpha=1$.}
	\label{cds-tbl}
\end{table}
 
Jumping into the very bottom of the table, one sees that the full algorithm gives the best performance (compare the bottom two rows to entries above them). In addition to error-based performance,  the most consistent neuron constructions require CDS. That is, the number of dendrites is the same -- 10 out of 10 simulations -- both for concurrent,  with its two or three-dendrite  solution and for progressive training, with its four dendrite solution.

Just  above these bottom two rows  are results produced without dendritogenesis but using neurons endowed, \textit{ab initio}, with four dendrites. Here too, error-free performance obtains, but only if CDS is present (cf. the top-half of the table). However, in obtaining the error-free performance, it is often the case that one  of the pre-endowed dendrites recognizes two prototypes while the fourth pre-endowed dendrite is non-functional (redundant) in the sense that it never wins a single competition for most excited dendrite.

In sum,  overall, reliable error-free performance -- and in the case of progressive training, full unmixing of each class's contribution to the conditional mixture distributions -- requires CDS to be included with dendritogenesis and supervised adaptive synaptogenesis. 

\subsubsection{Effect of CDS and dendritogenesis on stability}

\begin{figure}[H]
	\centering
	\begin{subfigure}{0.48\textwidth}
		\includegraphics[width=\textwidth]{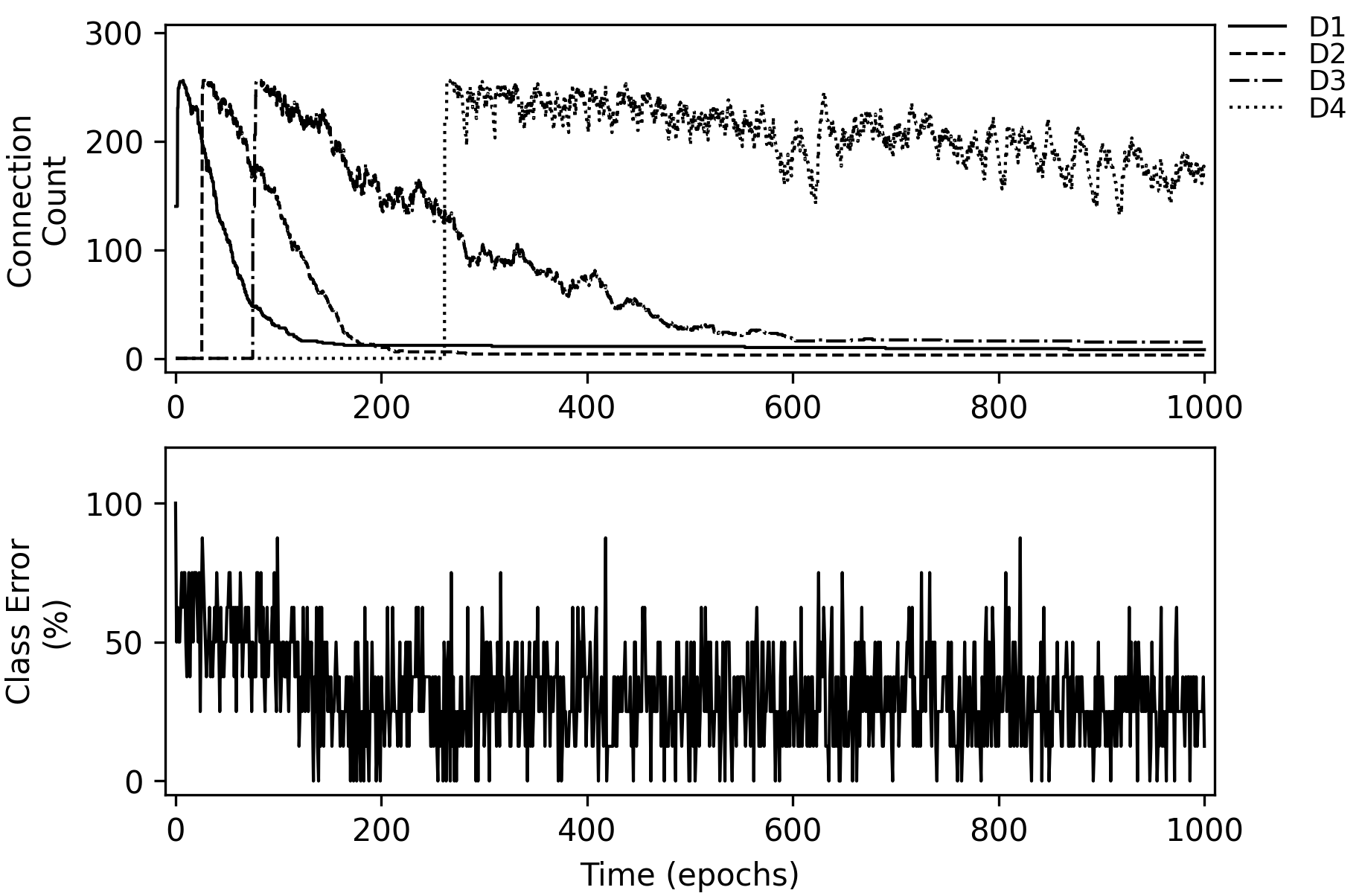}
		\caption{Without CDS}
	\end{subfigure}
	\hfill
	\begin{subfigure}{0.48\textwidth}
		\includegraphics[width=\textwidth]{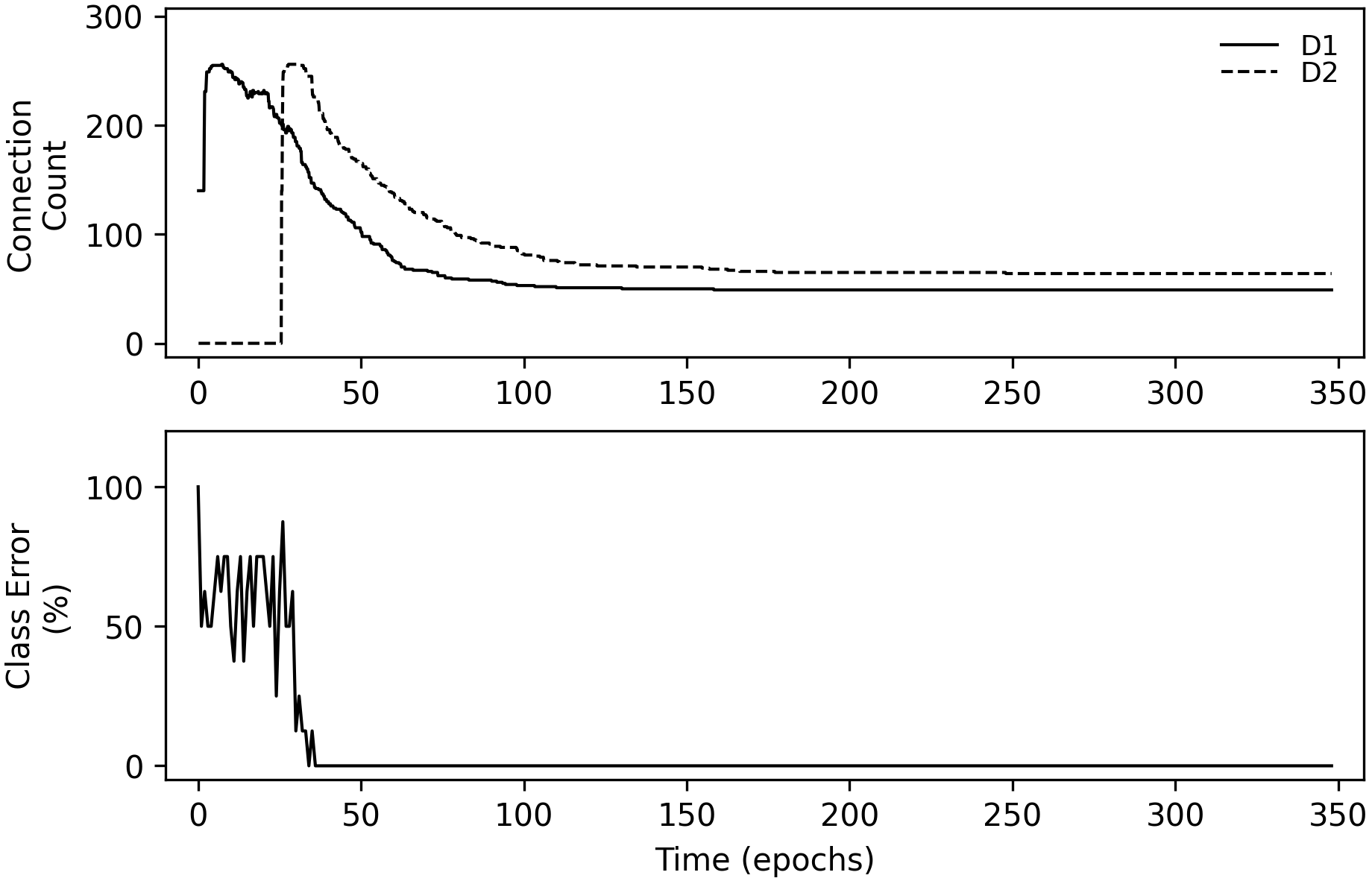}
		\caption{With CDS}
	\end{subfigure}
	\caption{
		CDS is necessary for concurrent training to produce zero errors and to reach a stable connectivity count.
		Without CDS connection (left two figures), counts and classification errors never stabilize ), 
		but with CDS (right two figures), errors stabilize within 40 epochs and connectivity stabilizes within 250 epochs.
		$4|4$ problem set with  20/30 prototype perturbations.$\epsilon_w=0.025;\;\epsilon_\gamma=0.03;\;\alpha=1.0$ 
	}
	\label{c2}
\end{figure}
Time-dependent visualizations underscore the necessity of CDS for stabilization.
Fig \ref{c2} illustrates examples of certain underlying dynamics that further distinguish simulations (as in Table \ref{error-stability-onset-cds-tbl}), with vs without CDS, under concurrent training.  Comparing connection counts when CDS is present (top right), one sees two dendrites developing  with a large, initial connectivity overshoot, followed by a  gradual connection decrease until epoch 248, at which time the youngest dendrite stabilizes  with 64 connections. On the other hand without CDS (top left),  a third and fourth dendrite are created. Moreover, the youngest dendrite never seems to stabilize while the third dendrite requires more than 600 epochs to stabilize.  Comparing the  two  error-rate graphs, with CDS (bottom right) reaches zero error around training epoch 40 while, without CDS (bottom left), there is no hint that error-rate will ever stay below 20\%.

All three training methods require CDS to stabilize. See appendix figure \ref{shed-cds-fig} which compares shedding over time for each of the the three training procedures with vs without CDS. There is no hint of connectivity stabilizing without CDS.

In sum, when learning under modest prototype perturbations (20/30 or 30/20), the observed simulations  demonstrate that removing CDS from the algorithm leads to three undesirable outcomes: (i) non-functional and occasionally redundant dendrites; (ii) a prolongation of time to stability, or more often an inability to demonstrate stable connectivity; and (iii) suboptimal, i.e., non-zero, error-rates. Supposing that there exists parameter settings that avoid all three of these issues, our search of parameter space indicates the difficulty in finding these hypothetical settings when CDS is not part of the algorithm. Thus our minimum conclusion is that without CDS the algorithm is not robust and easily succumbs to one or more of the three issues just mentioned. 

\subsection{Dynamics of dendritogenesis and stabilization of connectivity}
\subsubsection{Developmental onset  of dendritogenesis distinguishes the different training paradigms and the time  for overall stabilization, $\epsilon_w=0.025;\; \epsilon_\gamma=0.03$ or $\epsilon_\gamma=0.05;\;\alpha=1$}

Having just investigated  empirical  stabilization with and without CDS,  a description of  the dynamics of dendritogenesis and descriptions of various measures of empirical convergence is owed to the reader. Empirical convergence to  stabilization is measured in a variety of ways: (i) error-rate, (ii) connection counts, (iii) shedding, and (iv) some form of synaptic weight stabilization occurring after (ii) stabilizes and (iii) is zero. In what follows, (a) the problem set is the two-class 4$|$4 with 20/30 prototype perturbations, and (b) all assertions of stabilization are limited to prototype  perturbations no larger than 20/40 and 40/20.

For each illustrated neuron, Figure \ref{quad-err-gam-fig} juxtaposes error-rate (i.e., missed-detections)   with  $\gamma_{dj}(t)$ and with the  dendritogenesis events (blue equilateral triangles at the top of each error graph). In general for each neuron, dendritogenesis is suppressed until the newest dendrite has been sufficiently successful. Sufficient success is visualized by the declining  $\gamma_{dj}(t)$ value crossing the dashed line, i.e., $\gamma_{dj}(t)\leq \theta_\gamma$. Having reached or crossed this threshold,  dendritogenesis awaits only an increase in error-rates. 

\begin{figure}[H]
	\centering
	\begin{subfigure}{0.48\textwidth}
		\centering
		\includegraphics[width=\textwidth]{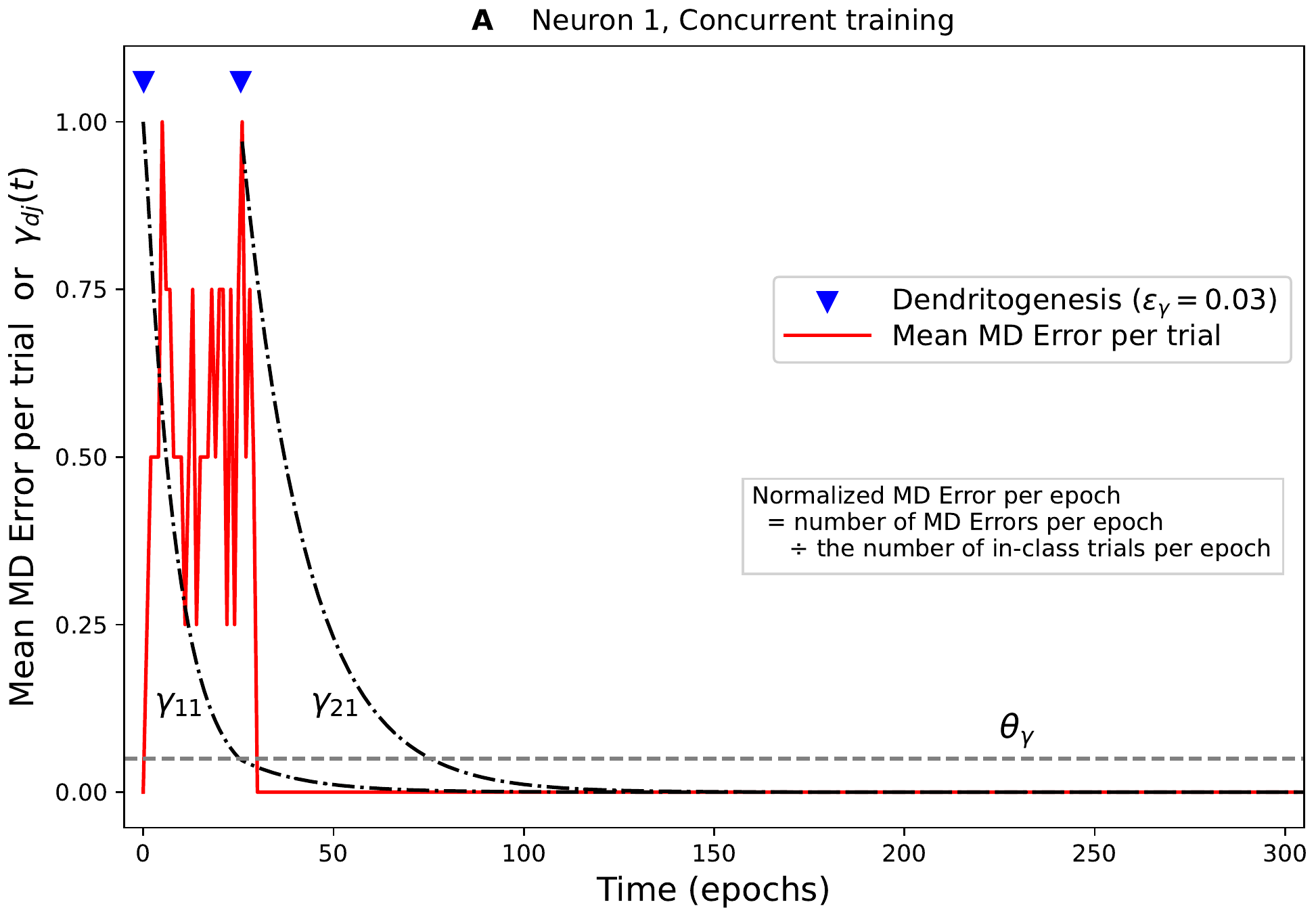}
	\end{subfigure}
	\hfill
	\begin{subfigure}{0.48\textwidth}
		\centering
		\includegraphics[width=\textwidth]{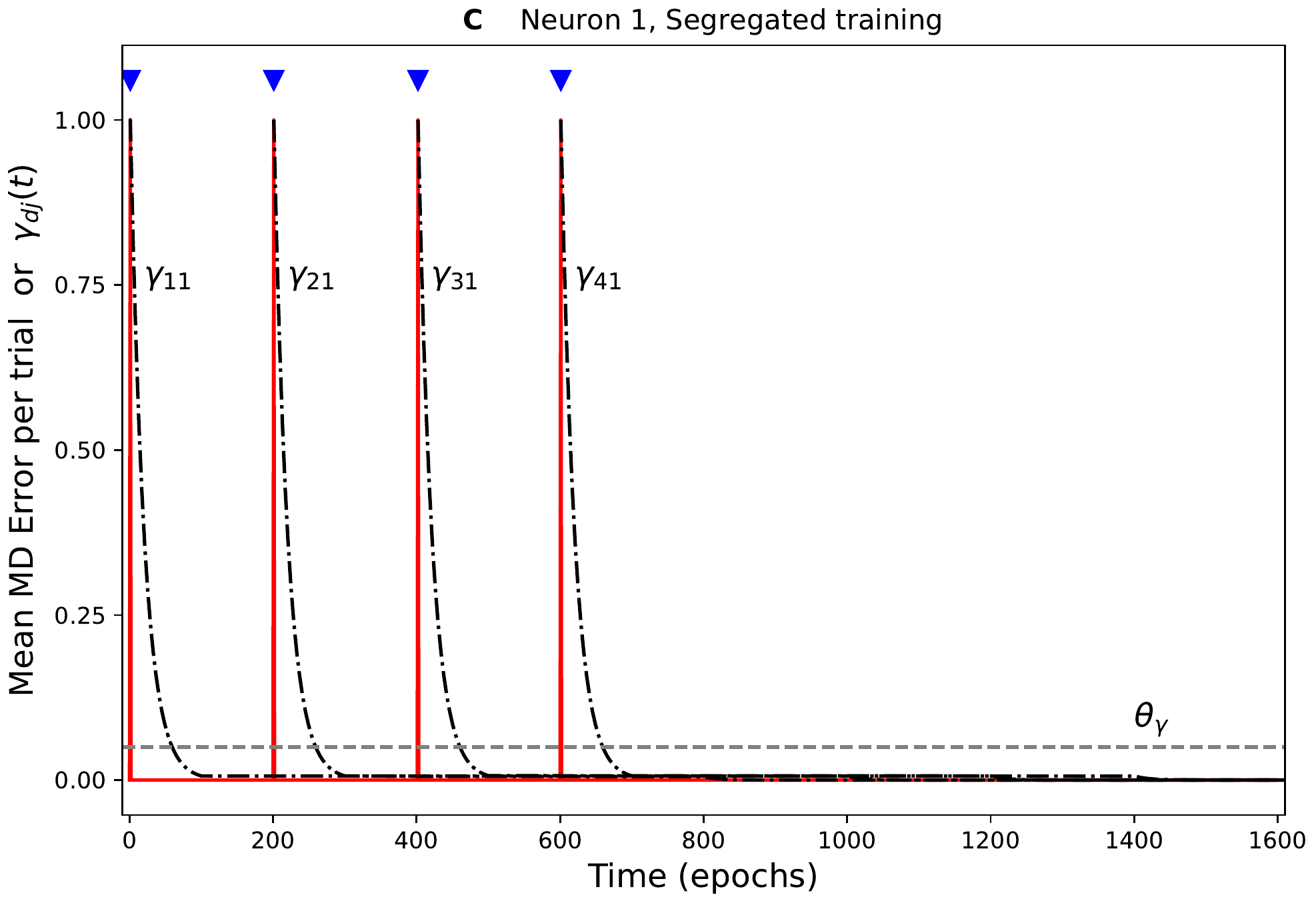}
	\end{subfigure}
	\vskip\baselineskip
	\begin{subfigure}{0.48\textwidth}
		\centering
		\includegraphics[width=\textwidth]{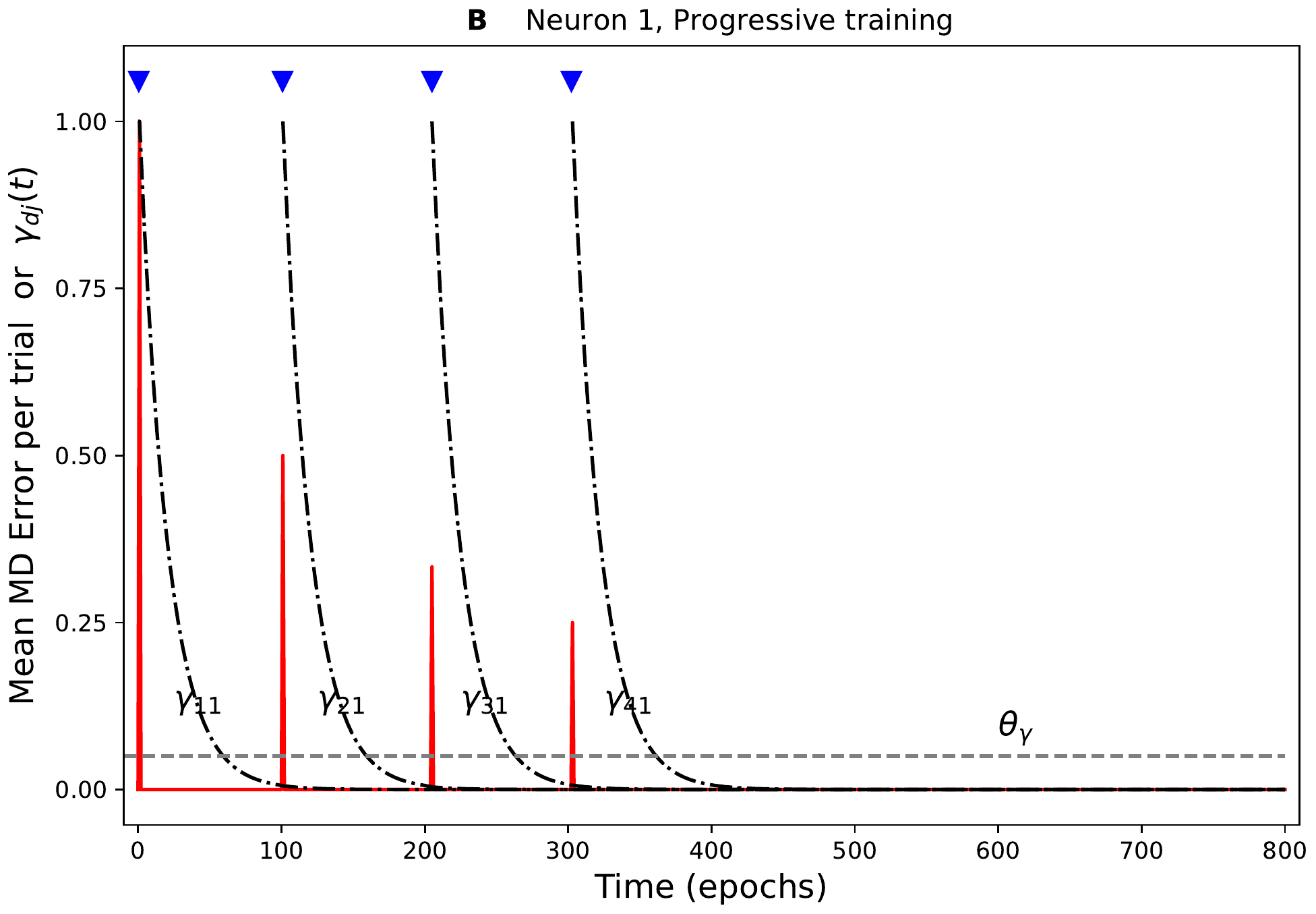}
	\end{subfigure}
	\hfill
	\begin{subfigure}{0.48\textwidth}
		\centering
		\includegraphics[width=\textwidth]{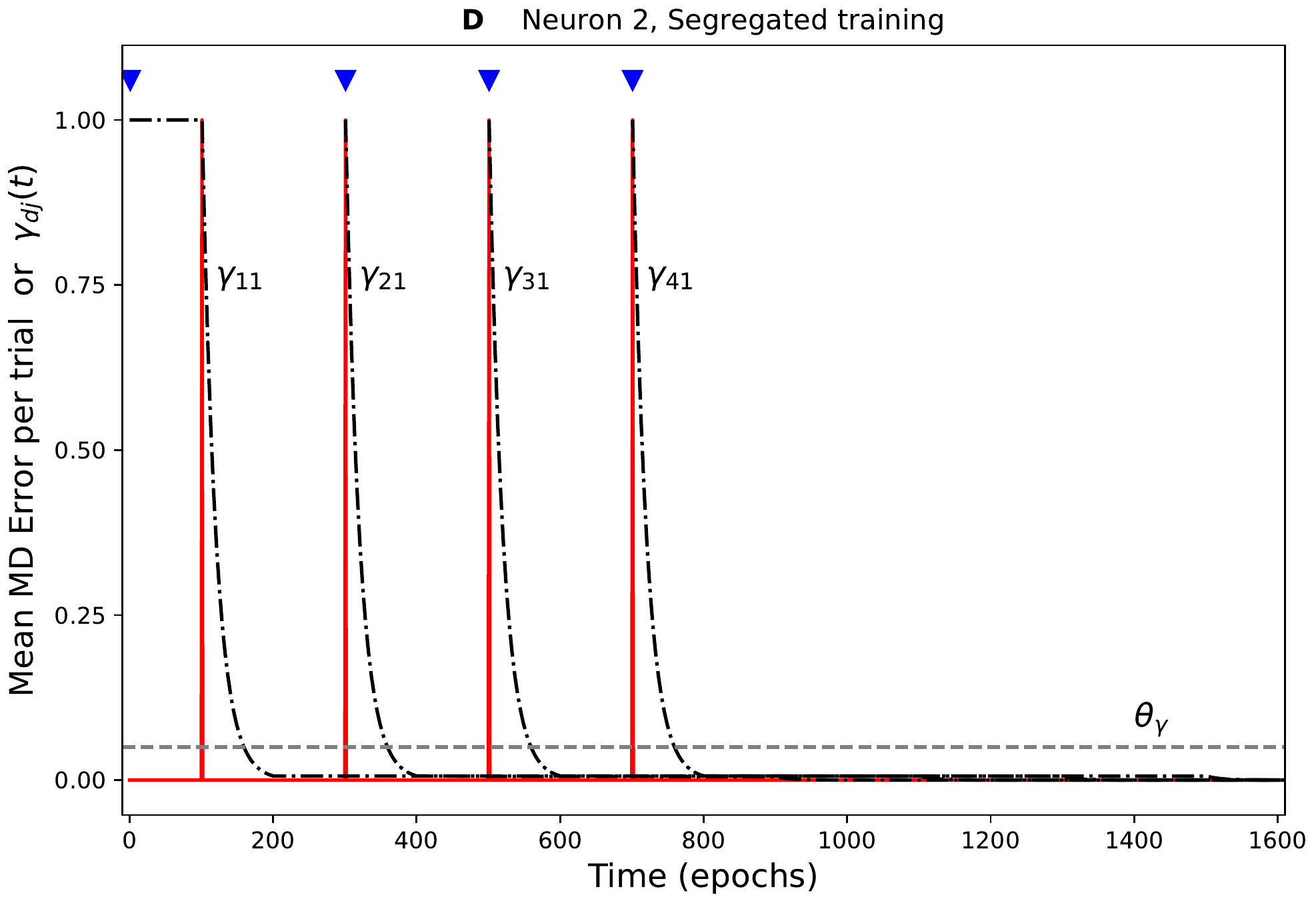}
	\end{subfigure}
	\caption{Developmental dynamics. Dendritogenesis (blue inverted triangles) is driven by missed detection errors (red lines) and $\gamma_{dj}(t)$ (black lines). 
		Dendritogenesis requires  $\gamma_{dj}(t) < \theta_\gamma$ and  one missed detection error when $\alpha$ is set to one. 
		At the very beginning of training, each neuron is endowed with a single, unconnected dendrite.
		With concurrent training, only two dendrites are required but the second dendrite does not appear until  $\gamma_{dj}(t) < \theta_\gamma$ for the first dendrite. In the case of progressive training, a new dendrite is created
		on each neuron sometime within the first epoch of each new phase. In the case of segregated training, a new dendrite is created between the first and second exemplar of each new phase.
		All training is for the  $4|4$ problem set and  20/30 prototype perturbations.
		($\epsilon_w = 0.025$ for all training methods; $\epsilon_\gamma = 0.03$ for concurrent training
		to avoid a non-functional dendrite while $\epsilon_\gamma = 0.05$ for progressive and segregated training.)
	}
	\label{quad-err-gam-fig}
\end{figure}

The main differences in the sub-figures  is a result  of the way prototypes are mixed during training. A secondary, and parameter dependent difference, is that concurrent training here is producing a two dendrite per-neuron solution while the phased methods find a four dendrite per-neuron solution. 

(Comparisons that follow are made in terms of epochs. Such comparisons are straightforward for any one type of training. However, if the reader would rather compare across training methods, then a decision must be made. One might continue to compare in terms of number of epochs, but there is a good argument to compare in terms of number of training trials (number of exemplars). The issue is that the trials per epoch differ with training style.

Note that the x-axis for each training method is different and that time is in epochs rather than trials. The following figure visualizes the relationship between epochs  and training trials. For concurrent training, there are eight trials per epoch, one for each prototype. For segregated training, there is just one trial per epoch. In the case of progressive, the number of trials per epoch is increasing over phases. With each successive phase, two new prototypes, one for each class, are added into the defined epoch. Thus, from the first through the fourth phase of progressive training, there are two, four, six, and finally eight trials per epoch.) 

In any event, Fig \ref{quad-err-gam-fig} illustrates that the parameter settings are producing rapid convergence within a phase, where the parameter settings are the same for all three training methods.

The table \ref{birth-tbl} gives the exact time of the critical events defining the dendriteogenesis dynamics illustrated in  Fig \ref{quad-err-gam-fig}. This table includes, precisely, each epoch when dendritogenesis occurs (Birth column) and when dendritogenesis is permitted to occur based on $\gamma_{dj}(t)$. For example, one compares the first time $\gamma_{dj} \leq \theta_\gamma$, for dendrite one of progressive this is epoch 60, to the birth of the second dendrite, which is epoch 101. The table also supplies the number of training exemplars per epoch, thus allowing the reader to convert epoch values to total training events. For the phased training methods, the numerical difference between the permissive $\gamma_{dj}(t)<\theta_\gamma$ and the next dendritogenesis event is indicative of the conservative developmental settings; that is, the parameter settings can be tweaked to produce faster development. For example under progressive training, 60 vs 101 is a delay of 41, 160 vs 205 for delay of 45, and 200 vs 303 is a delay of 103 epochs. Under segregated training, the delay is even longer, about 140 epochs. Such longer delays are due to both phase durations and to the single prototype training that completes training for one class before introducing a new prototype of the other class. In sum, such large developmental delays are not necessary for zero error development but are useful to illustrate within phase convergence. 

\begin{table}[h]
\centering
\ra{1.2}
\begin{tabular}{@{} c c c c @{}}
Dendrite    & \ Birth      & \  $\underset {t}{\min} \, \gamma_{dj}(t) \leq \theta_\gamma$ & Trials        \\
                 &  (epoch)   &   (epoch)                                                                                  & per epoch \\
\toprule

\multicolumn{4}{c}{CONCURRENT}  \\ \hline

 First         &     0    &  26  &  8  \\ 
 Second   &     26   &  77  &  8  \\  \hline
 
\multicolumn{4}{c}{PROGRESSIVE}    \\  \hline

 First  &     0    &  60  &   2  \\ 
 Second &     101  &  160  &  4  \\
 Third  &     205  &  260  &  6  \\
 Fourth &     303  &  360      & 8 \\ \hline

\multicolumn{4}{c}{SEGREGATED NEURON 1}  \\  \hline
 
 First  &     0    &  60  &   1  \\ 
 Second &     201  &  260 & 1 \\
 Third  &     402  &  461 & 1 \\
 Fourth &     602  &   660 & 1 \\ \hline

\multicolumn{4}{c}{SEGREGATED NEURON 2}  \\  \hline
 
 First  &     0    &  160  &  1  \\ 
 Second &     301   &  360  & 1 \\
 Third  &     501  &  560  & 1 \\
 Fourth &     701  &  760  & 1 \\ 
 \bottomrule

\end{tabular}
\caption{Important epochs of figure \ref{quad-err-gam-fig}, i.e., birth epoch and $\gamma_{dj}(t)$ permitting dendritogenesis, for the three different training paradigms.
$4|4$ problem set and 20/30 random prototype perturbations.
              }
          \label{birth-tbl}
\end{table}

An empirical convergence criterion, specifies "no change" of quantified variable , or variables, over some specific but arbitrary period of time. As pointed out earlier, the most demanding criterion for convergence here is no change in connections. Inevitably this criterion can be replaced by the cessation of  shedding since it is always the case that shedding continues beyond the last occurrence of synaptogenesis. As also noted earlier, the appendix contains a figure illustrating the dynamics of connectivity, including synaptogenesis and shedding with and without CDS. Without CDS, there is no stable connectivity.  Fig \ref{quad-conn-shed-fig} further illustrates the  temporal dynamic of shedding for all the training paradigms when CDS is present. This figure directly compares shedding (light gray bars) to connection counts (dark, seemingly  continuous lines). Two points are notable. First, comparing between Figs \ref{quad-err-gam-fig} and \ref{quad-conn-shed-fig},  error-rates are zero well before the no-connectivity-change criterion ($\equiv$no-shedding criterion) is satisfied. Related to the first, is a second difference arising from the conservative nature of using the cessation of shedding as the convergence criterion.  Every time even a single shedding event occurs, the convergence-criterion clock starts over. Here, as in most of the simulations, empirical convergence requires 800 epochs of continuously zero changes of connectivity. Note that everything else measured in this and the previous figure has stabilized. (Because the data of Fig \ref{quad-conn-shed-fig} come from exactly the same neurons, and plotted with the same timescales as in  the previous figure, this new figure allows comparisons of the shedding dynamic to the error-rate dynamic and dendritogenesis times in Fig \ref{quad-err-gam-fig}.)  Note the late, isolated shedding event under concurrent training occurring just before epoch 250 while the connection count has hardly changed since epoch 150. This single last shedding event required the simulation to continue well beyond the illustrated timeline, extending to more than 1000 epochs just to be certain that there were no more connection changes (see Fig \ref{shed-cds-fig}).

\begin{figure}[H]
	\centering
	\begin{subfigure}{0.48\textwidth}
		\centering
		\includegraphics[width=\textwidth]{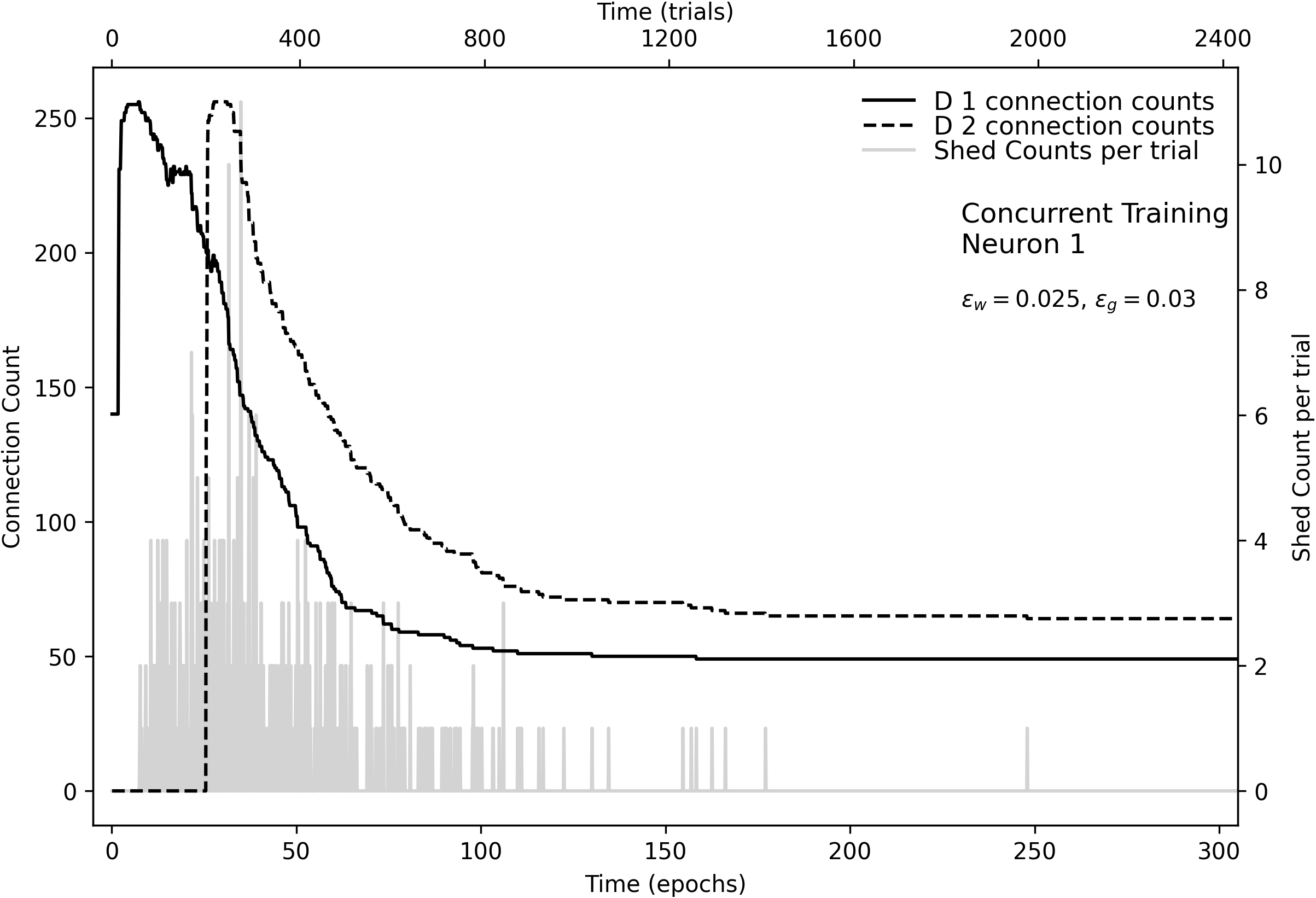}
		\caption{}
	\end{subfigure}
	\hfill
	\begin{subfigure}{0.48\textwidth}
		\centering
		\includegraphics[width=\textwidth]{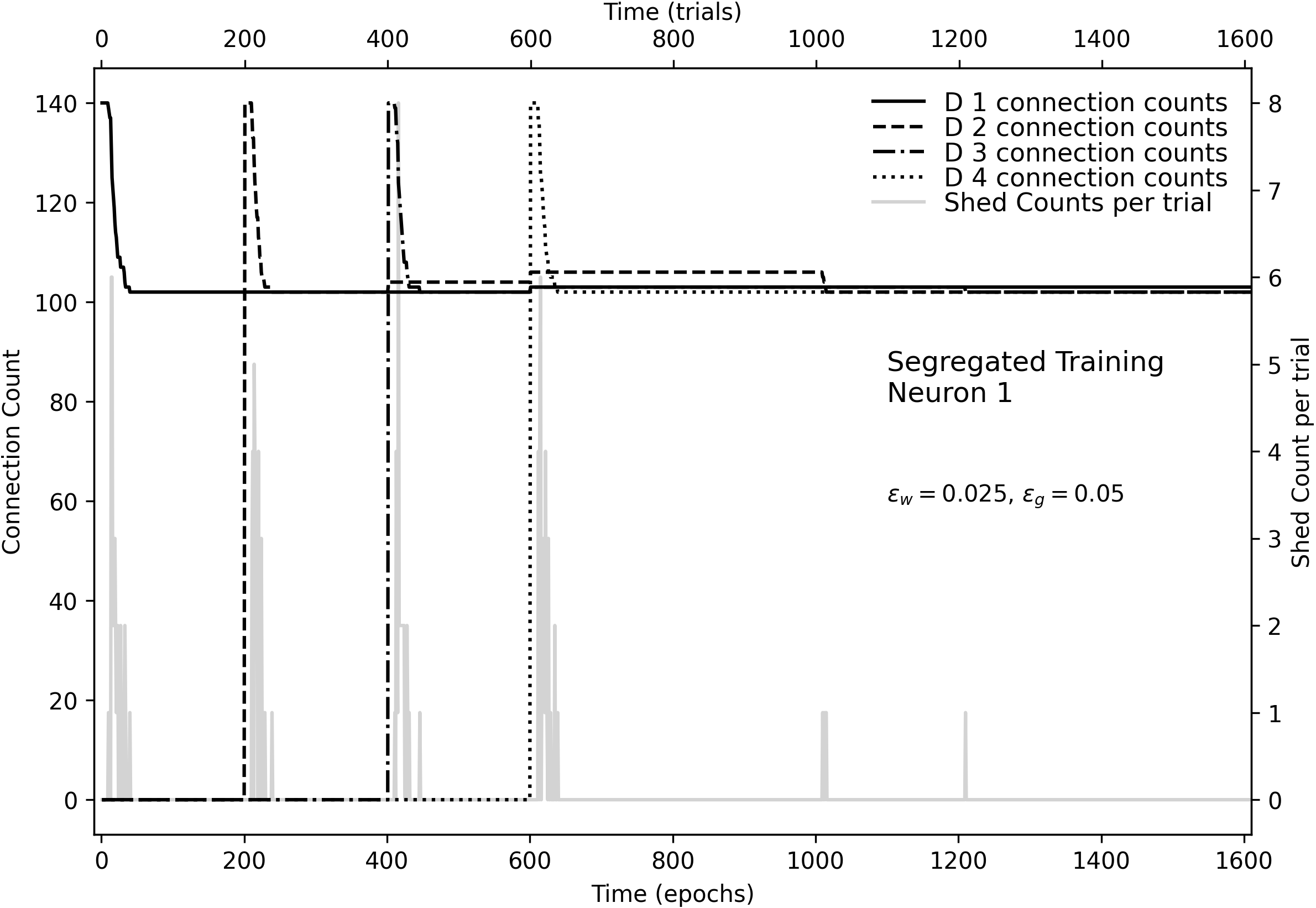}
		\caption{}
	\end{subfigure}
	\vskip\baselineskip
	\begin{subfigure}{0.48\textwidth}
		\centering
		\includegraphics[width=\textwidth]{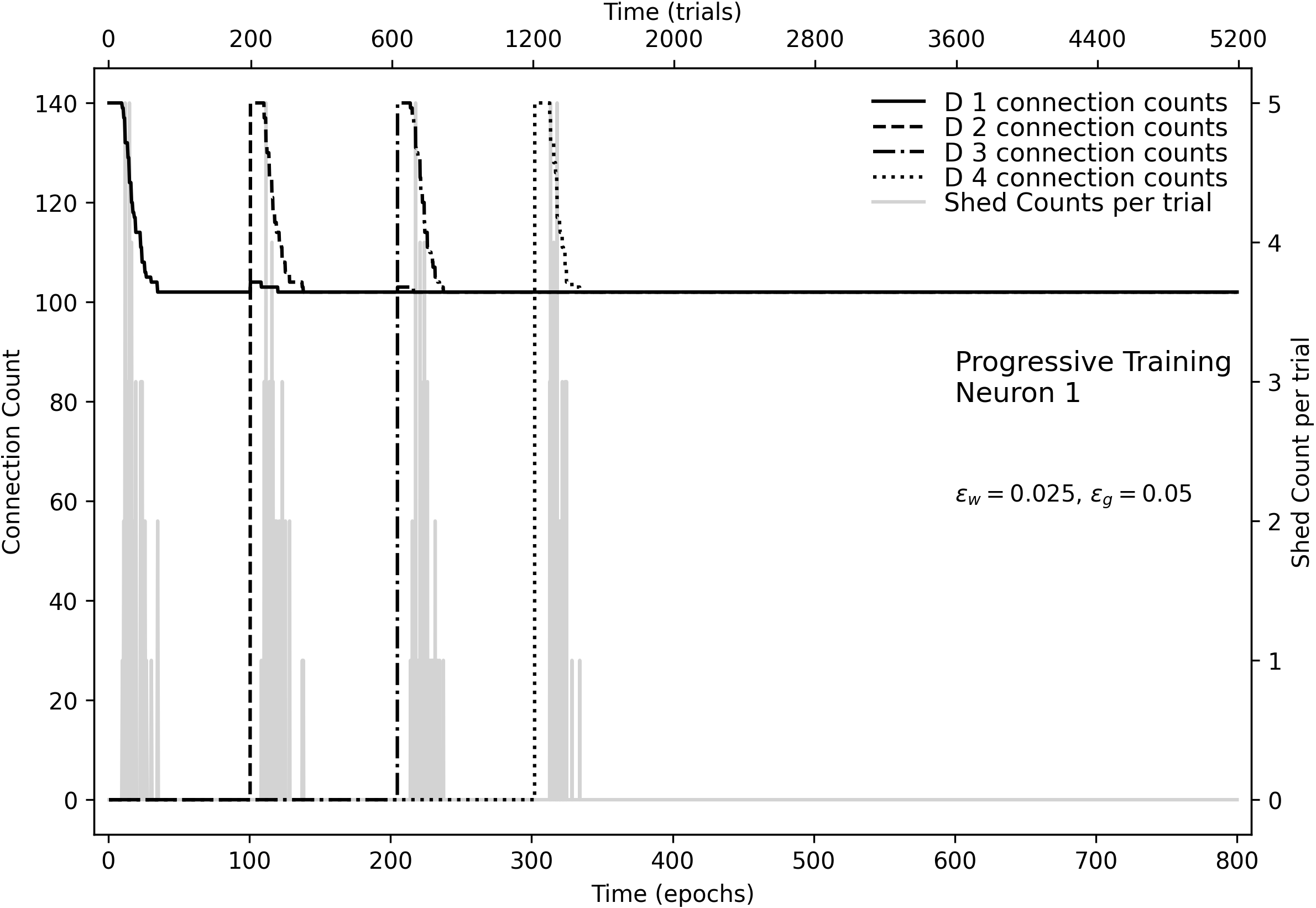}
		\caption{}
	\end{subfigure}
	\hfill
	\begin{subfigure}{0.48\textwidth}
		\centering
		\includegraphics[width=\textwidth]{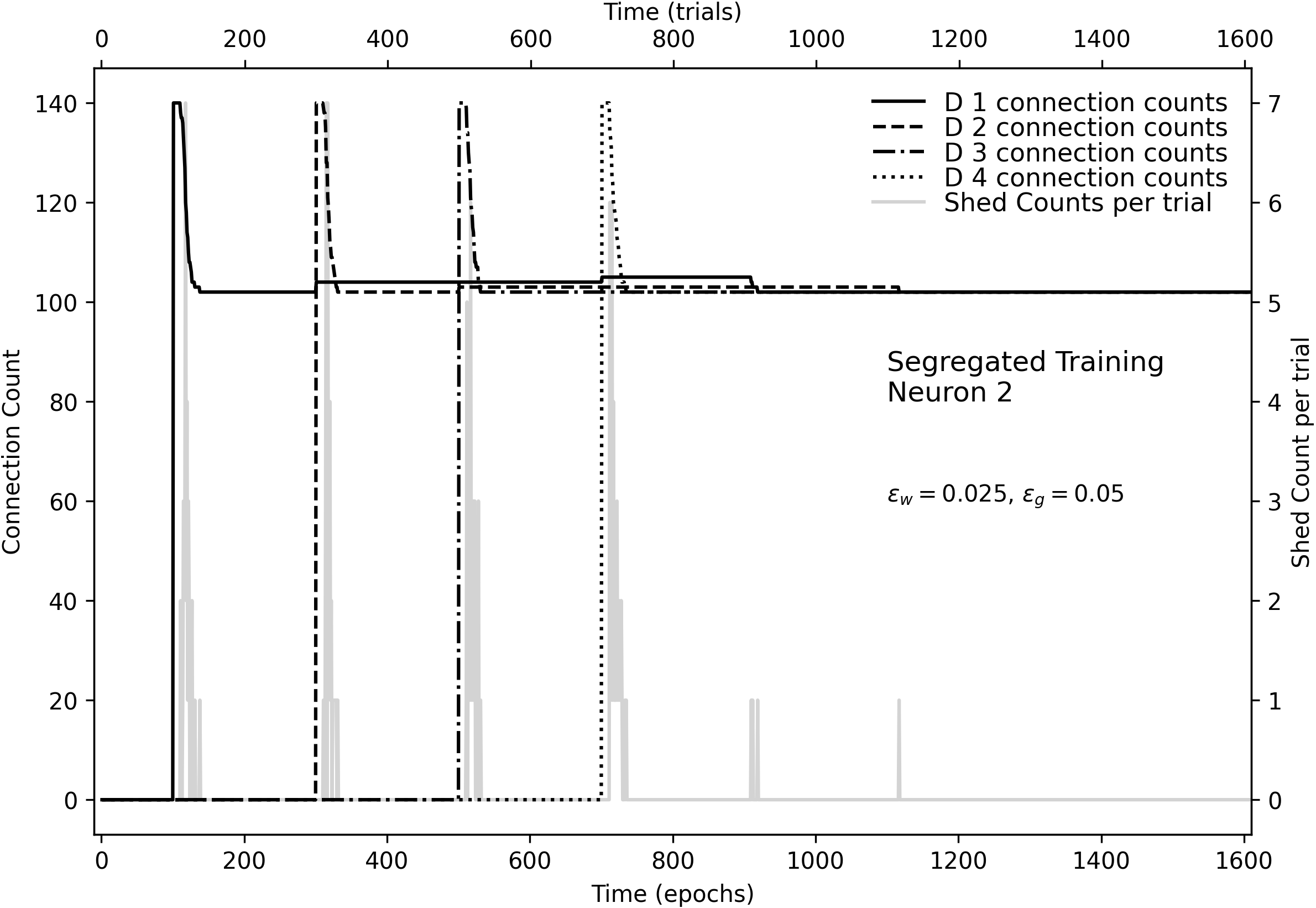}
		\caption{}
	\end{subfigure}
	\caption{
		Most of the connection counts and shed counts  stabilize quickly for concurrent (upper left), progressive (lower left), and segregated training (right figures, one for each neuron). For this graph training is for a fixed number of epochs. Concurrent is trained for 1300 epochs. Progressive is trained for 800 epochs. Segregated is trained for 1900 epoch.
		These data are from the same simulations generating Figure~\ref{quad-err-gam-fig}. Note different x-axes needed for different training paradigms. Parameters as in Fig \ref{quad-err-gam-fig}.
	}
	\label{quad-conn-shed-fig}
\end{figure}

In terms of epochs to stabilize, the convergence of segregated training appears particularly tedious, having to wait until ca. epoch 1200 for a single, last shedding event to occur and then simulating for another 800 epochs (not illustrated) to ascertain the absence of another shedding event. However, the epochs timescale is misleading. Segregated training employs only a single trial per epoch. Thus comparing number of trials to converge (note alternate time-axis on the top of each subfigure of Fig \ref{quad-conn-shed-fig}), segregated is just time consuming, if not a little faster, than the other two training paradigms.

However, there are two reasons for this apparently slow stabilization. First, there is only one trial per epoch, and second, segregated introduces only one class per phase while progressive introduces a new prototype for each class per new phase of training.

The connection counts over time also reveal that all three training procedures are tuned to produce an overshoot of synaptic connections, which speeds up development \citep{JuColbertLevy2017}. In the case of concurrent, the overshoot is extreme,  as large as is possible. In the case of progressive and segregated, the connectivity overshoots are a more modest fraction of final count per dendrite, about 40 \% larger than the stable final value 102 connections per dendrite. In regard to the stable connection count values for perturbation levels of 30/20 and 20/30 or below and $\gamma_{dj}(0)=1$, there is a reliable relation between the occlusion fraction and number of connections per dendrite, specifically, number of stable connections equals occlusion fraction times 128 where 128 is the number of ones in a parent prototype. However, the reader will recall Table \ref{gam0-tbl} showing that the number of stable connections decreases linearly with decreasing $\gamma_{dj}(0)$, down to as small as 63 (median) connections when $\gamma_{dj}(0)=0.6$.

\subsubsection{Other settings and convergence}

Recall that concurrent training can produce a two-dendrite per neuron or a four-dendrite per neuron solution to the $4|4$ problem depending on the parameter settings. The convergence of the previous figure corresponds to the two dendrite per neuron solution. Here on the other hand, Fig \ref{convergence_3-in-a-row-fig} illustrates convergence for a neuron that develops four functional dendrites, produced by altering parameters to $\alpha=0.01$ instead of one and $\epsilon_\gamma=0.999$ instead of 0.03 or 0.05. Because dendritogenesis is so rapid, (a) uses a stretched time-axis. (Note: only the first four dendrites are functional.) Although useful for viewing the time separation of dendritogenesis, this stretched axis is not useful for convergence, so Fig \ref{convergence_3-in-a-row-fig}(b) illustrates the same data over a longer time. Note the nearly ten-fold faster convergence  compared to the earlier Fig \ref{quad-conn-shed-fig}(a), which illustrates convergence of the concurrent paradigm with settings that produce a two-dendrite per neuron solution. The faster convergence here  is due to the rapid decrease in the rate of synaptogenesis, as controlled by $\epsilon_\gamma$. Note that with $\alpha=0.01$,  it appears that errors continue beyond the time of the last shedding event. In fact there are no errors  occurring after the last shedding event. The stretched out, no-zero error rate is due to the slow decay of the exponential moving averager that is tracking missed detections.

\begin{figure}[H]
	\centering
	\begin{subfigure}{0.48\textwidth}
		\centering
		\includegraphics[width=\textwidth]{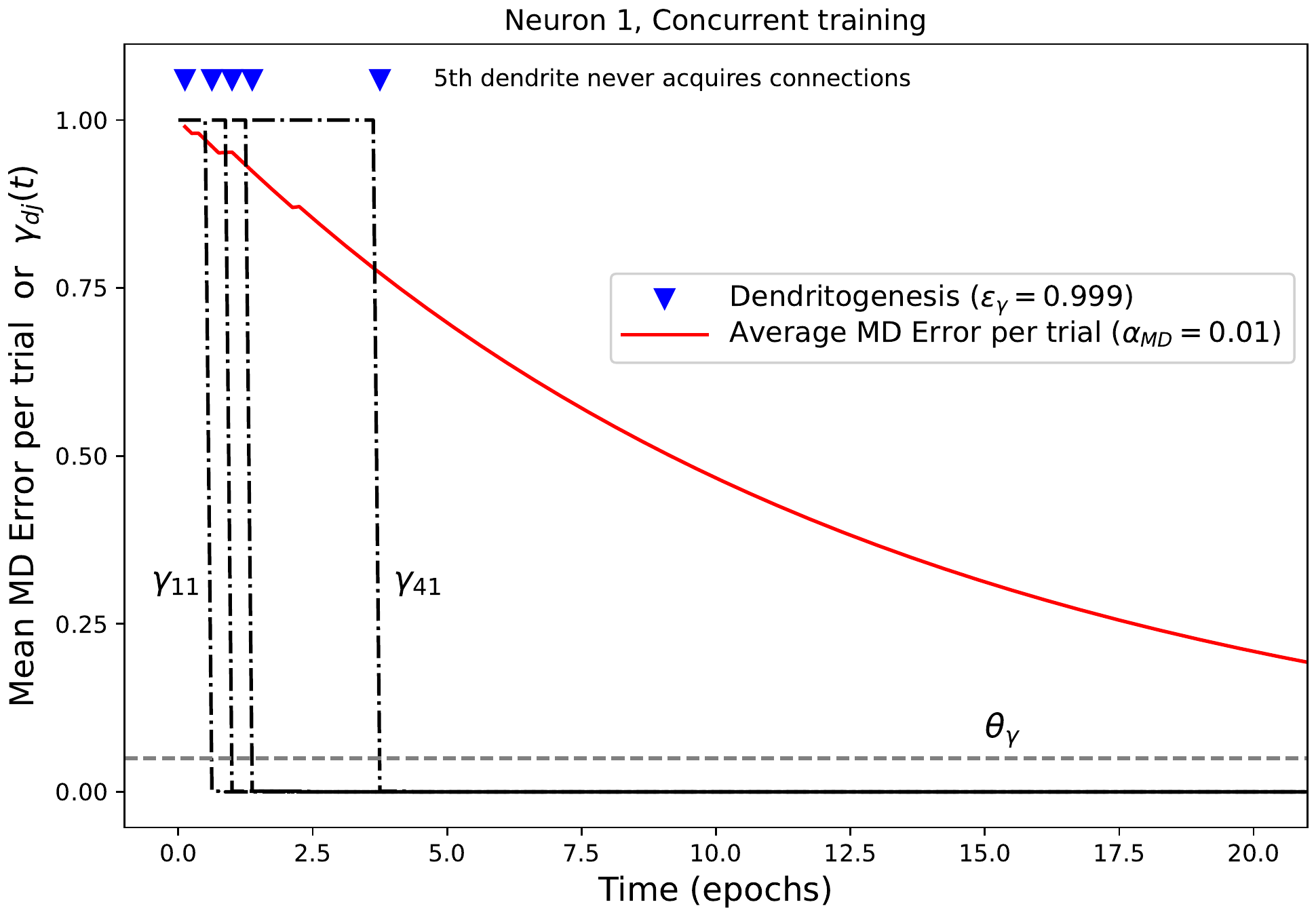}
		\caption{}
	\end{subfigure}
	\hspace*{-0.1in}
	\begin{subfigure}{0.49\textwidth}
		\centering
		\includegraphics[width=\textwidth]{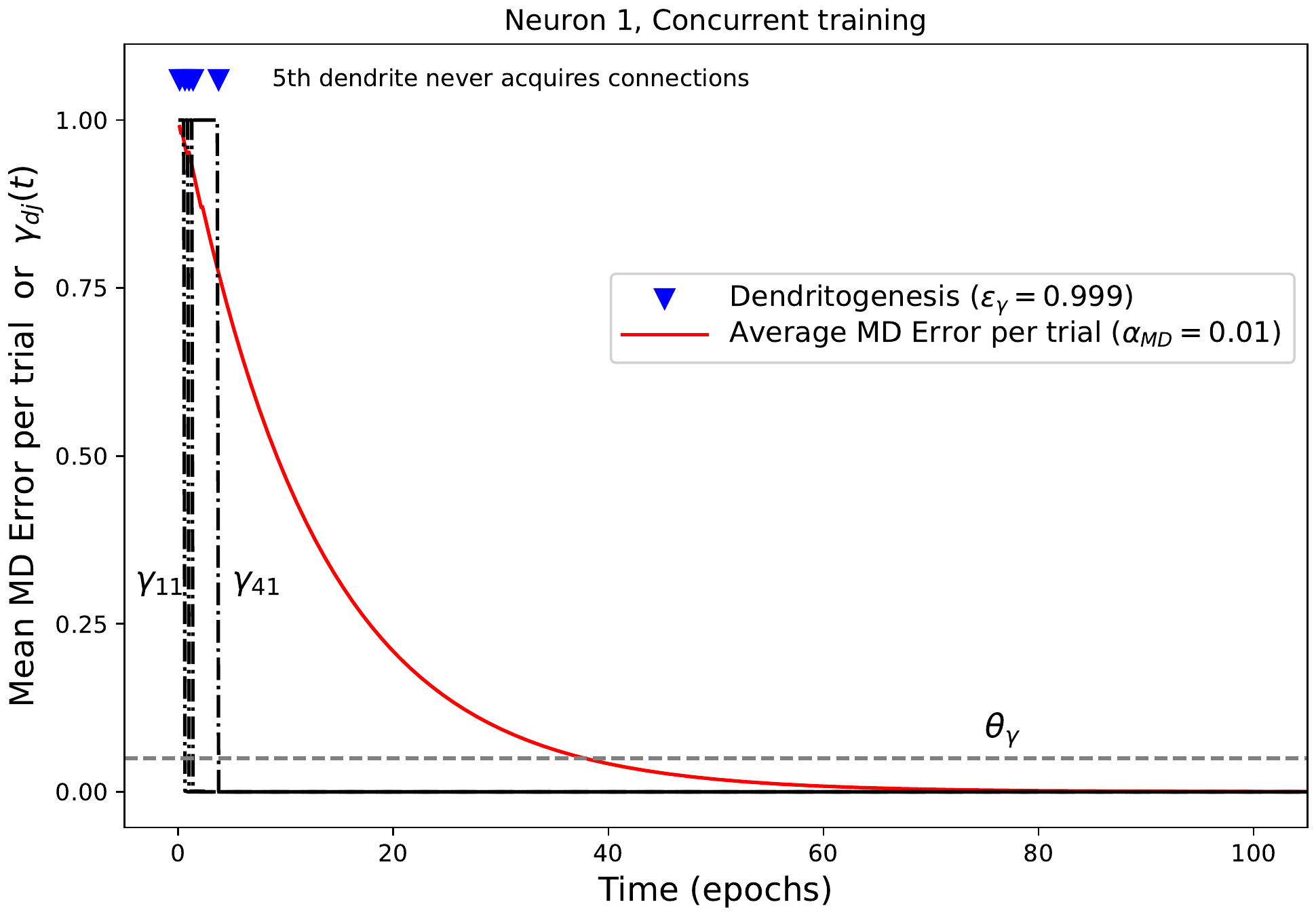}
		\caption{}
	\end{subfigure}
	\begin{subfigure}{0.49\textwidth}
		\centering
		\includegraphics[width=\textwidth]{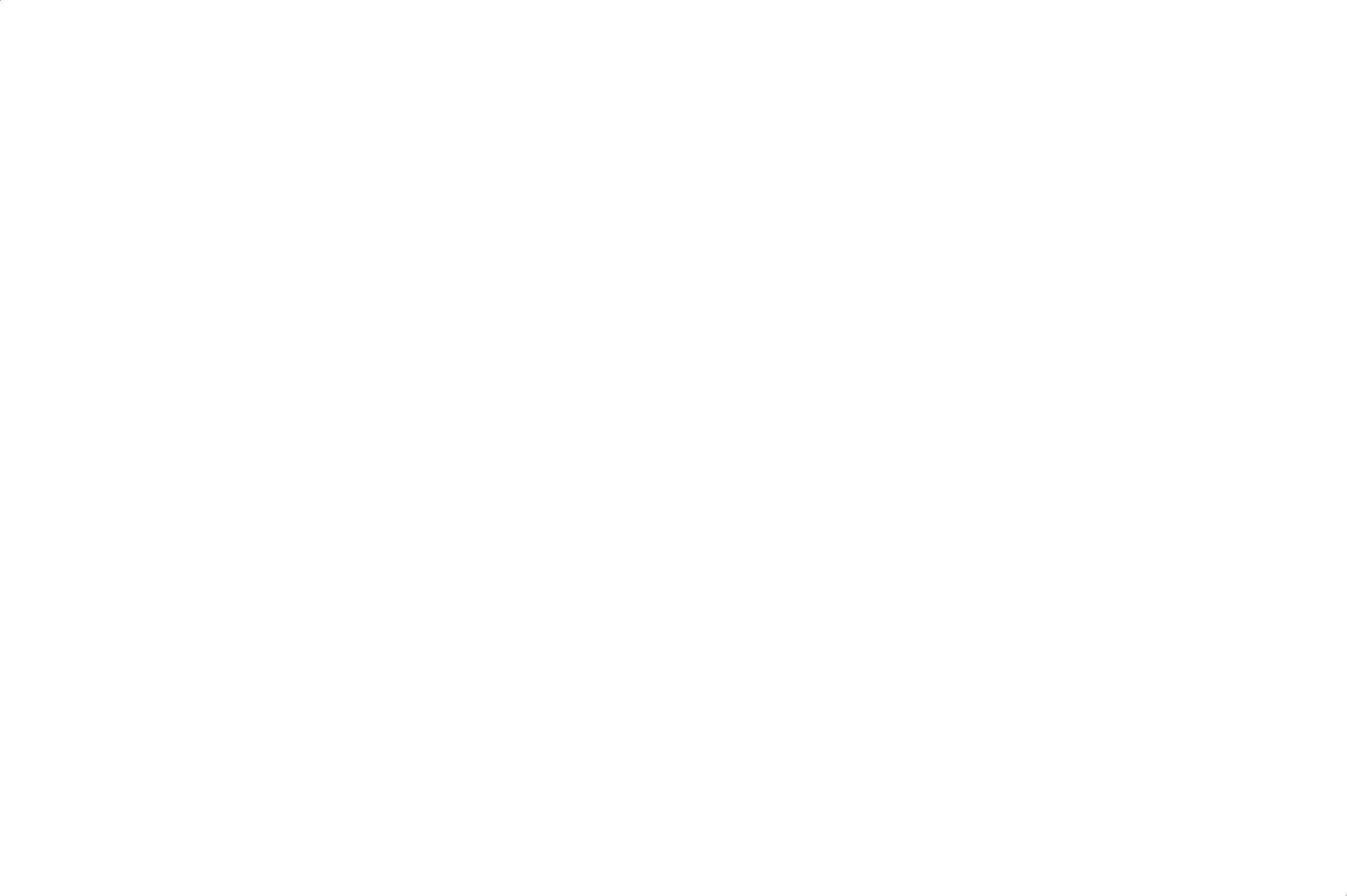}
	\end{subfigure}
	\hspace*{0.03in}
	\begin{subfigure}{0.49\textwidth}
		\centering
		\includegraphics[width=\textwidth]{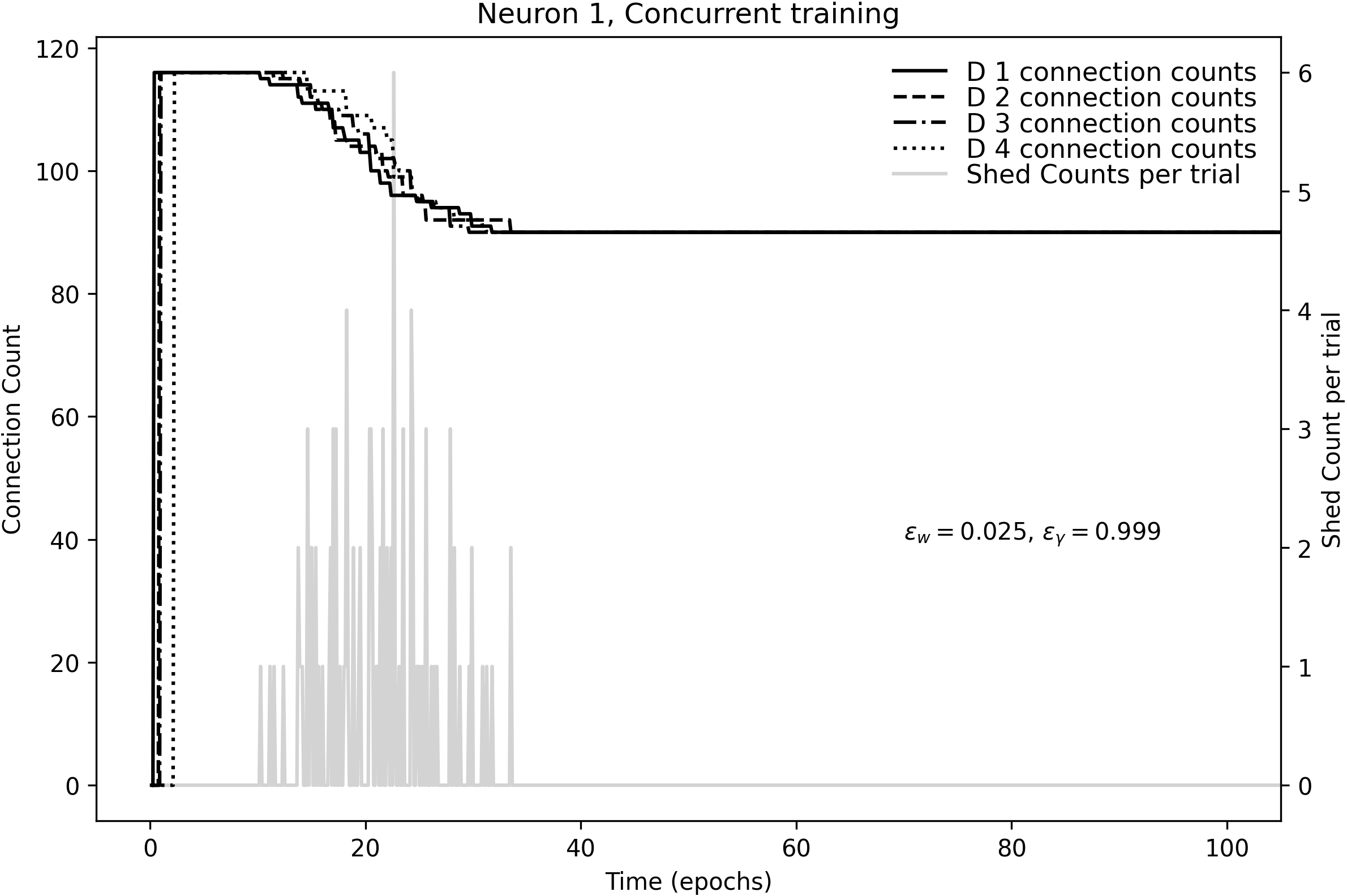}
		\caption{}
	\end{subfigure}
	\vspace*{-0.15in}
	\caption{The concurrent training dynamic when a four-dendrite, prototype matching neuron develops. Dendrite development occurs very quickly using the parameter settings conducive to four dendrites per neuron trained on the $4|4$ problem set. To see the successive timepoints of dendrite births, the timescale of (a) is stretched compared to (b) and (c).  Although convergence to zero errors begins its descent quickly, as in (b), reaching zero, ca. at epoch 70, this time to converge  is slow compared to the convergence of  shedding to zero (see c). Even though onset of shedding is delayed, around epoch 10 in (c), shedding halts very soon after, around epoch 34. Average error is the mean value of the missed-detections per epoch. Equilateral triangles in a and b indicate each dendritogenesis event. Dotted line in (a) and (b) indicated dendritogenesis threshold $\theta_\gamma=0.05$.  $\alpha=0.01$, $\epsilon_w=0.025$, $\epsilon_\gamma=0.999$.  }
	\label{convergence_3-in-a-row-fig}
\end{figure}
In the case of phased training, convergence is relatively robust and predictable to parameter settings, particularly $\epsilon_w$. Smaller values of this parameter slows down convergence, but it has other advantages (see below). Fig \ref{conn-stability-fig} compares time to connectivity convergence for $\epsilon_w$ equals 0.025 vs 0.002, a 12.5-fold range. The most important difference is that given the same f phase durations, the slowed convergence of the 0.002 setting results in connectivity stabilizations that occur a little beyond the phase that a new prototype is introduced. Note that connection counts themselves achieve their final value within (or just about within) each phase.

\begin{figure}[H]
	\begin{subfigure}{0.49\textwidth}
		\includegraphics[width=\textwidth]{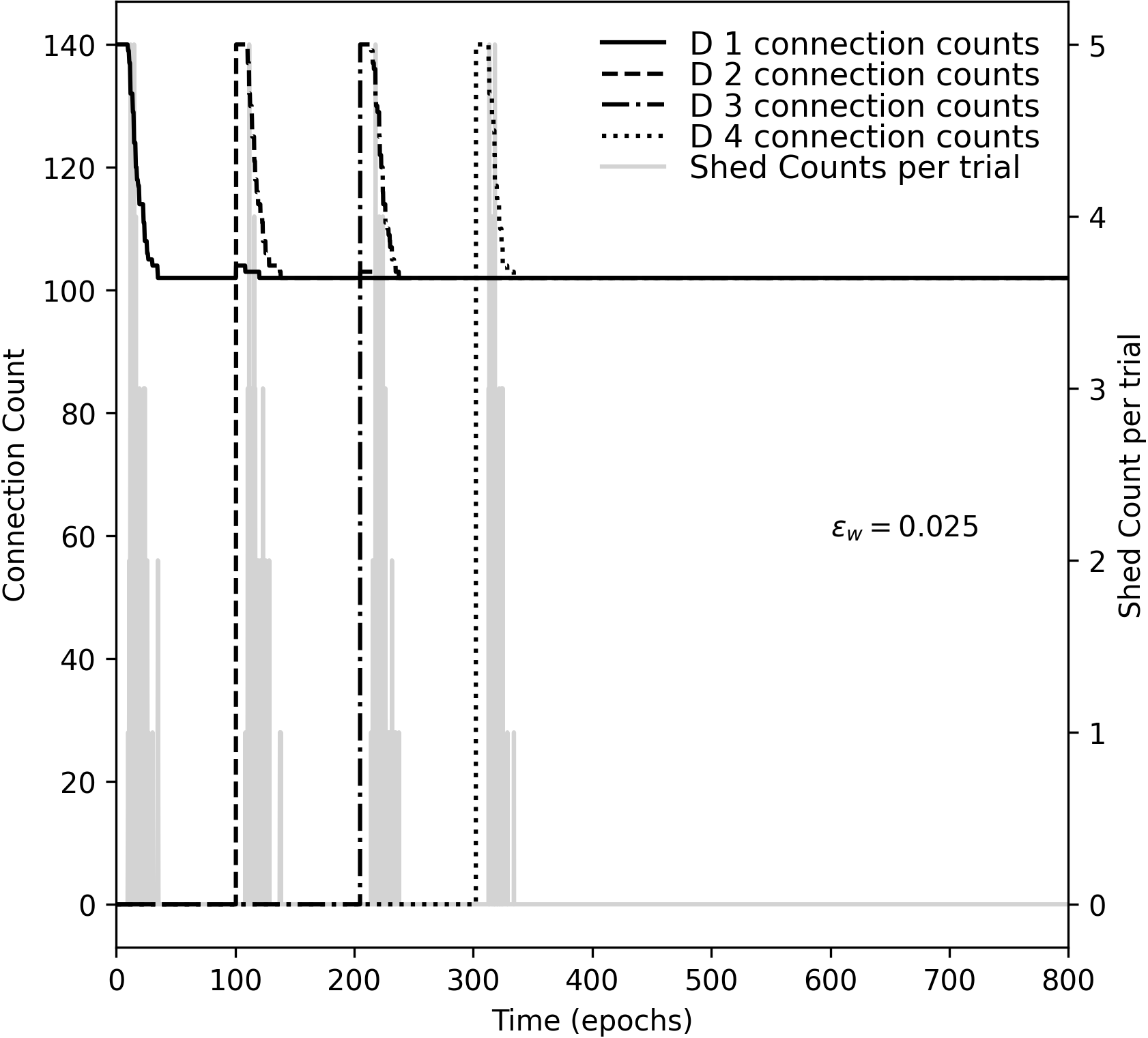}
	\end{subfigure}
	\hfill
	\begin{subfigure}{0.49\textwidth}
		\includegraphics[width=\textwidth]{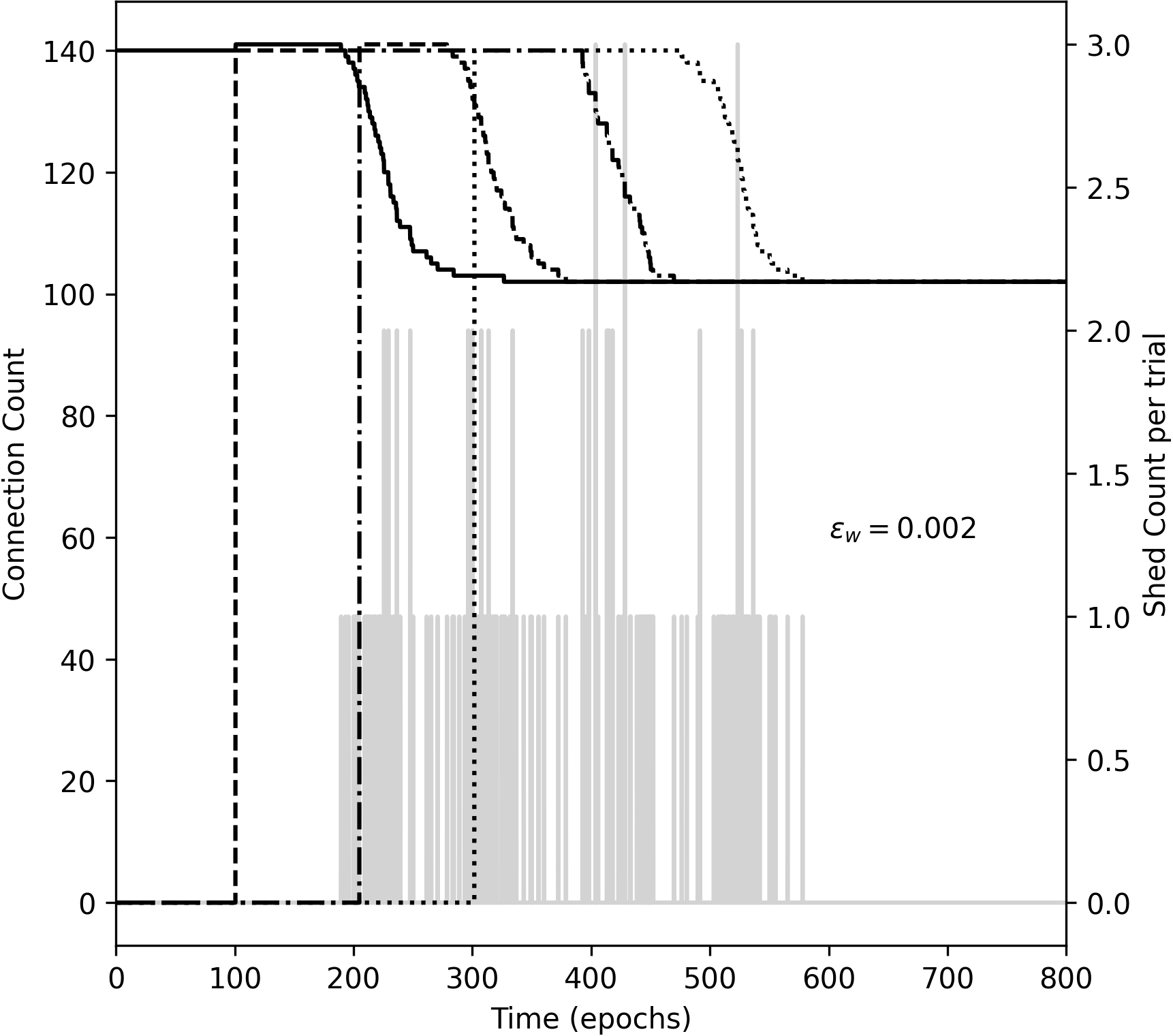}
	
	\end{subfigure}
	\centering
	\caption{
		The synaptic modification rate parameter $\epsilon_w$ has a strong influence on the connectivity stability.
		In (a) with $\epsilon_w = 0.025$, the connection counts are observed to stabilize within the first 25 epochs of each phase,
		and connectivity does not change after epoch 336.
		In (b) with $\epsilon_w = 0.002$, the connection counts do not stabilize until epoch 582.
		The errors dropped to 0 by epoch 302 for both values of $\epsilon_w$.
		(Progressive training, $\epsilon_\gamma = 0.05$, $\alpha = 1$.)
	}
	\label{conn-stability-fig}
\end{figure}

\subsubsection{Convergence and stabilization of synaptic weight values}
Because $\epsilon_w$ is a non-adaptive, small  positive constant, the synaptic weight values never stop fluctuating. However, while $\Delta w$'s do not converge to zero,  their expectation and long term empirical average do not change  \citep{LevyGeman1982}.  Figure \ref{weight_angle_convergence} illustrates such stabilizations for  representative dendrites each from a  representative neuron (all trained with the progressive paradigm on the 4$|$4 problem with 20\% occlusion and on-noise varied from 10 to 40\% in steps of ten).
 
Note that all eight dendritic weight vectors converge  to an average angle relative to its learned  prototype. The greater the prototype perturbations the greater the average angle against the dendrites preferred prototype and the greater the fluctuations of this angle around the average. For both values of $\epsilon_w$, both of these increases are monotonic.

The numerical values will make it easier to compare across $\epsilon_w$'s.
Using the last 400 epochs for the measurements and noting values in the sequence of increasing perturbations (20/10, 20/20, 20/30, 20/40):  for the average angle, $\epsilon_w=0.025$ 27.5, 27.8, 28.2, and 29.5 degrees which are larger than the same averages when $\epsilon_w=0.002$ 26.88, 26.90, 26.94, 27.05 degrees.  Comparing the standard deviation in these angles in the same way, for $\epsilon_w=0.025$ 0.08, 0.11, 0.21, 0.23 degrees. For $\epsilon=0.002$ the standard deviations are about ten times smaller, 0.01, 0.01, 0.01, 0.02. 

\begin{figure}[H]
	\centering
	\begin{subfigure}{0.48\textwidth}
		\includegraphics[width=\textwidth]{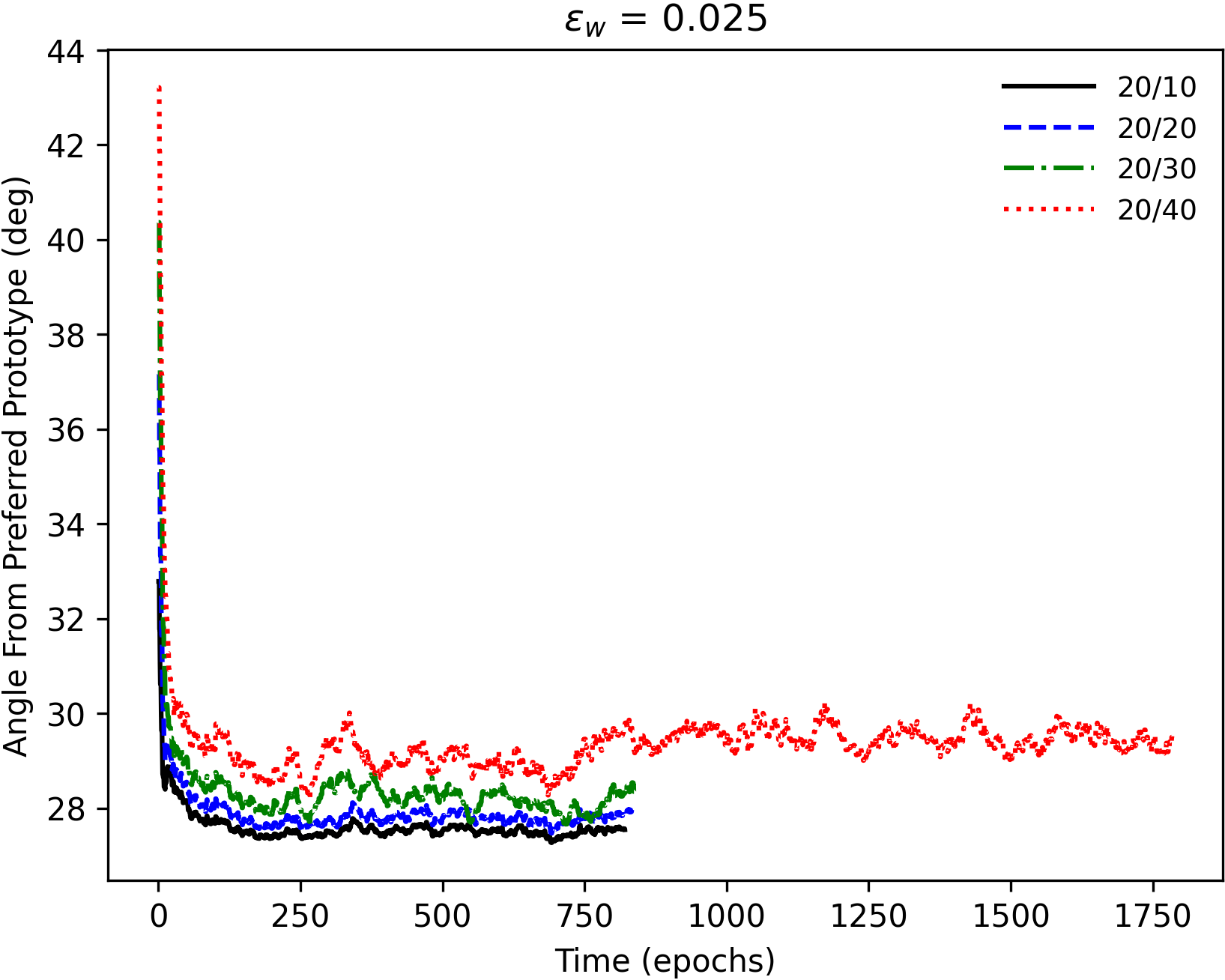}
	\end{subfigure}
	\hfill
	\begin{subfigure}{0.48\textwidth}
		\includegraphics[width=\textwidth]{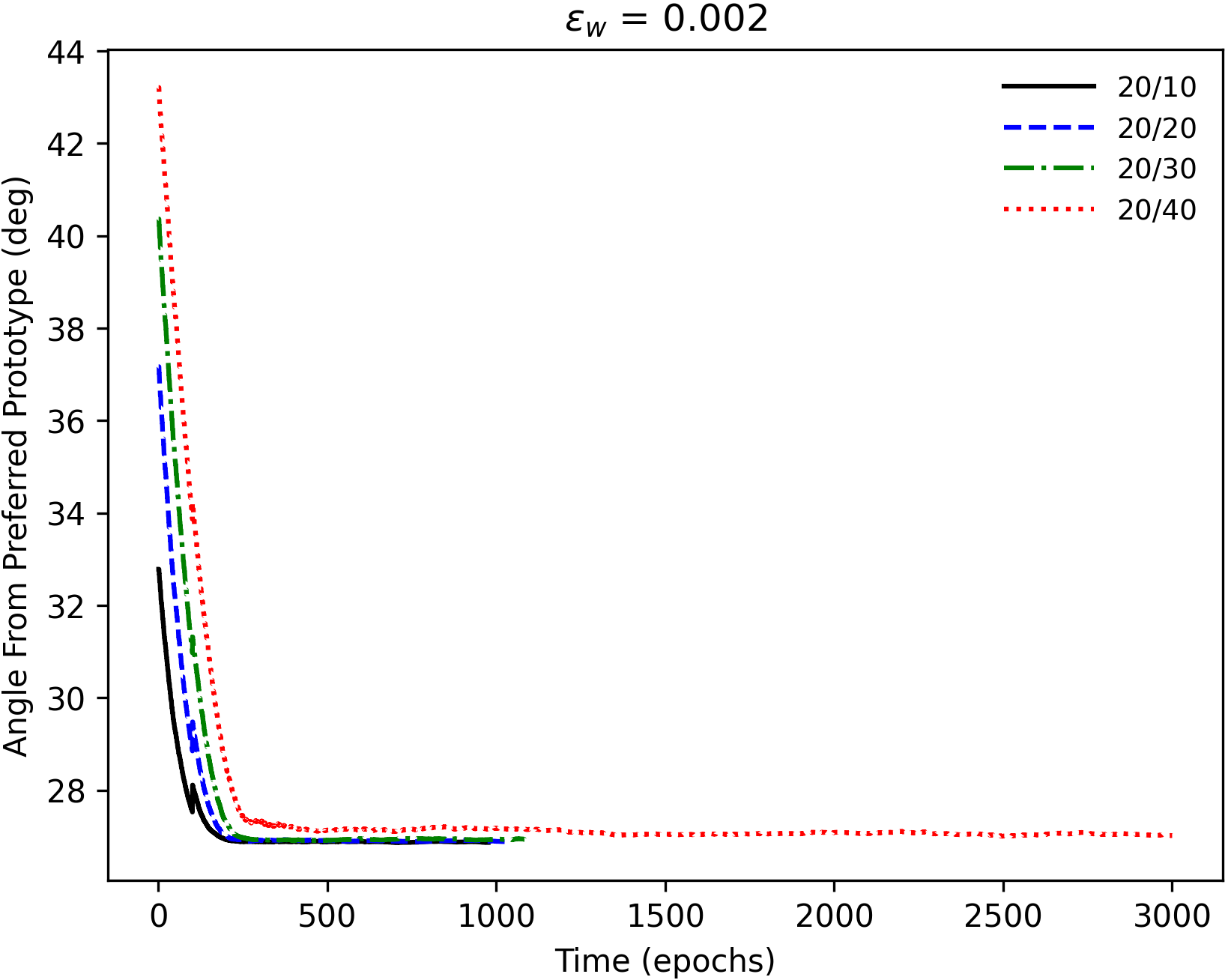}\\
	\end{subfigure}
	\caption{Stochastic convergence of the average angle between a dendrite and its preferred prototype. Comparing across figures, smaller $\epsilon_w$ slows convergence but decreases variance around each stable point. Increasing on-noise modestly slow convergence and produces greater variance around stable points. The small but noticeable differences in stable-point angles. Although individual synaptic weights are fluctuating above and below their on-average stable point, the fluctuating angle is consistently moving farther away as it responds to successively higher levels of on-noise.
	Progressive training on the $4|4$ problem set with the same prototype perturbations for training and testing.  All simulations are running at zero error. All simulations but the 20/40 achieve stable connectivity using the 500 epoch criterion. The 20/40 simulation ran for 3000 training epochs with no connections changes during the final 302 epochs.}
	\label{weight_angle_convergence}
\end{figure}

In the appendix the reader will find more data concerning the time course of various convergences to stable values including Fig \ref{tts-vs-phase-duration-fig},  stability as a function of (comparing two) phase duration, and  comments on convergence in parameter settings Appendix E.

In sum, the important point of this section is that quantitative dynamics depends on the training method but all three methods display convergence to stable connectivity when prototype perturbations are moderate (30/20 or 20/30) or even smaller.

\section{Discussion}
\subsection{Overview of results}
The results demonstrate that a neocortex-inspired and biologically plausible (i.e., local) neuron algorithm solves certain computational problems that simpler neural-like elements find difficult, costly, or impossible. 

The ingredients to the algorithm are the following:

(i) There is at least one neuron per class. At each trial, a feature vector occurs, then each neuron transmits its dominant dendritic excitation to the interneuron, whose feedback determines the winner (or winners if there are more than one neuron per class); the winner(s) is the class prediction based on the featurers present. 

(ii) Then  the class signal arrives. The neuron-localized in-class signal is a necessary permissive event for various modifications at such a neuron; the in-class signal also allows the determination of a missed detection by an in-class neuron.

(iii) Modification of existing weights, including shedding, occurs only at the most excited dendrite (via CDS) of an in-class neuron.

(iv) If youngest dendrite has been sufficiently successful (enough correct predictions) and if the missed-detection error-rate is sufficiently large (as small as a single error as  controlled by the  $\alpha$ setting), then one new dendrite appears.

(v) If this in-class neuron committed a missed detection, all dendrites of this neuron are eligible to participate in synaptogenesis.

The algorithm endows neurons with five notable characteristics: (A) The algorithm resists catastrophic interference. (B) The algorithm produces a one-layer, two  neuron solution to the XOR and to related, but harder non-linear discriminations. (C) As a function of increasing input dimension and number of coactive features, the memory capacity seems unlimited (but see Section 4.3.3). (D) The algorithm can unmix the components of a mixture distribution, and within a class, (E) the algorithm performs this unmixing in an unsupervised manner.

The simulations in Results are designed not just to demonstrate the five preceding functions but are also designed to understand (i) how the CD\&SAS algorithm works (see also Section 4.5) and (ii) in what senses  the algorithm is extendable beyond the low dimensional simulations, i.e., 256-D inputs and the minimal two or three primary neurons, of the Results. 

Regarding this last point, the variety of 256-D problem sets imply  the feasibility of successfully extending the algorithm  to input worlds with larger and possibly less well defined  structures and statistics. The results demonstrate that the algorithm can be expanded to any number of classes and that these classes can have varying probabilities as in the 2/2/4 problem where the third class is twice as likely as each of the other two and the 2/6 problem where the second class is three times as likely as the first class. Individual prototypes can  vary widely in probability (Table \ref{unequal-prob-tbl}) where one prototype is one-tenth as likely as any other prototype in the same class. 

Coming at extensions of learnable problem sets from the other direction, a substantial fraction of Results is dedicated to pointing out feature vectors and the associated class that are learnable versus those that  are not learnable in the sense of achieving  zero or nearly errors. Note that in the 256-D world,  even before any prototype perturbations, the eight prototypes are severely overlapped (fifty percent in most every direction, Fig \ref{angle-matrix-4d-vs-8d-fig}b). See Fig \ref{exemplars-with-diff-occ-fig} for examples with randomizations); nevertheless, the 20/40 and 40/20 prototype perturbations are still learnable to the level of  error-free performance. However, when exemplars are the result of further perturbations (e.g., 30/30 prototype perturbations during training and testing), errors begin to creep in. If the exemplars from  different classes become too similar, which is what occurs at large prototype perturbations, as in Figs \ref{diff-test-noise-concurrent-fig} 60/20 and 20/60, then discrimination performance deteriorates to unacceptable levels. Of course, such deterioration can be expected of most any paradigm faced with both extreme amounts of uniform randomization and, probabilistically speaking, never a repeated exemplar. That is, recall that at the larger prototype perturbations, the generated exemplars of training and testing can be closer to a non-parent prototype than  parent prototypes; therefore, even if this particular exemplar is associated with the dendrite that points at its parent exemplary, two aspects of the future are note worthy: this exemplar will never occur again (with high probability), and there will be exemplars from another parent prototype appearing very close to this "learned" exemplar, leading to a classification error.

Most influential to the research here is the work of Mel and colleagues. In fact, we have put into practice some suggestions in \citealt{Behabadi2014,Mel2017,Caze2013}. They demonstrate via simulations that neurons with dendrites can provide much greater functionality, memory capacity, and classification accuracy than neurons with a single set of weights \cite{Mel1991,Poirazi2001}. Mel's model has two commonalities with the model proposed here: the neuron's output is determined by multiple dendrites each having a set of synaptic weights, and these dendritic weights can grow or shrink or shed. However, the adaptive flavor of Mel's model is different from the model proposed here primarily because (a) it uses gradient descent with annealing to determine the dendritic weights, (b) the dendrites do not compete to determine the neuron's output, (c) the synaptic modification rule is driven by derivatives and error signals, (d) synaptogenesis is not driven by missed detection errors, and (e) the number of dendritic branches is fixed; i.e., there is no explicit dendritogenesis rule driven by missed detection errors.
 
\subsection{Understanding the algorithm as biological}

\subsubsection{The current algorithm evolved from AS and SAS}
Adaptive synaptogenesis was created as a synaptogenic algorithm consistent with Hebbian synaptic modification \citep{Levy1985,Levy1990}.
A key part of the algorithm is the biased random formation of synapses and the selective shedding (see also \cite{Scholl2021}). In the AS algorithm synaptogenesis of excitatory synapses is designed to create, homeostatically, a desired activity level by favoring neurons with low activity and ignoring neurons with appropriate activity. In the SAS algorithm \cite{BaxterLevy2019,BaxterLevy2020}, this desired activity level is conditional on the class. Here the conditioning is also extended to a dendrite of that class.

As will be covered in detail below, the $w_{idj}$ synaptic modification  equation encodes a positive correlation at a synapse via a stochastic approximation (averaging) algorithm. The shedding rule enforces this restriction to positive correlations and, working in concert with synaptogenesis, provides a random sampling mechanism that functionally searches, via biased random synaptogenesis, for local positive correlations. 

In the absence of dendritogenesis and with an endowment of just one dendrite per neuron, the synaptic modification equation confers a undesirable susceptibility to catastrophic interference (Fig \ref{8D_bar_graph}), particularly when synaptogenesis produces new synapses on this solo dendrite. When these new synapses are sufficiently co-activated by a novel prototype, the anti-Hebbian  aspect of the synaptic modification rule (subtraction terms $E[X_i]$ and $w_{idj}$) weaken and eventually discard  synapses acquired during earlier learning. Indeed  the single dendrite neuron, using the $\Delta w_{idj}$  modification equation, is not only subject to catastrophic interference given the appropriate sequence of exemplars but it is not saved by AS or SAS in the case of the prototypes used here. 
However, enhancing SAS (i) with dendritogenesis, (ii) the inhibitory feedback interneuron, and (iii) CDS leads to  encodings, i.e., dendritic weight vectors, corresponding to  distinct exemplar clusters generated around individual prototypes. Moreover, disruptive synaptogenesis and weight development on previously successful dendrites is now avoided, ultimately because there are no missed detections by a neuron but even before that synaptogenesis becomes highly unlikely due to the decrementing $\gamma_{dj}$ driven by a large number of successful, in-class discriminations by a dendrite. 

\subsubsection{Restating the algorithm for biological feasibility }
This section translates the DC\&AS algorithm  into a more biological form. To begin, there are primary neurons (e.g., pyramidal neurons), which are excitatory in their output. One target of this output is very fast inhibitory neuron that feeds back to all the primary neurons that excite this one special neuron.

In the biological setting, processing occurs continuously in time with a start signal, e.g., the ending of a saccade and the beginning of a fixation. With the start signal, neurons are reset and then  feature-based excitation builds up continuously in time. A summed, somatic depolarization affects all electronically proximal dendrites to the same excitation, but each such dendrite has its own localized excitation added to this general depolarization. As the excitatory signals are building up on each dendrite, one dendrite will be the first to reach the NMDA-receptor mediated   threshold (\cite{Schiller2000}, which might also be augmented by a voltage-activated Na-conductance \cite{Ariav2003}), which has two effects. First, there is the local communication from the spiking dendrite to the other perisomatic dendrites of this neuron. This spike provides the CDS signal, perhaps activating K$^+$ conductances, which prevents the other dendrites from firing and modifying; events which correspond to the within neuron dendritic competition. The  second effect is initial segment spike-initiation \cite{Mel2017}. It is this event that is communicated to the neuron's recipients including the inhibitory interneuron. In turn, the  inhibitory interneuron is excited and feeds it inhibition to all neurons suppressing further spikes. In the event that no neuron spikes in a particular time-window (especially during very early development before many synapses are formed), the threshold to fire is lowered uniformly across all neurons (or neurons are uniformly excited by GABA-A synapses)  until one dendrite (and thus just one neuron) fires. For even more realism, there are multiple neurons coding for a prototype-class and the feedback is tuned to allow an appropriate number of neurons to fire.

Also relevant to the biological perspective are the interpretations of the different training paradigms; for such interpretations in the context of life-long learning, see Appendix G.

\subsubsection{Robustness versus energy efficiency}
In regard to parameterizing for more or fewer connections per dendrite, there is a trade-off. With more connections specific to one prototype, the angle between dendrites increases, which is desirable because there is greater resistance to prototype perturbations. On the other hand, one loses the implicit energy efficiency of fewer synapses.

\subsubsection{Limitations on encodings and memory}
It might seem that the memory capacity is infinite. Consider the following extension of the present model in the context of the perturbation-free 8-D problem set and the $4|4$ problem set, which will produce four dendrites per neuron. Double the input space to sixteen dimensional and double the number of prototypes by just replicating the original eight prototypes but shifted from the original first eight dimensions to the newly created dimensions nine through sixteen. Note that the eight new prototypes are orthogonal to the first eight prototypes. These new prototypes are divided up evenly between the two classes, and their will be no difficulty creating eight more dendrites, four more on each neuron. Continue this prototype enhancement by increasing input dimension for a long as you like, and the new prototypes will be individually encoded via new dendrites. In this manner, one might argue for unlimited memory capacity.  

Of course, physical limitations inevitably appear, so memory capacity is not infinite. As a computer program, the memory capacity of this algorithm  is limited by computer memory. From the biological perspective, there is also a limit on memory. This limit occurs in terms of space occupied by the dendrites, synapses and axons. There are profound spatial and energy consumption problems if one trying to increase the number of dendrites and synapses beyond some finite value. Time is also limiting, particularly in the context of biological development, where this developmental time maps to number of  training trials.

There are also training experiences that a single layer network  can never learn. A well known behavioral/cognitive problem is learning reversals while still retaining original memories. Again the $4|4$ problem will do as an example. First, train to perfect performance with prototypes one through four associated with class I and prototypes five through eight associated with class II. Then reverse the prototype-class association (e.g., prototypes five through eight are now associated with class I). Repeat such contingency reversal  many times as segregated trainings. A mammal can learn both contingencies and use the initial results of a single test to determine its responses on that particular day. That is, mammals can learn higher order context-dependent associations. Undoubtedly it takes multiple brain regions to solve such a problem, and certainly it is not solvable by the single layer model used here.

\subsection{Stabilization of learned encodings}

Because dendritogenesis and synaptogenesis require a missed-detection, any simulation that achieves error-free performance will not gain another synapse although connections can still be shed  whenever the $\Delta w_{idj}$ equation drives a synaptic weight down to the shedding threshold. However, under the error-free assumption such shedding is merely part of the weight stabilization process. That is, the weight modification equation produces  on-average zero change  in a stationary environment when connectivity is fixed (see  Fig \ref{weight_angle_convergence}; see also \cite{Fauth2015,Fauth2016} ). Such stationary inputs include concurrent training and the last phase of progressive training. In the case of segregated training, each phase is locally stationary so here too there is convergence in terms of the average value of synaptic weights. 

A back of the envelope calculation supports the empirically observed error-free performance and implied no new connections.  First note that, under these moderate perturbations, the developed dendrites point rather specifically in the  direction of the prototype (in the case of progressive and segregated training) or pair of prototypes (in the case of concurrent training at standard parameter settings). The connections that are made are precisely limited input lines that define the appropriate prototype (or pair of prototypes). Then once such specific connectivity is established, the issue becomes how far can  generated exemplars move beyond a parent prototype into the vicinity of a non-parent prototype. Consider the case, a dendrite developed with concurrent training that contains 64 connections (see Fig \ref{c2}). All of these 64 connections belong exactly to the intersection of two prototypes; such a dendrite recognizes  32 of these 64 connections which are shared by five of the other six prototypes. Recalling that the prototype perturbation method is exact (i.e., constant) in the number of complemented input lines, the 30/20 perturbation produces an exemplar that shares seventy percent of the one-valued lines of the parent prototype, so there will be this percentage of activated connections on the prototype appropriate dendrite, i.e., 45 synapses. On the other hand, the on-noise of 20\% only activates 26 input lines, all of which may possibly activate synapses on a single \textit{in}appropriate dendrite. Then so long as the weights are all about the same, the dendrite with 45 active input lines will be more excited than the dendrite getting the 26 active inputs. Thus the appropriate dendrite is more excited by a wide margin; i.e., no errors occur under this worst possible 30/20 randomization of input lines. The calculation produces an even wider margin of safety for the phase training methods and the parameter settings used here since each dendrite contains 90 or more connections belonging to it recognized prototype.       

By design, parameters, e.g., $\epsilon_w,\; \epsilon_\gamma$, and $\gamma(0)$ are selected to produce weight convergence  within each phase (or not long after) when using a phased training method. 

Convergence of synaptic weight values is in the sense of stochastic convergence, i.e., spending most of the time near an on-average stable point \citep{Ljung1978,Kushner2003,Kushner2010}. Examples are seen in Fig \ref{weight_angle_convergence}.  When the settings are appropriate for convergence, the  $\Delta w$ algorithm converges in the limit of many samples to a small region with a standard deviation of about the size of $\epsilon_w$. The center of this region is a particular conditional statistic; the form of this conditional statistic depends on the particular $\Delta w$ equation in use. 

\subsection{Understanding the algorithm from the perspective of machine learning and statistical inference}
 
\subsubsection{The algorithm develops latent variables that correspond to the true latent, vector variables}
Having established at least empirical, on-average stability of dendrite weight vectors (Fig \ref{weight_angle_convergence}), it is appropriate to understand and interpret what each dendrite is encoding when reaching such a state. Such insights include both the vector statistic encoded by a dendrite and also  latent variable interpretations of this vector and the scalar excitations mediated by such a vector. For the latent variable interpretation, we will call (i) the prototype (or in some cases intersection of prototypes)  "the true" latent variable versus (ii) a dendrite's developmentally learned latent variable, i.e., its weight \textit{vector}, and finally (iii) the \textit{scalar} latent variable that is the projection of an exemplar, $x(t)$, onto the normalized, learned  dendritic weight vector, $\frac {w_{dj}} {\sum_i w_{idj}}$, yielding a $y_{dj}(t)$, i.e., equation (1) using the stable weight values which are not time-dependent.  

First, because the true structure that generates exemplars is known to us, the researchers, one can compare the created latent vector variables, the dendritic vectors, to the true latent vector, the exemplar-generating prototypes. In fact even under moderate levels of prototype perturbations, the algorithm allows dendrites to develop in a manner that successfully creates  latent variables, which allow perfect discrimination performance, but do not precisely match the true latent variables.

Second, once at or near convergence to a stable connectivity, the excitation of each dendrite by an exemplar is also a meaningful latent variable. This scalar latent variable is a dendrite's  estimate of the resemblance between the current exemplar and a dendrite's learned weight vector. With this perspective, such a scalar latent variable is just a projection upon a constructed basis. In the case of the $4|4$ problem with no more than moderate prototype perturbations and phased training, there are eight dendrites  (e.g., Figs \ref{dp-angles-44-pro-fig} and \ref{c1}) and therefore eight such projections -- one for each dendrite, valued $y_{dj}(t)$. Thus, such a set of dendrites can be interpreted as a basis set. For the inputs used here, these dendrites are approximately equiangular  (see Figs \ref{dendrite-angles-44-pro-fig} and \ref{dendrite-angles-26-pro-fig}). Then one is encouraged to note (i) this eight dimensional encoding is a significant compression from the 256-D input vector and (ii) the equiangular approximation implies an approximation of the optimal Welch-bound for such non-orthogonal non-negative bases.
 
When there is not a one-to-one correspondence between prototypes and functional dendrites, it is still sensible to discuss learned latent variables. Under concurrent training with $\alpha=1$ and $\epsilon_\gamma=0.03$ or 0.05  and under any training paradigm at larger prototype perturbations, a dendrite's latent variable might be a combination of prototypes depending on the number of synapses per dendrite. This combination is an intersection, not a union, with the resulting stabilized synapses per dendrite relatively smaller than for a dendrite that responds strongly to only a single prototype. For example,
a dendrite with 64 synapses can recognize the two prototypes that share exactly these 64 out of the 128 active input lines of each of these two prototypes. (Recall that  a dendrite that is selective to a single prototype might have from 90 to nearly 128 synapses.) Given a dendrite that is the intersection of prototype, one can still consider such a dendrite's weight vector  to be a latent variable; this latent variable just does not to any single one of the true latent variables.

Regardless of the existence of the one-to-one correspondence between dendrites and prototypes, the stable point of each dendrite's weight vector can be interpreted as the examples in Table \ref{dw equations}. This table presents four similar modification equations, all of which are compatible with the dendritogenesis  algorithm and two of which, (c) and (d), generated some of the data already presented here. (The list of suitable forms is by no means limited to those illustrated; these are just the ones we have examined.) In all cases where there is conditioning, the conditioning state limits the expectation to the in-class neuron and its most excited dendrite; that is, $z^*_j\cdot \one_{d_{\max}}(d_j)=1$.

\begin{table}[h]
	
	\renewcommand{\arraystretch}{2.0}
	
	\begin{center} 
		\begin{tabular}{l l l l }
			
			\hspace{0.4in}&&Modification equation with CDS \ & Converges to \\
			\hline 
			(a)&\qquad\qquad& $\Delta w_{idj}(t)\propto (X_i(t)-w_{idj})z^*_j(t)\one_{d_{\max}}(d_j) $ &
			$E[X_i |z^*_j(t) \one_{d_{\max}}(d_j)]$\\
			
			(b)&& $\Delta w_{idj}(t)\propto (X_i(t)-E[X_i]-w_{idj})z^*_j(t)\one_{d_{\max}}(d_j) $ &$E[X_i|z^*_j(t) \one_{d_{\max}}(d_j)] - E[X_i]$\\
			
			(c)&& $\Delta w_{idj}(t)\propto (X_i(t)-w_{idj})z^*_j(t)\one_{d_{\max}}(d_j)y_{dj} $ & $\frac{\displaystyle E[X_{i}Y_{dj} | z^*_j(t) \one_{d_{\max}}(d_j)]}{\displaystyle E[Y_{dj}| z^*_j(t) \one_{d_{\max}}(d_j)]}$\\
			
			(d)&& $\Delta w_{idj}(t)\propto (X_i(t)-E[X_i]-w_{idj})z^*_j(t) \one_{d_{\max}}(d_j)y_{dj} $& 
			$\frac{\displaystyle E[X_{i}Y_{dj} |  z^*_j(t) \one_{d_{\max}}(d_j)]}{\displaystyle E[Y_{dj}| z^*_j(t) \one_{d_{\max}}(d_j)]}- E[X_i]$\\
			\hline 
		\end{tabular}
	\end{center}
	\begin{center}	
		Eigen-equation interpretation of stable weight vector \quad ($\wvec_{dj}:=\wvec_{dj}(\infty)$) \\
		\begin{tabular} {l r l l}
			& Matrix$\,\cdot \wvec_{dj}$&$=$&$\wvec_{dj}\cdot$ Scaler\\
			(c')& $E[XX^T| z^*_j(t) \one_{d_{\max}}(d_j)]\cdot \wvec_{dj}$&$=$&$\wvec_{dj}\cdot E[X^T\wvec_{dj}| z^*_j(t) \one_{d_{\max}}(d_j)]$\\
			(d')& $(E[XX^T| z^*_j(t) \one_{d_{\max}}(d_j)]-  E[X]\cdot E[X^T| z^*_j(t) \one_{d_{\max}}(d_j)])\cdot \wvec_{dj}$&$=$&$\wvec_{dj}\cdot E[X^T\wvec_{dj}|z^*_j(t) \one_{d_{\max}}(d_j)] $ \\
			\hline
			
		\end{tabular}
	\end{center}
	\caption{Some of the weight modification algorithms compatible with DC\&SAS algorithm and their stable attractor}
	\label{dw equations}
\end{table}
 
Each of the four modification equations can be interpreted in terms of an encoded conditional statistic to which they converge. Versions (a) and (c) are recommended for segregated training where the mathematical and implemented meaning of $E[X_i]$, which appears in (b) and (d), is,  at the very least, troublesome; for example, we do not know what this expectation means  given the non-stationary nature of the inputs when using segregated training.  Versions (a) and (b) are best understood in terms of  expectations of individual input  line activity conditioned on $z_j^*=1$ and the winner of the local dendrite competition. These two equations, and their encoded statistic, only differ by the unconditional offset $E[X_i]$. On the other hand, the conditioning of the learned expectation in (c) and (d) not only depends on the same two binary conditioning variables but also depends on $Y_{d_j}$, which is a continuous rather than a binary variable. Bringing the $y_{dj}(t)$ multiplier into the modification equation yields a single synapse learned ratio of two statistics, which is less easy to understand without further assumptions. Instead of assuming a particular linear relationship, we interpret the stable expectation of the dendrite's entire weight vector. 

Re-expressing a dendrite's stable synapses  as a weight vector  in the form of (c') and (d') provides an alternative, and for some, a more intuitive statistical interpretation of the synaptic encodings. Specifically, the weight vector converges, in alignment, to the direction of dominant eigenvector (call it $e_{1dj}$) of one of the indicated matrices. Although there is an explicit form of the length of such a $w_{dj}$, here the normalized form ($w_{dj}\div \sum_i w_{idj}  = e_{1dj}\div \sum_i e_{i1dj} $)  is used in the calculation of a $y_{dj}(t)$, so the un-normalized length is of less interest here then when not using such normalization in the excitation calculation. 

Such a stable weight vector, either of case (c') or (d') is a statistic in the sense that each is based on a particular matrix expectation, which is never calculated by a dendrite. The implicit matrix statistic in (c') is the conditional correlation matrix for the one dendrite with conditioning also dependent on class presence. Unfortunately,  as far as we know, in the case of (d') the implicit matrix statistic does not have a name. This matrix  is almost, but not quite, the conditional  covariance matrix of input vector to the dendrite. The problem is the centering term. Note that this term is the product of an unconditional expectation times the appropriate conditional one. To be the conditional covariance matrix requires both of these means to be conditional in the appropriate way, i.e., just like the conditioning of the correlation matrix. To fix this, the obvious modification to the associated $\Delta w$ equation (d) does not work. Using the conditional expectation,$E[X_i| z^*_j(t) \one_{d_{\max}}(d_j)]$, as the offset term instead of the unconditional expectation is incompatible with the DC\&SAS algorithm. Indeed,  subtraction of the conditional expectation $E[X_i|Z_j^*=1]$ is incompatible with SAS even without dendritogenesis.

\subsubsection{The algorithm is in the generative class}

From our knowledge of cortical physiology and anatomy, we speculate that this structure's computations are primarily in the generative class. Thus our algorithms are designed to capture information useful to a generative system. Here generative is meant to be distinguished from discriminative in a manner consistent with  \cite{Rubinstein1997},  \cite{NgJordan2001}, and      \cite{ShalevShwartzS2014} (chapter 24) . That is, as opposed to  a purely discriminative approach that constructs half-space boundaries leading to a tessellation of feature space into class specific regions, here the algorithm  encodes statistics (and from these statistics and certain assumptions, implicitly encodes  probability distributions). There are at least two ways to use such statistics.

An arguably desirable implementation using such  synaptically stored, dendritically localized  conditional expectations  is a  naive Bayes log-odds calculation for each dendrite (see Levy et al 1990 which gives a neural conjecture of this approach for a compartmentalized dendrite to produce a   class supervised neuron calculating as a naive Bayesian).  For example consider the stable synaptic encoding of (a) of Table \ref{dw equations}. For any one dendrite $d$, the set of weights (the dendrite's weight vector) 
$\{ \wvec_{dj} = E[\Xvec_{idj}|\; z^*_j(t) \one_{d_{\max}}(d_j)]\}$ implies a set of conditional $\{0,1\}$ Bernoulli distributions $\{p(X_i=x_i|\; z^*_j(t) \one_{d_{\max}}(d_j )) \,\forall \, i \in dj \}$. This set of probabilities, along with a locally learned class prior,  is enough information to imply a log-odds, naive Bayesian inference at each dendrite. However, such a classic computation is not implemented. Instead, here, a cluster-similarity approach is used, which approach allows the presentation to concentrate on what is innovative here, the extension of SAS to DC\&SAS and the dendritic latent encodings. 

With the cluster similarity approach, each functional dendrite ends up developing its representation of some latent cluster center. For discrimination purposes,  the similarity measure in use resembles, but is not, a cosine comparison. The actual comparison used for dendritic excitation in the competition is a bit of a compromise.   We could have used two different definitions for dendritic excitation, the one  used in the $\Delta w$ equations (2) and (3) corresponding to the excitation defined by equation (1) while  using a different one for the excitation of a dendrite used in excitation competitions.
That is, for the competitions  compute excitation as $\frac  {x(t)^Tw_{dj}} {\sqrt{w_{dj}^Tw_{dj}}}  $ (zero padding of weight vectors as needed). Since this form  uses local information, it is at least biologically plausible, and this form allows the more intuitive cosine comparison for the competitions. (The $x(t)$ lengths can be ignored as this input at full dimension is  the same everywhere.) Instead we opted, mainly for simplicity, that both excitations be the same as the $y_{dj}(t)$ calculation of equation (1).

The list in Table \ref{discrim vs gen} points to the distinctiveness of DC\&SAS particularly in regard to the  more familiar and eminently successful discriminative algorithms of machine learning. In comparison to machine learning methods, it is notable that the  error-free performance here occurs without any quantitative error-correction, without calculating derivatives or matrix inverses, and without any explicit objective-function that is to be optimized.  Because there is no explicit objective function being optimized, there is no need for regularization and its induced biases. Nor is over-training a problem.
As the  algorithm here hews to a biological perspective and motivations, the algorithm is local by nature. Only the local averaging property of a stochastic approximation (e.g., Kushner and Yin, 2003) algorithm is used to encode the necessary statistics.

\begin{table}[ht]
      
\centering
 \caption{\textbf{DC\&SASD differs from discriminative algorithms}.}
	\begin{tabular}{|cl|}
		  \hline

& \hspace*{-1.08 cm} Special aspects of the algorithm \\ 

(1) & Synaptic changes limited to a neuron committing a missed-detection \\
(2) & Synaptogenesis limited to active input lines \\
(3)& CDS allows anti-correlated connections to exist on the same neuron but on different dendrites \\
(4) &Modification of existing synapses selects for inputs  positively correlated with any one class \\
(5)& A dendrite corresponds to a prototype (or the intersection of multiple prototypes) \\ \hline

&  \hspace*{-1.08 cm} Notably absent from algorithm \\ 
(6) & No backpropagation between neurons \\
(7) & No  derivatives or gradients to calculate  \\
(8) & No  explicit optimization  criterion \\
(9) & No   regularization \\ \hline

	\end{tabular}
\label{discrim vs gen}
\end{table}  
  
Although of pedagogical utility, the unmistakable distinctions between generative and discriminative algorithms does not imply that they are mutually exclusive. Indeed, the machine learning community recognizes certain advantages in combining the two approaches and such combinations can also be argued to exist as part of evolved brain function. Based on little more than differences between synaptic modification equations and dominating synaptic interactions, one can argue which brain regions specialize in generative computations, the hippocampal formation and the neocortex, and which regions  appear to use discriminative-style  encoding methods and computations,  the cerebellum and the basal ganglia.  Given the current states of knowledge, we can only await a detailed understanding of how the two styles of neurocomputation should be interfaced.

In closing, the multi-dendrite interpretation presented here is not the last word in translating biological neurons into computation neurons. Just as the McCulloch-Pitts single, one-compartment dendrite serves as a useful simplified neuron both for biological and for AI simulations, so too for the neuron here in the future.  That is, the non-branching, multi-dendrite neuron presented here is a simplification, but it still should  be a useful  computational model of a neuron to incorporate into applications that include   understanding the functionality of biological neural networks and for practical engineering  applications.

\end{large}
\newpage 
\appendix

\renewcommand{\thefigure}{A.\arabic{table}}
\setcounter{table}{0}

\renewcommand{\thefigure}{A.\arabic{figure}}
\setcounter{figure}{0}

\section{Relationship between prototypes and exemplars}

The input vectors used in the 256-D simulations, exemplars, are based on prototype vectors with four groups of consecutive, active input lines, i.e., 128 active input lines out of 256. There are two types of randomization: occlusion (or off-noise) and on-noise. The number of occlusion bits is the number of bits turned off in the set of active input lines (i.e., all input lines in the prototypes that are set to 1). The number of on-noise bits is the number of bits turned on in the set of inactive input lines (i.e., all input lines in the prototypes that are set to 0). The randomization, the bits turned on and the bits turned off, differs from epoch to epoch.

\begin{figure}[H]
	\begin{subfigure}{0.49\textwidth}
		\includegraphics[width=\textwidth]{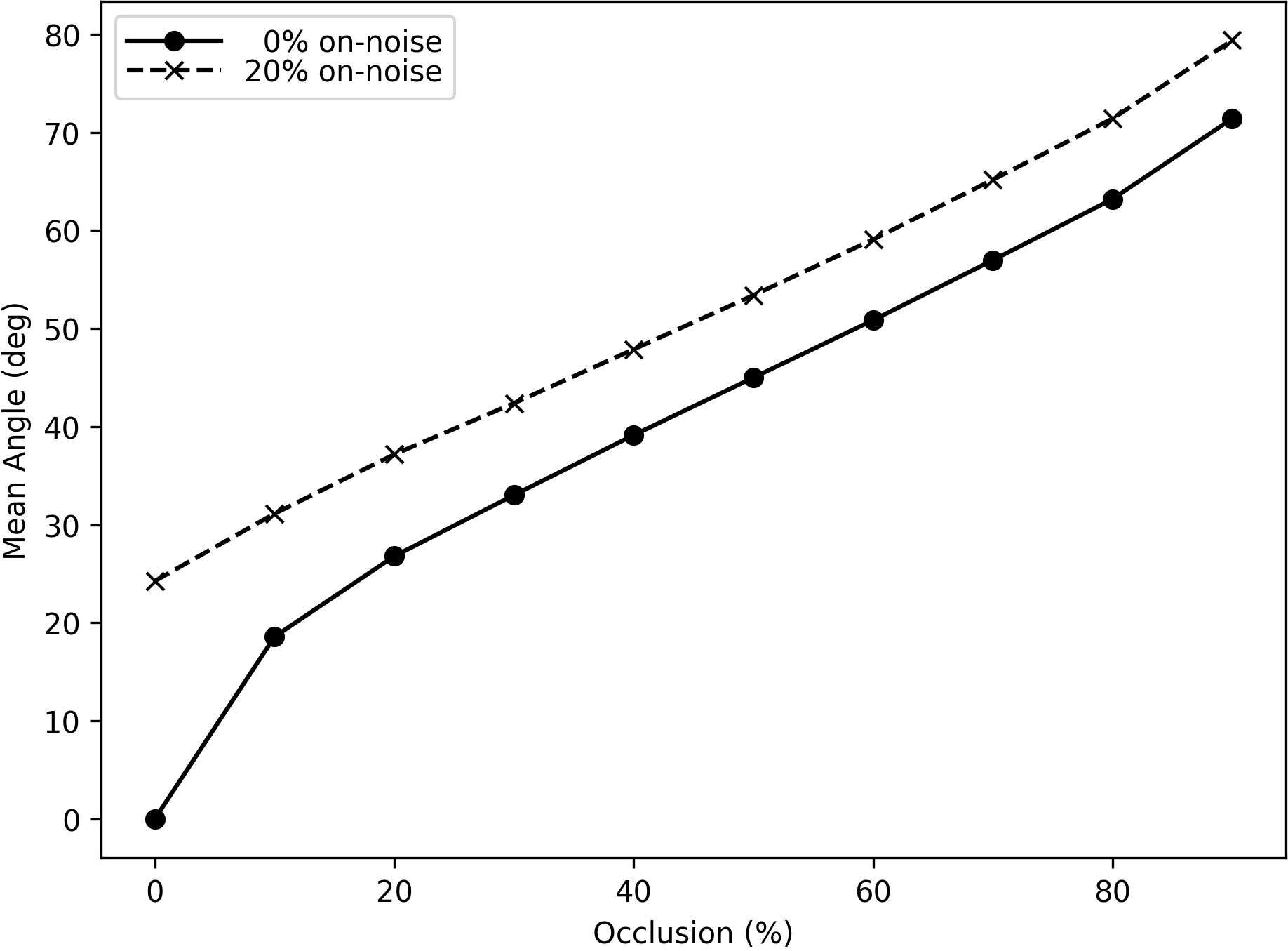}
		\caption{}
	\end{subfigure}
	\hfill
	\begin{subfigure}{0.49\textwidth}
		\includegraphics[width=\textwidth]{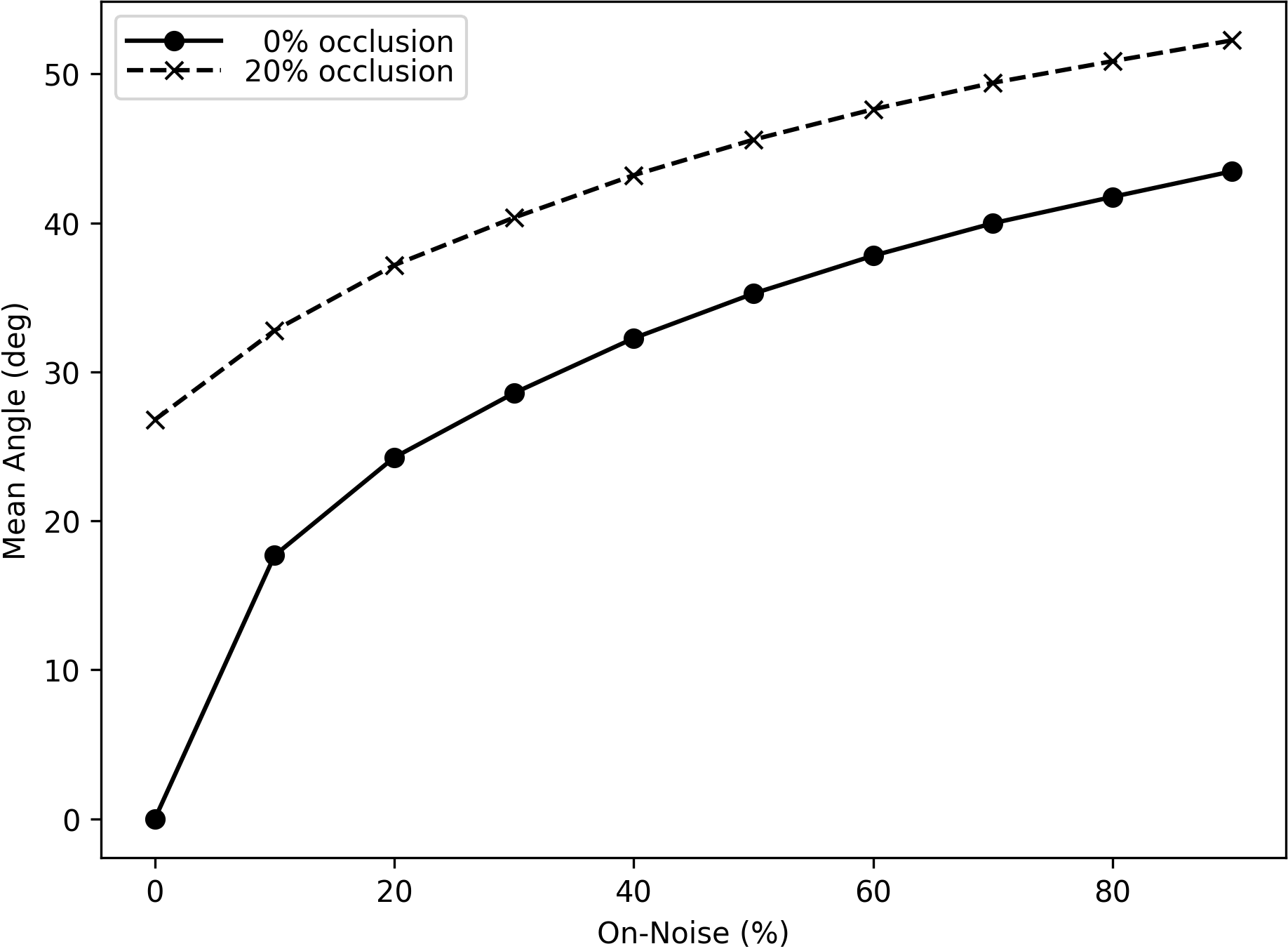}
		\caption{}
	\end{subfigure}
	\centering
	\caption{
		Angles between exemplars and the underlying prototype increase with increasing randomization,
		and the occlusion angles are larger than on-noise angles at high levels of randomization.
		(a) Mean angles as a function of occlusion, (b) mean angles as a function of on-noise.
		Means were computed across 1000 exemplars per prototype.
	}
	\label{exemplars-prototypes-fig}
\end{figure}

With no randomization, the exemplars are equal to the prototype. As the amount of randomization increases, the angles between the exemplars and the underlying prototype increases. Figure~\ref{exemplars-prototypes-fig} graphs how the angles increase as a function of the amount of occlusion in (a) and as a function of the amount of on-noise in (b). A comparison of Figure~\ref{exemplars-prototypes-fig} (a) and (b) shows that occlusion causes larger angular differences from the prototype than on-noise, so it is not surprising that occlusion makes classification slightly more difficult than on-noise with high levels of randomization.

\section{Effect of prototype perturbations on the angle between dendrites}

\begin{figure}[H]
	\centering
	\begin{subfigure}{0.48\textwidth}
		\includegraphics[width=\textwidth]{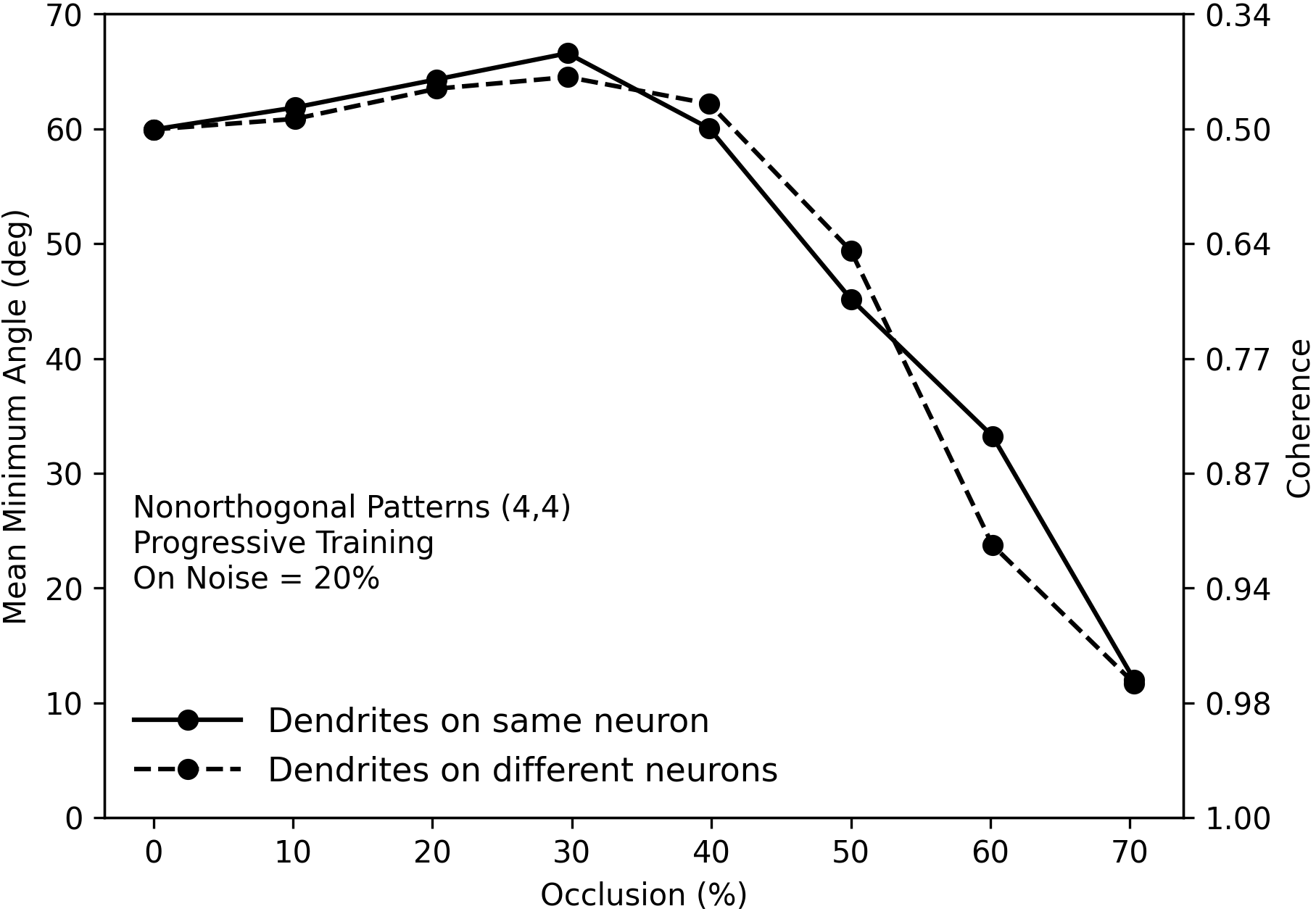}
		\caption{}
	\end{subfigure}
	\hfill
	\begin{subfigure}{0.48\textwidth}
		\includegraphics[width=\textwidth]{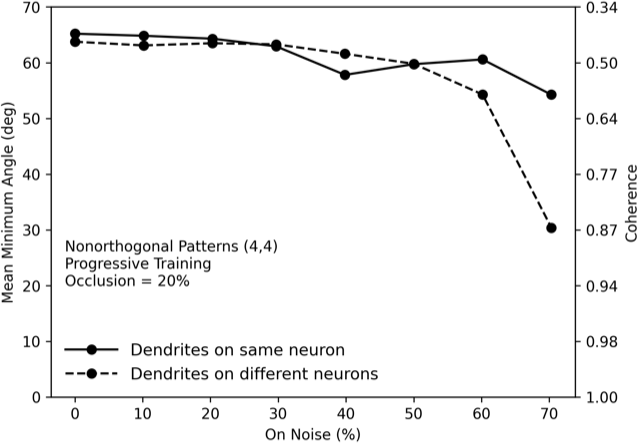}\\
		\caption{}
	\end{subfigure}
	\caption{
		Increasing occlusion decreases the minimum cross-neuron dendritic angle beginning at 40/20 but the decrease with increasing on-noise perturbations is much less significant. Recall that error-rates move away from zero when the occlusion values rise from 40/20 to 50/20 and when on-noise rises from 20/40 to 20/50, but error-rates are more severe for occlusion versus on-noise.
		Based on these graphs, the more severe error-rates for occlusion can be attributed to the smaller minimum angles between dendrites, 
		which can be attributed to the larger angles between the exemplars and prototypes (see Figure~\ref{exemplars-prototypes-fig}).
		The coherence values (i.e., the maximum inner product) are shown on the right y-axes.
		The means are across both neurons and 10 simulations, i.e., across 20 neurons.
		(Progressive training on the 4$|$4 problem set with $\epsilon_w = 0.002$, $\epsilon_\gamma = 0.05$.)
	}
	\label{min-angle-pro}
\end{figure}

Figure~\ref{min-angle-pro} shows how prototype perturbations levels affect the angle between dendrites
for progressive training with the 4$|$4 problem set.
The solid curve in both graphs represent the mean minimum angle between dendrites on the same neuron,
and the dashed curve represents the mean minimum angle between dendrites on different neurons.
The means are computed across both neurons and ten simulations.
The slight increase from 0 to 30\% in Figure~\ref{min-angle-pro} (a) is expected with increasing occlusion;
however, beyond 30\% occlusion, with the 20\% on-noise, the randomization is severe enough to cause the 
minimum angle to decrease. In Figure~\ref{min-angle-pro} (b), the minimum angle does not decrease significantly
until the on-noise exceeds 50\%. Comparing the two graphs leads one to conclude that performance might be
worse for 50/20 versus 20/50, which is confirmed by simulations (see Figure~\ref{error-vs-occ-and-on-fig} - fig 5).

local covariance 
\section{Shedding dynamics with and without CDS}

Connectivity counts are not definitive for stabilization because there could be a ``treadmilling effect'' with shedding and synaptogenesis running at equal rates. Tracking the amount of on-going shedding is revealing because if shedding does not stabilize over time, then synaptogenesis is still creating new synapses, some of which are being shed.

\begin{figure}[H]
	\centering
	\begin{subfigure}{0.32\textwidth}
		\includegraphics[width=\textwidth]{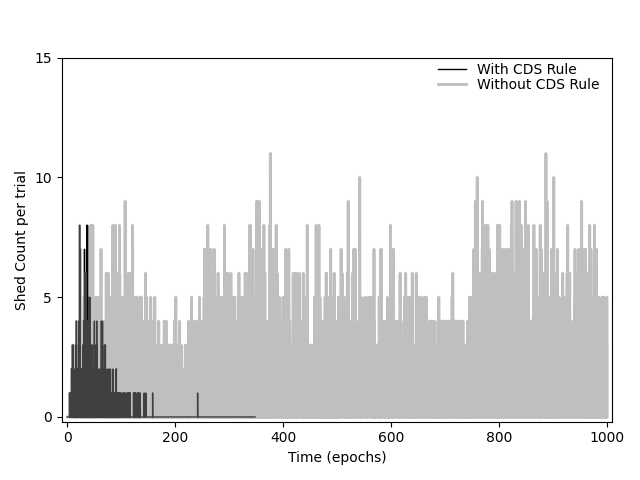}
		\caption{Concurrent}
	\end{subfigure}
	\hfill
	\begin{subfigure}{0.32\textwidth}
		\includegraphics[width=\textwidth]{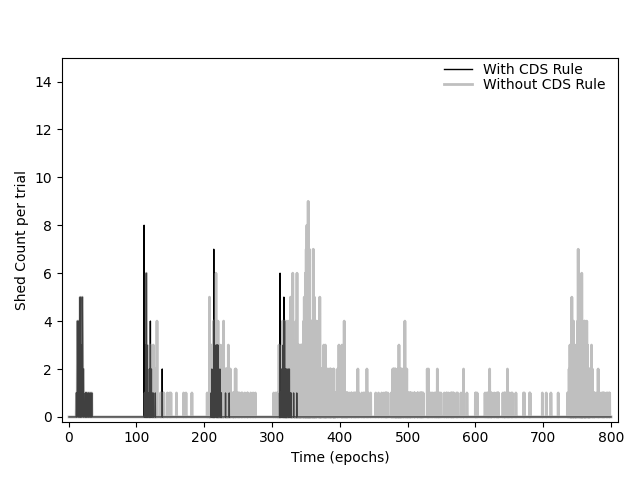}
		\caption{Progressive}
	\end{subfigure}
	\hfill
	\begin{subfigure}{0.32\textwidth}
		\includegraphics[width=\textwidth]{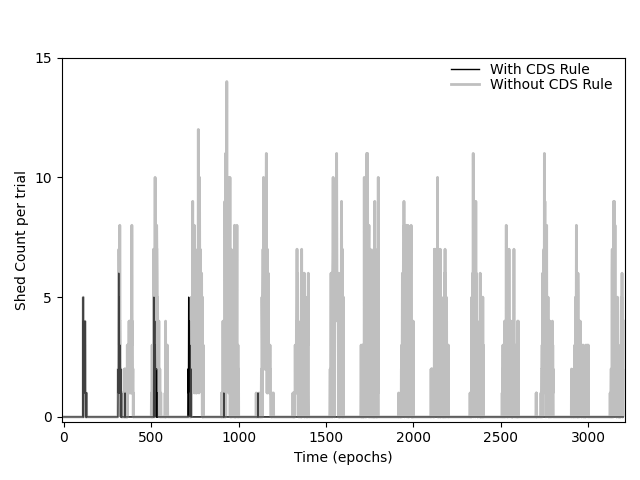}
		\caption{Segregated}
	\end{subfigure}
	\caption{With CDS (black lines), shedding, and implicitly synaptogenesis, stop for all three training methods. 
		Without CDS (light gray lines), shedding persists throughout the simulations; moreover, there is some tendency for increased  shedding with prolonged training.
		With CDS, (black line), the connection counts stabilize by epoch 242 for concurrent training and
		shortly after exemplars based on the last two prototypes are introduced at epoch 301 for progressive training.
		For segregated training, a second pass through the eight phases are necessary for shedding to stop
		(which happens after epoch 1110).
		Note the different timescales on each graph.
		(Problem set $4|4$ with  20/30 prototype perturbations.)
	}
	\label{shed-cds-fig}
\end{figure}

Figure~\ref{shed-cds-fig} graphs shedding over time, with and without CDS, for each of the three training paradigms, using the 4$|$4 problem set with 20/30 prototype-perturbations. For all three training paradigms, shedding without CDS (light gray lines) continues throughout the simulations; thus, connectivity does not stabilize. On the other hand when CDS is present (black lines), shedding stops and connectivity stabilizes. Shedding stops within 242 epochs for concurrent training. For progressive training, shedding stops by epoch 337 shortly after exemplars based on the last two prototypes are introduced at epoch 301 (the beginning of the fourth phase). For segregated training, most of the shedding stops soon after the last training phase, which  begins at epoch 701, phase eight. However, two  shedding events occur during the second pass through the eight training phases.  It is then confirmed that  shedding stops after epoch 1110. 

\section{Time to stability versus phase duration}

\begin{figure}[H]
	\centering
	\includegraphics[width=0.52\textwidth]{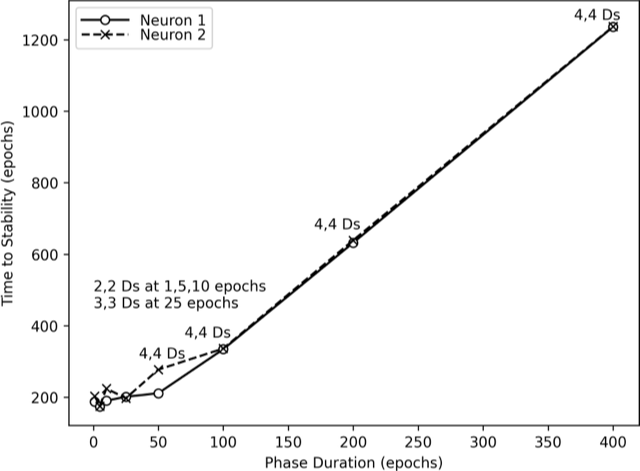}
	\caption{The time it takes connectivity to stabilize increases with increasing phase duration for progressive training.
		With very short phase durations, the connectivity is similar to concurrent training: 
		the solution consists of two dendrites on each neuron. 
		With phase durations of 50 epochs or more, the solution consists of four dendrites on each neuron.
		Classification performance was perfect (0 errors) for all simulations.
	}\label{tts-vs-phase-duration-fig}
\end{figure}

Figure~\ref{tts-vs-phase-duration-fig} graphs the time it takes for the connections to stabilize as a function of the phase duration.
For very short phase durations, the solution consists of two dendrites per neuron which is the same solution observed 
for concurrent training. For phase durations of 50 epochs or more,
the solution consists of four dendrites per neuron. The minimum phase duration with a four-dendrite solution is 50 epochs; 
phase durations longer than 100 epochs have longer development times without any performance benefit.

\section{Parameter settings affect rate of development and convergence}

The purpose of this appendix is to explain how the parameters affect development, connectivity, and classification performance,
and to provide some guidelines for setting parameters.
There are four key parameters: the synaptic modification rate parameter, $\epsilon_w$; 
the parameter controlling the decrement rate of synaptogenesis, $\epsilon_\gamma$;
the averaging rate of the missed detection error signal, $\alpha$;
and the initial value of the synaptogenesis rate parameter, $\gamma_{dj}(0)$.

The synaptic modification rate parameter, $\epsilon_w$, controls the update rate of the synaptic weights on each dendrite. 
If $\epsilon_w$ is too large, the synaptic weights will tend to oscillate and fail to stabilize; 
if $\epsilon_w$ is too small, the development time, i.e., the time it takes for the connectivity to stabilize, may be unnecessarily lengthened. For concurrent training, a value of $\epsilon_w$ of 0.002 tends to yield acceptable results and development times for a variety of circumstances.
For progressive and segregated training, the phase duration and the value of $\epsilon_w$ are recommended to be set so that the weights approach their asymptotic values within each phase. Figures~\ref{conn-stability-fig} (a) and (b) compare the stabilization of connectivity for $\epsilon_w$ values of 0.025 and 0.002, respectively. In Figure\ref{conn-stability-fig}(a) with $\epsilon_w = 0.025$, the connectivity is observed to stabilize within the first 25 epochs of each phase; whereas in Figure\ref{conn-stability-fig}(b) with $\epsilon_w = 0.002$, the connectivity does not stabilize until epoch 582. The classification performance was perfect for both of these simulations, and the errors dropped to 0 by epoch 302 (just after the creation of the fourth dendrite) for both values of $\epsilon_w$.
Note that smaller values of $\epsilon_w$ tend to slow down synaptic shedding, but smaller values can provide better performance at high levels of randomization.

The rate of synaptogenesis is determined by the parameter $\gamma_{dj}(t)$, which is local to each dendrite. This synaptogenesis rate parameter is decremented as the dendrite successfully responds to exemplars associated with the latent prototype the dendrite has learned. A successful response means that the dendrite was the most excited dendrite and the neuron to which the dendrite belongs did not make a missed detection error. The decrement rate of $\gamma_{dj}(t)$, $1 - \epsilon_\gamma$, along with the initial value $\gamma_{dj}(0)$ and the threshold $\theta_\gamma$, determines how many dendritic successes are required before a new dendrite can be created. If $\epsilon_\gamma$ is too large, the existing dendrite(s) may not fully develop before new dendrites are created; if $\epsilon_\gamma$ is too small, existing dendrites may not be able to become selectively sensitive to one or more latent prototypes, or the development time may be unnecessarily delayed. For phase durations of 100 epochs with progressive and segregated training, a value of $\epsilon_\gamma = 0.05$ tends to yield acceptable performance. For concurrent training, a smaller value such as 0.03 helps prevent extra, non-functional dendrites from being created. Note that using a large value of $\epsilon_\gamma$, such as 0.999, essentially requires only a single dendritic success for the creation of a new dendrite; such a large value of $\epsilon_\gamma$ will tend to cause a proliferation of dendrites early in development. We used this approach, along with $\alpha = 0.01$, to cause concurrent training to produce four dendrites per neuron (corresponding to the four latent prototypes per class) with a relatively high ($\approx$80\%) probability.

The missed detection error signal, $\ebar_j^{MD}(t)$, is used for dendritogenesis. Using the instantaneous MD error signal 
(i.e., setting $\alpha = 1$) often yields acceptable performance. However, in environments with high levels of noise, instantaneous missed detections may tend to cause dendritogenesis to create unnecessary dendrites. Using an averaged missed detection error signal (equation~\ref{MD-ra-error-eqn}) tends to avoid the creation of such unnecessary dendrites. The parameter $\alpha$ controls the extent of averaging. We have examined several values of $\alpha$; a value of 0.01 along with $\theta_{MD} = 0.05$ tends to yield acceptable results for a variety of noisy environments.

The remaining four parameters, listed at the bottom of Table~\ref{parameter_table}, were constant across all of the results presented.
The initial value of a synaptic weight, $W_0$, must be set to a value sufficiently larger than the synaptic shedding threshold, $\theta_w$;
otherwise, new synapses will be shed soon after they are formed. As a guideline, the value of $W_0$ should also be set at least 50\% below the estimated asymptotic or final values of the synaptic weights. From a biological viewpoint, the synaptic weights should increase as new patterns are learned so $W_0$ should be set such that the final weight values are greater than $W_0$. However, the algorithm does not require that the final weight values end up above $W_0$, and the algorithm will function as long as $W_0$ is sufficiently larger than $\theta_w$. The dendritogenesis threshold, $\theta_\gamma$, along with $\epsilon_\gamma$ and $\gamma_{dj}(0)$, control how often dendrites can form. When $\gamma_{dj}(t)$ drops below $\theta_\gamma$, dendritogenesis will occur when $\ebar_j^{MD}(t) > \theta_{MD}$. If $\alpha = 1$, the error signal is binary so $\theta_{MD}$ can be set to any value less than 1; when $\alpha < 1$, the value of $\theta_{MD}$ determines the dendritogenesis sensitivity to missed detection errors.

\section{Miscellaneous implementation considerations}

The datasets considered herein consist of binary input vectors with identical $L_1$ norms across a given set of input vectors. While there is nothing inherent in the algorithm that requires binary input vectors or identical $L_1$ norms, training with non-binary vectors or with vectors having significantly different $L_1$ norms can lead to poor performance without preprocessing. The following preprocessing steps are suggested for input vectors that are non-binary or that have significantly different $L_1$ norms.

Angles between pairs of prototypes, pairs of dendrites, and prototype vs dendrite pairs are calculated as arc cosine of the usual vector pair definition of sine including the usual Euclidean distance. When these is a space (dimensional) mismatch between vector pairs, zero padding is used.

\begin{enumerate}
	\item If $x_i(t) < 0$ or $x_i(t) > 1$ for any $i$ or $t$, transform the N-dimensional input vectors to $[0,1]^N$.
	
	\item Let $L_1(t) = \sum_i x_i(t)$ and let $E[L_1(t)]$ be the mean value of $L_1(t)$ across all $t$ ($t$ is assumed to be discretized). If $( \max_t ( L_1(t) ) - \min_t ( L_1(t) ) ) / E[ L_1(t) ] > 0.1$, then transform the input vectors using complement coding or $L_1$ normalization. Complement coding expands the input vector dimensionality by 2; complement coding an N-dimensional input vector forms the 2N-dimensional vector $[x_1, x_2, \ldots, x_N, 1 - x_1, 1 - x_2, \ldots, 1 - x_N]$. $L_1$ normalization consists of dividing the input vectors by $L_1(t)$. Complement coding tends to be a more reliable approach to dealing with input vectors not satisfying the $L_1$ criterion versus $L_1$ normalization, and complement coding can provide improved performance even if $L_1(t)$ is constant because large and small inputs are weighted similarly.
	
\end{enumerate}

\section{Interpreting the different training paradigms as reflecting different stages  of life-long learning}
Recently the machine learning community has become interested in distinctive training  procedures, e.g., mini-batch and transfer learning, that improve learning. One approach to simulating lifelong learning that has been used recently is to create multiple tasks using permuted inputs and then attempt to learn the different tasks. As shown in Section~\ref{two-tasks}, the proposed network is able to learn two tasks with negligible degradation in performance.

In the Results section, progressive training was demonstrated to be superior to concurrent training. Our knowledge that progressive training is superior to concurrent training comes from the behavioral psychology literature (\cite{Alvarado1992}).  This knowledge is put to use in training recurrent, spiking neural networks including showing that the superiority of progressive training over concurrent training \citep{Shon2000}, just as occurs in animals.  In addition to searching for superior training methods in a general context, there are a biologically relevant contexts that encourage comparisons between training methods.

From the dual perspective of biology and life-long learning, we hypothesize that an organism's sensory experiences, from prenatal to adult, will correlate with a change in the manner that new experiences occur. For example prenatal learning will be more like concurrent training with its random presentations. Immediately postnatal, the complexities of the sensed world grows gradually with much repetition, corresponding to progressive training. Finally, as one grows older, new experiences occur in single batches without the repetition of progressive training. After a while one-shot experiences predominate but here we must defer to the complexity of the mammalian brain with its multiple memory systems including a hippocampal formation that supports long-term neocortical encoding  \citep{Levy1989,AugustLevy1996,Scoville1957,Wilson1993,Wilson1994}.

\clearpage
\bibliography{refs}

\begin{thebibliography}{63}
\expandafter\ifx\csname natexlab\endcsname\relax\def\natexlab#1{#1}\fi
\providecommand{\url}[1]{\texttt{#1}}
\providecommand{\href}[2]{#2}
\providecommand{\path}[1]{#1}
\providecommand{\DOIprefix}{doi:}
\providecommand{\ArXivprefix}{arXiv:}
\providecommand{\URLprefix}{URL: }
\providecommand{\Pubmedprefix}{pmid:}
\providecommand{\doi}[1]{\href{http://dx.doi.org/#1}{\path{#1}}}
\providecommand{\Pubmed}[1]{\href{pmid:#1}{\path{#1}}}
\providecommand{\bibinfo}[2]{#2}
\ifx\xfnm\relax \def\xfnm[#1]{\unskip,\space#1}\fi
\bibitem[{Adelsberger-Mangan and Levy(1993)}]{Adelsberger-Mangan1993B}
\bibinfo{author}{Adelsberger-Mangan, D.}, \bibinfo{author}{Levy, W.},
  \bibinfo{year}{1993}.
\newblock \bibinfo{title}{Adaptive synaptogenesis constructs networks that
  maintain information and reduce statistical dependence}.
\newblock \bibinfo{journal}{Biological cybernetics} \bibinfo{volume}{70},
  \bibinfo{pages}{81--7}.
\bibitem[{Adelsberger-Mangan and Levy(1994a)}]{Adelsberger-Mangan1994B}
\bibinfo{author}{Adelsberger-Mangan, D.}, \bibinfo{author}{Levy, W.},
  \bibinfo{year}{1994}a.
\newblock \bibinfo{title}{Adaptive synaptogenesis constructs networks which
  allocate network resources by category frequency}.
\newblock \bibinfo{journal}{1994 IEEE International Conference on Neural
  Networks, IEEEWorld Congress on Computational Intelligence}
  \bibinfo{volume}{4}, \bibinfo{pages}{2245--9}.
\bibitem[{Adelsberger-Mangan and Levy(1994b)}]{Adelsberger-Mangan1994A}
\bibinfo{author}{Adelsberger-Mangan, D.}, \bibinfo{author}{Levy, W.},
  \bibinfo{year}{1994}b.
\newblock \bibinfo{title}{The influence of limited presynaptic growth and
  synapse removal on adaptive synaptogenesis}.
\newblock \bibinfo{journal}{Biological cybernetics} \bibinfo{volume}{71},
  \bibinfo{pages}{461--8}.
\bibitem[{Alvarado and Rudy(1992)}]{Alvarado1992}
\bibinfo{author}{Alvarado, M.C.}, \bibinfo{author}{Rudy, J.W.},
  \bibinfo{year}{1992}.
\newblock \bibinfo{title}{Some properties of configural learning: an
  investigation of the transverse-patterning problem.}
\newblock \bibinfo{journal}{Journal of Experimental Psychology: Animal Behavior
  Processes} \bibinfo{volume}{18}, \bibinfo{pages}{145}.
\bibitem[{Ariav et~al.(2003)Ariav, Polsky and Schiller}]{Ariav2003}
\bibinfo{author}{Ariav, G.}, \bibinfo{author}{Polsky, A.},
  \bibinfo{author}{Schiller, J.}, \bibinfo{year}{2003}.
\newblock \bibinfo{title}{Submillisecond precision of the input-output
  transformation function mediated by fast sodium dendritic spikes in basal
  dendrites of ca1 pyramidal neurons}.
\newblock \bibinfo{journal}{Journal of Neuroscience} \bibinfo{volume}{23},
  \bibinfo{pages}{7750--7758}.
\bibitem[{August and Levy(1996)}]{AugustLevy1996}
\bibinfo{author}{August, D.A.}, \bibinfo{author}{Levy, W.B.},
  \bibinfo{year}{1996}.
\newblock \bibinfo{title}{A simple spike train decoder inspired by the sampling
  theorem}.
\newblock \bibinfo{journal}{Neural computation} \bibinfo{volume}{8},
  \bibinfo{pages}{67--84}.
\bibitem[{Baxter and Levy(2019)}]{BaxterLevy2019}
\bibinfo{author}{Baxter, R.A.}, \bibinfo{author}{Levy, W.B.},
  \bibinfo{year}{2019}.
\newblock \bibinfo{title}{Multilayered neural networks with sparse, data-driven
  connectivity and balanced information and energy efficiency}, in:
  \bibinfo{booktitle}{2019 53rd Annual Conference on Information Sciences and
  Systems (CISS)}, \bibinfo{organization}{IEEE}. pp. \bibinfo{pages}{1--6}.
\bibitem[{Baxter and Levy(2020)}]{BaxterLevy2020}
\bibinfo{author}{Baxter, R.A.}, \bibinfo{author}{Levy, W.B.},
  \bibinfo{year}{2020}.
\newblock \bibinfo{title}{Constructing multilayered neural networks with
  sparse, data-driven connectivity using biologically-inspired, complementary,
  homeostatic mechanisms}.
\newblock \bibinfo{journal}{Neural Networks} \bibinfo{volume}{122},
  \bibinfo{pages}{68--93}.
\bibitem[{Behabadi and Mel(2014)}]{Behabadi2014}
\bibinfo{author}{Behabadi, B.F.}, \bibinfo{author}{Mel, B.W.},
  \bibinfo{year}{2014}.
\newblock \bibinfo{title}{Mechanisms underlying subunit independence in
  pyramidal neuron dendrites}.
\newblock \bibinfo{journal}{Proceedings of the National Academy of Sciences}
  \bibinfo{volume}{111}, \bibinfo{pages}{498--503}.
\bibitem[{Carpenter and Grossberg(1987)}]{Carpenter1987}
\bibinfo{author}{Carpenter, G.A.}, \bibinfo{author}{Grossberg, S.},
  \bibinfo{year}{1987}.
\newblock \bibinfo{title}{Art 2: Self-organization of stable category
  recognition codes for analog input patterns}.
\newblock \bibinfo{journal}{Applied optics} \bibinfo{volume}{26},
  \bibinfo{pages}{4919--4930}.
\bibitem[{Carpenter et~al.(1991)Carpenter, Grossberg and Rosen}]{Carpenter1991}
\bibinfo{author}{Carpenter, G.A.}, \bibinfo{author}{Grossberg, S.},
  \bibinfo{author}{Rosen, D.B.}, \bibinfo{year}{1991}.
\newblock \bibinfo{title}{Fuzzy art: Fast stable learning and categorization of
  analog patterns by an adaptive resonance system}.
\newblock \bibinfo{journal}{Neural networks} \bibinfo{volume}{4},
  \bibinfo{pages}{759--771}.
\bibitem[{Caz{\'e} et~al.(2013)Caz{\'e}, Humphries and Gutkin}]{Caze2013}
\bibinfo{author}{Caz{\'e}, R.D.}, \bibinfo{author}{Humphries, M.},
  \bibinfo{author}{Gutkin, B.}, \bibinfo{year}{2013}.
\newblock \bibinfo{title}{Passive dendrites enable single neurons to compute
  linearly non-separable functions}.
\newblock \bibinfo{journal}{PLoS computational biology} \bibinfo{volume}{9},
  \bibinfo{pages}{e1002867}.
\bibitem[{Colbert et~al.(1994)Colbert, Fall and Levy}]{Colbert1994}
\bibinfo{author}{Colbert, C.}, \bibinfo{author}{Fall, C.},
  \bibinfo{author}{Levy, W.}, \bibinfo{year}{1994}.
\newblock \bibinfo{title}{Using adaptive synaptogenesis to model the
  development of ocular dominance in kitten visual cortex}, in:
  \bibinfo{editor}{Eeckman, F.} (Ed.), \bibinfo{booktitle}{Computation in
  neurons and neural systems}. \bibinfo{publisher}{Kluwer},
  \bibinfo{address}{Boston}, pp. \bibinfo{pages}{139--44}.
\bibitem[{Colbert and Levy(1993)}]{Colbert1993}
\bibinfo{author}{Colbert, C.M.}, \bibinfo{author}{Levy, W.B.},
  \bibinfo{year}{1993}.
\newblock \bibinfo{title}{Long-term potentiation of perforant path synapses in
  hippocampll ca1 in vitro}.
\newblock \bibinfo{journal}{Brain research} \bibinfo{volume}{606},
  \bibinfo{pages}{87--91}.
\bibitem[{Delange et~al.(2021)Delange, Aljundi, Masana, Parisot, Jia,
  Leonardis, Slabaugh and Tuytelaars}]{Delange2021}
\bibinfo{author}{Delange, M.}, \bibinfo{author}{Aljundi, R.},
  \bibinfo{author}{Masana, M.}, \bibinfo{author}{Parisot, S.},
  \bibinfo{author}{Jia, X.}, \bibinfo{author}{Leonardis, A.},
  \bibinfo{author}{Slabaugh, G.}, \bibinfo{author}{Tuytelaars, T.},
  \bibinfo{year}{2021}.
\newblock \bibinfo{title}{A continual learning survey: Defying forgetting in
  classification tasks}.
\newblock \bibinfo{journal}{IEEE Transactions on Pattern Analysis and Machine
  Intelligence} .
\bibitem[{Fauth and Tetzlaff(2016)}]{Fauth2016}
\bibinfo{author}{Fauth, M.}, \bibinfo{author}{Tetzlaff, C.},
  \bibinfo{year}{2016}.
\newblock \bibinfo{title}{Opposing effects of neuronal activity on structural
  plasticity}.
\newblock \bibinfo{journal}{Frontiers in neuroanatomy} \bibinfo{volume}{10},
  \bibinfo{pages}{75}.
\bibitem[{Fauth et~al.(2015)Fauth, W{\"o}rg{\"o}tter and Tetzlaff}]{Fauth2015}
\bibinfo{author}{Fauth, M.}, \bibinfo{author}{W{\"o}rg{\"o}tter, F.},
  \bibinfo{author}{Tetzlaff, C.}, \bibinfo{year}{2015}.
\newblock \bibinfo{title}{The formation of multi-synaptic connections by the
  interaction of synaptic and structural plasticity and their functional
  consequences}.
\newblock \bibinfo{journal}{PLoS computational biology} \bibinfo{volume}{11},
  \bibinfo{pages}{e1004031}.
\bibitem[{F{\"o}ldiak(1990)}]{Foldiak1990}
\bibinfo{author}{F{\"o}ldiak, P.}, \bibinfo{year}{1990}.
\newblock \bibinfo{title}{Forming sparse representations by local anti-hebbian
  learning}.
\newblock \bibinfo{journal}{Biological cybernetics} \bibinfo{volume}{64},
  \bibinfo{pages}{165--170}.
\bibitem[{French(1991)}]{French1991}
\bibinfo{author}{French, R.M.}, \bibinfo{year}{1991}.
\newblock \bibinfo{title}{Using semi-distributed representations to overcome
  catastrophic forgetting in connectionist networks}, in:
  \bibinfo{booktitle}{Proceedings of the 13th Annual Cognitive Science Society
  Conference}, pp. \bibinfo{pages}{173--178}.
\bibitem[{French(1999)}]{French1999}
\bibinfo{author}{French, R.M.}, \bibinfo{year}{1999}.
\newblock \bibinfo{title}{Catastrophic forgetting in connectionist networks}.
\newblock \bibinfo{journal}{Trends in Cognitive Sciences} \bibinfo{volume}{3},
  \bibinfo{pages}{128--135}.
\bibitem[{Grossberg(1976)}]{Grossberg1976}
\bibinfo{author}{Grossberg, S.}, \bibinfo{year}{1976}.
\newblock \bibinfo{title}{Adaptive pattern classification and universal
  recoding: I. parallel development and coding of neural feature detectors}.
\newblock \bibinfo{journal}{Biological cybernetics} \bibinfo{volume}{23},
  \bibinfo{pages}{121--134}.
\bibitem[{Jadi et~al.(2012)Jadi, Polsky, Schiller and Mel}]{Jadi2012}
\bibinfo{author}{Jadi, M.}, \bibinfo{author}{Polsky, A.},
  \bibinfo{author}{Schiller, J.}, \bibinfo{author}{Mel, B.W.},
  \bibinfo{year}{2012}.
\newblock \bibinfo{title}{Location-dependent effects of inhibition on local
  spiking in pyramidal neuron dendrites}.
\newblock \bibinfo{journal}{PLoS computational biology} \bibinfo{volume}{8}.
\bibitem[{Ju et~al.(2017)Ju, Colbert and Levy}]{JuColbertLevy2017}
\bibinfo{author}{Ju, H.}, \bibinfo{author}{Colbert, C.M.},
  \bibinfo{author}{Levy, W.B.}, \bibinfo{year}{2017}.
\newblock \bibinfo{title}{Limited synapse overproduction can speed development
  but sometimes with long-term energy and discrimination penalties}.
\newblock \bibinfo{journal}{PLOS Computational Biology} \bibinfo{volume}{13}.
\newblock \DOIprefix\doi{http://dx.doi.org/10.1371/journal.pcbi.1005750}.
\bibitem[{Kirkpatrick et~al.(2017)Kirkpatrick, Pascanu, Rabinowitz, Veness,
  Desjardins, Rusu, Milan, Quan, Ramalho, Grabska-Barwinska
  et~al.}]{Kirkpatrick2017}
\bibinfo{author}{Kirkpatrick, J.}, \bibinfo{author}{Pascanu, R.},
  \bibinfo{author}{Rabinowitz, N.}, \bibinfo{author}{Veness, J.},
  \bibinfo{author}{Desjardins, G.}, \bibinfo{author}{Rusu, A.A.},
  \bibinfo{author}{Milan, K.}, \bibinfo{author}{Quan, J.},
  \bibinfo{author}{Ramalho, T.}, \bibinfo{author}{Grabska-Barwinska, A.},
  et~al., \bibinfo{year}{2017}.
\newblock \bibinfo{title}{Overcoming catastrophic forgetting in neural
  networks}.
\newblock \bibinfo{journal}{Proceedings of the national academy of sciences}
  \bibinfo{volume}{114}, \bibinfo{pages}{3521--3526}.
\bibitem[{Kushner(2010)}]{Kushner2010}
\bibinfo{author}{Kushner, H.}, \bibinfo{year}{2010}.
\newblock \bibinfo{title}{Stochastic approximation: a survey}.
\newblock \bibinfo{journal}{Wiley Interdisciplinary Reviews: Computational
  Statistics} \bibinfo{volume}{2}, \bibinfo{pages}{87--96}.
\bibitem[{Kushner and Yin(2003)}]{Kushner2003}
\bibinfo{author}{Kushner, H.}, \bibinfo{author}{Yin, G.G.},
  \bibinfo{year}{2003}.
\newblock \bibinfo{title}{Stochastic approximation and recursive algorithms and
  applications}. volume~\bibinfo{volume}{35}.
\newblock \bibinfo{publisher}{Springer Science \& Business Media}.
\bibitem[{Levy and Colbert(1991)}]{LevyColbert1991}
\bibinfo{author}{Levy, W.}, \bibinfo{author}{Colbert, C.},
  \bibinfo{year}{1991}.
\newblock \bibinfo{title}{Adaptive synaptogenesis can complement associative
  potentiation/depression}, in: \bibinfo{editor}{Commons, M.L.},
  \bibinfo{editor}{Grossberg, S.}, \bibinfo{editor}{Staddon, J.E.R.} (Eds.),
  \bibinfo{booktitle}{Neural Network Models of Conditioning: Quantitative
  Analysis of Behavior}. \bibinfo{publisher}{Erlbaum},
  \bibinfo{address}{Hillsdale,NJ}, pp. \bibinfo{pages}{53--68}.
\bibitem[{Levy et~al.(1990)Levy, Colbert and Desmond}]{Levy1990}
\bibinfo{author}{Levy, W.}, \bibinfo{author}{Colbert, C.},
  \bibinfo{author}{Desmond, N.}, \bibinfo{year}{1990}.
\newblock \bibinfo{title}{Elemental adaptive processes of neurons and synapses:
  a statistical/computational perspective}, in: \bibinfo{editor}{Gluck, M.},
  \bibinfo{editor}{Rumelhart, D.} (Eds.), \bibinfo{booktitle}{Neuroscience and
  ConnectionistTheory}. \bibinfo{publisher}{Erlbaum},
  \bibinfo{address}{Hillsdale, NJ}, pp. \bibinfo{pages}{187--235}.
\bibitem[{Levy and Desmond(1985)}]{Levy1985}
\bibinfo{author}{Levy, W.}, \bibinfo{author}{Desmond, N.},
  \bibinfo{year}{1985}.
\newblock \bibinfo{title}{The rules of elemental synaptic plasticity}, in:
  \bibinfo{editor}{Levy, W.}, \bibinfo{editor}{Anderson, J.},
  \bibinfo{editor}{S, L.} (Eds.), \bibinfo{booktitle}{Synaptic modification,
  neuron selectivity, and nervous system organization}.
  \bibinfo{publisher}{Erlbaum}, \bibinfo{address}{Hillsdale,NJ}, pp.
  \bibinfo{pages}{105--121}.
\bibitem[{Levy and Geman(1982)}]{LevyGeman1982}
\bibinfo{author}{Levy, W.}, \bibinfo{author}{Geman, S.}, \bibinfo{year}{1982}.
\newblock \bibinfo{title}{Limit behavior of experimentally derived synaptic
  modification rules}.
\newblock \bibinfo{journal}{Reports in pattern analysis} .
\bibitem[{Levy et~al.(2016)Levy, Ju, Baxter and Colbert}]{Levy2016}
\bibinfo{author}{Levy, W.}, \bibinfo{author}{Ju, H.}, \bibinfo{author}{Baxter,
  R.}, \bibinfo{author}{Colbert, C.}, \bibinfo{year}{2016}.
\newblock \bibinfo{title}{Controlling information flow and energy use via
  adaptive synaptogenesis}, in: \bibinfo{booktitle}{Information Science and
  Systems (CISS), 2016 Annual Conference on}, \bibinfo{organization}{IEEE}. pp.
  \bibinfo{pages}{535--538}.
\bibitem[{Levy and Steward(1983)}]{Levy1983}
\bibinfo{author}{Levy, W.}, \bibinfo{author}{Steward, O.},
  \bibinfo{year}{1983}.
\newblock \bibinfo{title}{Temporal contiguity requirements for long-term
  associative potentiation/depression in the hippocampus}.
\newblock \bibinfo{journal}{Neuroscience} \bibinfo{volume}{8},
  \bibinfo{pages}{791--797}.
\bibitem[{Levy(1989)}]{Levy1989}
\bibinfo{author}{Levy, W.B.}, \bibinfo{year}{1989}.
\newblock \bibinfo{title}{A computational approach to hippocampal function},
  in: \bibinfo{editor}{Hawkins, R.D.}, \bibinfo{editor}{Bower, G.H.} (Eds.),
  \bibinfo{booktitle}{Psychology of learning and motivation}.
  \bibinfo{publisher}{Elsevier}. volume~\bibinfo{volume}{23}, pp.
  \bibinfo{pages}{243--305}.
\bibitem[{Levy and Calvert(2021)}]{Levy2021}
\bibinfo{author}{Levy, W.B.}, \bibinfo{author}{Calvert, V.G.},
  \bibinfo{year}{2021}.
\newblock \bibinfo{title}{Communication consumes 35 times more energy than
  computation in the human cortex, but both costs are needed to predict synapse
  number}.
\newblock \bibinfo{journal}{Proceedings of the National Academy of Sciences}
  \bibinfo{volume}{118}, \bibinfo{pages}{e2008173118}.
\bibitem[{Levy and Steward(1979)}]{LevySteward1979}
\bibinfo{author}{Levy, W.B.}, \bibinfo{author}{Steward, O.},
  \bibinfo{year}{1979}.
\newblock \bibinfo{title}{Synapses as associative memory elements in the
  hippocampal formation}.
\newblock \bibinfo{journal}{Brain research} \bibinfo{volume}{175},
  \bibinfo{pages}{233--245}.
\bibitem[{Limbacher and Legenstein(2020)}]{Limbacher2020}
\bibinfo{author}{Limbacher, T.}, \bibinfo{author}{Legenstein, R.},
  \bibinfo{year}{2020}.
\newblock \bibinfo{title}{Emergence of stable synaptic clusters on dendrites
  through synaptic rewiring}.
\newblock \bibinfo{journal}{Frontiers in computational neuroscience} ,
  \bibinfo{pages}{57}.
\bibitem[{Ljung and Box(1978)}]{Ljung1978}
\bibinfo{author}{Ljung, G.M.}, \bibinfo{author}{Box, G.E.},
  \bibinfo{year}{1978}.
\newblock \bibinfo{title}{On a measure of lack of fit in time series models}.
\newblock \bibinfo{journal}{Biometrika} \bibinfo{volume}{65},
  \bibinfo{pages}{297--303}.
\bibitem[{Masse et~al.(2018)Masse, Grant and Freedman}]{Masse2018}
\bibinfo{author}{Masse, N.Y.}, \bibinfo{author}{Grant, G.D.},
  \bibinfo{author}{Freedman, D.J.}, \bibinfo{year}{2018}.
\newblock \bibinfo{title}{Alleviating catastrophic forgetting using
  context-dependent gating and synaptic stabilization}.
\newblock \bibinfo{journal}{Proceedings of the National Academy of Sciences}
  \bibinfo{volume}{115}, \bibinfo{pages}{E10467--E10475}.
\bibitem[{McClelland et~al.(1995)McClelland, McNaughton and
  O'Reilly}]{McClelland1995}
\bibinfo{author}{McClelland, J.L.}, \bibinfo{author}{McNaughton, B.L.},
  \bibinfo{author}{O'Reilly, R.C.}, \bibinfo{year}{1995}.
\newblock \bibinfo{title}{Why there are complementary learning systems in the
  hippocampus and neocortex: insights from the successes and failures of
  connectionist models of learning and memory}.
\newblock \bibinfo{journal}{Psychological review} \bibinfo{volume}{102},
  \bibinfo{pages}{419}.
\bibitem[{McCloskey and Cohen(1989)}]{McCloskey1989}
\bibinfo{author}{McCloskey, M.}, \bibinfo{author}{Cohen, N.J.},
  \bibinfo{year}{1989}.
\newblock \bibinfo{title}{Catastrophic interference in connectionist networks:
  The sequential learning problem}, in: \bibinfo{booktitle}{Psychology of
  learning and motivation}. \bibinfo{publisher}{Elsevier}.
  volume~\bibinfo{volume}{24}, pp. \bibinfo{pages}{109--165}.
\bibitem[{Mel(1991)}]{Mel1991}
\bibinfo{author}{Mel, B.}, \bibinfo{year}{1991}.
\newblock \bibinfo{title}{The clusteron: toward a simple abstraction for a
  complex neuron}.
\newblock \bibinfo{journal}{Advances in neural information processing systems}
  \bibinfo{volume}{4}.
\bibitem[{Mel(1994)}]{Mel1994}
\bibinfo{author}{Mel, B.W.}, \bibinfo{year}{1994}.
\newblock \bibinfo{title}{Information processing in dendritic trees}.
\newblock \bibinfo{journal}{Neural computation} \bibinfo{volume}{6},
  \bibinfo{pages}{1031--1085}.
\bibitem[{Mel(1999)}]{Mel1999}
\bibinfo{author}{Mel, B.W.}, \bibinfo{year}{1999}.
\newblock \bibinfo{title}{Why have dendrites? a computational perspective}, in:
  \bibinfo{editor}{Stuart, G.}, \bibinfo{editor}{Spruston, N.},
  \bibinfo{editor}{H{\"a}usser, M.} (Eds.), \bibinfo{booktitle}{Dendrites}.
  \bibinfo{publisher}{Oxford University Press}.
\bibitem[{Mel et~al.(2017)Mel, Schiller and Poirazi}]{Mel2017}
\bibinfo{author}{Mel, B.W.}, \bibinfo{author}{Schiller, J.},
  \bibinfo{author}{Poirazi, P.}, \bibinfo{year}{2017}.
\newblock \bibinfo{title}{Synaptic plasticity in dendrites: complications and
  coping strategies}.
\newblock \bibinfo{journal}{Current opinion in neurobiology}
  \bibinfo{volume}{43}, \bibinfo{pages}{177--186}.
\bibitem[{Ng and Jordan(2001)}]{NgJordan2001}
\bibinfo{author}{Ng, A.}, \bibinfo{author}{Jordan, M.}, \bibinfo{year}{2001}.
\newblock \bibinfo{title}{On discriminative vs. generative classifiers: A
  comparison of logistic regression and naive bayes}.
\newblock \bibinfo{journal}{Advances in neural information processing systems}
  \bibinfo{volume}{14}.
\bibitem[{Petanjek et~al.(2008)Petanjek, Juda{\v{s}}, Kostovi{\'c} and
  Uylings}]{Petanjek2008}
\bibinfo{author}{Petanjek, Z.}, \bibinfo{author}{Juda{\v{s}}, M.},
  \bibinfo{author}{Kostovi{\'c}, I.}, \bibinfo{author}{Uylings, H.B.},
  \bibinfo{year}{2008}.
\newblock \bibinfo{title}{Lifespan alterations of basal dendritic trees of
  pyramidal neurons in the human prefrontal cortex: a layer-specific pattern}.
\newblock \bibinfo{journal}{Cerebral cortex} \bibinfo{volume}{18},
  \bibinfo{pages}{915--929}.
\bibitem[{Petanjek et~al.(2011)Petanjek, Juda{\v{s}}, {\v{S}}imi{\'c},
  Ra{\v{s}}in, Uylings, Rakic and Kostovi{\'c}}]{Petanjek2011}
\bibinfo{author}{Petanjek, Z.}, \bibinfo{author}{Juda{\v{s}}, M.},
  \bibinfo{author}{{\v{S}}imi{\'c}, G.}, \bibinfo{author}{Ra{\v{s}}in, M.R.},
  \bibinfo{author}{Uylings, H.B.}, \bibinfo{author}{Rakic, P.},
  \bibinfo{author}{Kostovi{\'c}, I.}, \bibinfo{year}{2011}.
\newblock \bibinfo{title}{Extraordinary neoteny of synaptic spines in the human
  prefrontal cortex}.
\newblock \bibinfo{journal}{Proceedings of the National Academy of Sciences}
  \bibinfo{volume}{108}, \bibinfo{pages}{13281--13286}.
\bibitem[{Petanjek et~al.(2019)Petanjek, Sedmak, D{\v{z}}aja, Hladnik,
  Ra{\v{s}}in and Jovanov-Milosevic}]{Petanjek2019}
\bibinfo{author}{Petanjek, Z.}, \bibinfo{author}{Sedmak, D.},
  \bibinfo{author}{D{\v{z}}aja, D.}, \bibinfo{author}{Hladnik, A.},
  \bibinfo{author}{Ra{\v{s}}in, M.R.}, \bibinfo{author}{Jovanov-Milosevic, N.},
  \bibinfo{year}{2019}.
\newblock \bibinfo{title}{The protracted maturation of associative layer iiic
  pyramidal neurons in the human prefrontal cortex during childhood: a major
  role in cognitive development and selective alteration in autism}.
\newblock \bibinfo{journal}{Frontiers in psychiatry} \bibinfo{volume}{10},
  \bibinfo{pages}{122}.
\bibitem[{Poirazi et~al.(2003)Poirazi, Brannon and Mel}]{Poirazi2003}
\bibinfo{author}{Poirazi, P.}, \bibinfo{author}{Brannon, T.},
  \bibinfo{author}{Mel, B.W.}, \bibinfo{year}{2003}.
\newblock \bibinfo{title}{Pyramidal neuron as two-layer neural network}.
\newblock \bibinfo{journal}{Neuron} \bibinfo{volume}{37},
  \bibinfo{pages}{989--999}.
\bibitem[{Poirazi and Mel(2001)}]{Poirazi2001}
\bibinfo{author}{Poirazi, P.}, \bibinfo{author}{Mel, B.W.},
  \bibinfo{year}{2001}.
\newblock \bibinfo{title}{Impact of active dendrites and structural plasticity
  on the memory capacity of neural tissue}.
\newblock \bibinfo{journal}{Neuron} \bibinfo{volume}{29},
  \bibinfo{pages}{779--796}.
\bibitem[{Polsky et~al.(2004)Polsky, Mel and Schiller}]{Polsky2004}
\bibinfo{author}{Polsky, A.}, \bibinfo{author}{Mel, B.W.},
  \bibinfo{author}{Schiller, J.}, \bibinfo{year}{2004}.
\newblock \bibinfo{title}{Computational subunits in thin dendrites of pyramidal
  cells}.
\newblock \bibinfo{journal}{Nature neuroscience} \bibinfo{volume}{7},
  \bibinfo{pages}{621--627}.
\bibitem[{Rubinstein et~al.(1997)Rubinstein, Hastie et~al.}]{Rubinstein1997}
\bibinfo{author}{Rubinstein, Y.D.}, \bibinfo{author}{Hastie, T.}, et~al.,
  \bibinfo{year}{1997}.
\newblock \bibinfo{title}{Discriminative vs informative learning.}, in:
  \bibinfo{booktitle}{KDD}, pp. \bibinfo{pages}{49--53}.
\bibitem[{Schiller et~al.(2000)Schiller, Major, Koester and
  Schiller}]{Schiller2000}
\bibinfo{author}{Schiller, J.}, \bibinfo{author}{Major, G.},
  \bibinfo{author}{Koester, H.J.}, \bibinfo{author}{Schiller, Y.},
  \bibinfo{year}{2000}.
\newblock \bibinfo{title}{Nmda spikes in basal dendrites of cortical pyramidal
  neurons}.
\newblock \bibinfo{journal}{Nature} \bibinfo{volume}{404},
  \bibinfo{pages}{285--289}.
\bibitem[{Scholl et~al.(2021)Scholl, Rule and Hennig}]{Scholl2021}
\bibinfo{author}{Scholl, C.}, \bibinfo{author}{Rule, M.E.},
  \bibinfo{author}{Hennig, M.H.}, \bibinfo{year}{2021}.
\newblock \bibinfo{title}{The information theory of developmental pruning:
  Optimizing global network architectures using local synaptic rules}.
\newblock \bibinfo{journal}{PLoS computational biology} \bibinfo{volume}{17},
  \bibinfo{pages}{e1009458}.
\bibitem[{Scoville and Milner(1957)}]{Scoville1957}
\bibinfo{author}{Scoville, W.B.}, \bibinfo{author}{Milner, B.},
  \bibinfo{year}{1957}.
\newblock \bibinfo{title}{Loss of recent memory after bilateral hippocampal
  lesions}.
\newblock \bibinfo{journal}{Journal of neurology, neurosurgery, and psychiatry}
  \bibinfo{volume}{20}, \bibinfo{pages}{11}.
\bibitem[{Shalev-Shwartz and Ben-David(2014)}]{ShalevShwartzS2014}
\bibinfo{author}{Shalev-Shwartz, S.}, \bibinfo{author}{Ben-David, S.},
  \bibinfo{year}{2014}.
\newblock \bibinfo{title}{Understanding machine learning: From theory to
  algorithms}.
\newblock \bibinfo{publisher}{Cambridge university press}.
\bibitem[{Shon et~al.(2000)Shon, Wu and Levy}]{Shon2000}
\bibinfo{author}{Shon, A.P.}, \bibinfo{author}{Wu, X.}, \bibinfo{author}{Levy,
  W.B.}, \bibinfo{year}{2000}.
\newblock \bibinfo{title}{Using computational simulations to discover optimal
  training paradigms}.
\newblock \bibinfo{journal}{Neurocomputing} \bibinfo{volume}{32},
  \bibinfo{pages}{995--1002}.
\bibitem[{Tazerart et~al.(2020)Tazerart, Mitchell, Miranda-Rottmann and
  Araya}]{Tazerart2020}
\bibinfo{author}{Tazerart, S.}, \bibinfo{author}{Mitchell, D.E.},
  \bibinfo{author}{Miranda-Rottmann, S.}, \bibinfo{author}{Araya, R.},
  \bibinfo{year}{2020}.
\newblock \bibinfo{title}{A spike-timing-dependent plasticity rule for
  dendritic spines}.
\newblock \bibinfo{journal}{Nature communications} \bibinfo{volume}{11},
  \bibinfo{pages}{1--16}.
\bibitem[{Thomas et~al.(2015)Thomas, Blalock and Levy}]{Thomas2015}
\bibinfo{author}{Thomas, B.T.}, \bibinfo{author}{Blalock, D.W.},
  \bibinfo{author}{Levy, W.B.}, \bibinfo{year}{2015}.
\newblock \bibinfo{title}{Adaptive synaptogenesis constructs neural codes that
  benefit discrimination}.
\newblock \bibinfo{journal}{PLOS Computational Biology} \bibinfo{volume}{11},
  \bibinfo{pages}{e1004299}.
\newblock \DOIprefix\doi{10.1371/journal.pcbi.1004299}.
\bibitem[{White et~al.(1990)White, Levy and Steward}]{White1990}
\bibinfo{author}{White, G.}, \bibinfo{author}{Levy, W.B.},
  \bibinfo{author}{Steward, O.}, \bibinfo{year}{1990}.
\newblock \bibinfo{title}{Spatial overlap between populations of synapses
  determines the extent of their associative interaction during the induction
  of long-term potentiation and depression}.
\newblock \bibinfo{journal}{Journal of neurophysiology} \bibinfo{volume}{64},
  \bibinfo{pages}{1186--1198}.
\bibitem[{Wilson and McNaughton(1993)}]{Wilson1993}
\bibinfo{author}{Wilson, M.A.}, \bibinfo{author}{McNaughton, B.L.},
  \bibinfo{year}{1993}.
\newblock \bibinfo{title}{Dynamics of the hippocampal ensemble code for space}.
\newblock \bibinfo{journal}{Science} \bibinfo{volume}{261},
  \bibinfo{pages}{1055--1058}.
\bibitem[{Wilson and McNaughton(1994)}]{Wilson1994}
\bibinfo{author}{Wilson, M.A.}, \bibinfo{author}{McNaughton, B.L.},
  \bibinfo{year}{1994}.
\newblock \bibinfo{title}{Reactivation of hippocampal ensemble memories during
  sleep}.
\newblock \bibinfo{journal}{Science} \bibinfo{volume}{265},
  \bibinfo{pages}{676--679}.
\bibitem[{Zenke et~al.(2017)Zenke, Poole and Ganguli}]{Zenke2017}
\bibinfo{author}{Zenke, F.}, \bibinfo{author}{Poole, B.},
  \bibinfo{author}{Ganguli, S.}, \bibinfo{year}{2017}.
\newblock \bibinfo{title}{Continual learning through synaptic intelligence},
  in: \bibinfo{booktitle}{International Conference on Machine Learning},
  \bibinfo{organization}{PMLR}. pp. \bibinfo{pages}{3987--3995}.

\end{thebibliography}

\end{document}